\documentclass[conference]{IEEEtran}
\IEEEoverridecommandlockouts

\usepackage{amsmath}
\usepackage{bm}
\usepackage{graphicx}
\usepackage{booktabs}
\usepackage{tabularx}
\usepackage{adjustbox}
\usepackage[ruled,linesnumbered]{algorithm2e}
\usepackage[normalem]{ulem}
\usepackage{makecell}
\usepackage{amsmath}
\usepackage{url}
\usepackage{cite}
\usepackage{enumitem}
\usepackage{multirow}
\usepackage{tikz}
\usepackage{afterpage}
\usepackage{subfig}
\usepackage{xcolor}
\usepackage[english]{babel}
\newtheorem{theorem}{Theorem}

\makeatletter
\@addtoreset{footnote}{respfootnote}
\@addtoreset{figure}{respfigure}
\@addtoreset{table}{resptable}
\makeatother

\newcommand{\SystemName}{LoHan}

\newcommand\blfootnote[1]{%
  \begingroup
  \renewcommand\thefootnote{}\footnote{#1}%
  \addtocounter{footnote}{-1}%
  \endgroup
}
\newcommand\narrow[1]{\scalebox{.93}[1.0]{#1}}
\newcommand\vnarrow[1]{\scalebox{.88}[1.0]{#1}}

\renewcommand{\textasciitilde}{\raisebox{0.5ex}{\texttildelow}} 

\begin{document}

\title{\SystemName: \underline{Lo}w-Cost \underline{H}igh-Perform\underline{an}ce Framework to Fine-Tune 100B Model on a Consumer GPU}

\author{
    \IEEEauthorblockN{Changyue Liao$^{*}$\dag, Mo Sun$^{*}$\dag, Zihan Yang$^{*}$\dag, Jun Xie\dag, Kaiqi Chen\dag, Binhang Yuan\ddag, Fei Wu\dag, Zeke Wang\dag}
    \IEEEauthorblockA{\dag Zhejiang University, China}
    \IEEEauthorblockA{\ddag HKUST, China}
}

\maketitle

\thispagestyle{plain}   
\pagestyle{plain}   

\blfootnote{$^*$~Equal contribution.}

\vspace{-2ex}
\begin{abstract}

Nowadays, AI researchers become more and more interested in fine-tuning a pre-trained LLM, whose size has grown to up to over 100B parameters, for their downstream tasks. One approach to fine-tune such huge models is to aggregate device memory from many GPUs. However, this approach introduces prohibitive costs for most data scientists with a limited budget for high-end GPU servers. In this paper, we focus on LLM fine-tuning on a single consumer-grade GPU in a commodity server with limited main memory capacity, which is accessible to most AI researchers. 
In such a scenario, existing offloading-based methods fail to fine-tune an LLM efficiently due to a lack of holistic intra-server tensor movement management. 
To this end, we present \SystemName{}, a low-cost, high-performance deep learning training framework that enables efficient 100B-scale model fine-tuning on a commodity server with a consumer-grade GPU and limited main memory capacity. The key idea is to add holistic offloading traffic as an optimization dimension for 1)~active gradient offloading, and 2)~holistic traffic-aware activation swapping mechanism. 
The experimental results show that 1)~\SystemName{} is the first to fine-tune a 175B model on an RTX 4090 and 256 GB main memory, 2)~\SystemName{} achieves 2.32$\times$ throughput than the state-of-the-art baselines when fine-tuning a small 13B model, and 3)~\SystemName{} enables a cheap low-end consumer GPU to have higher cost-effectiveness than a DGX\nobreakdash-A100 cluster when fine-tuning a 175B model. 
\end{abstract}


\vspace{-1px}
\section{Introduction}
\label{sec:intro}

Large language models (LLMs) have achieved impressive accuracy in natural language processing jobs~\cite{bert, opt, gpt2, gpt3, bloom} and data management tasks~\cite{trummer2021case, fernandez2023large}. There is a strong demand for data scientists to fine-tune a pre-trained LLM to be used for downstream AI tasks~\cite{sft, llama2}.
However, the model sizes of LLMs are growing fast. The largest open-source transformer models for fine-tuning in recent years have grown to over 100 billion (100B) parameters~\cite{llama3}.\footnote{\label{footnote_model_size}Size of an LLM is defined by the number of parameters. We use ``100B model'' to represent a model with 100 billion parameters in this paper. }
Fine-tuning a 100B model requires storing \textasciitilde 2.6 TB of temporary and persistent tensors at peak times,
while the latest on-market data-center GPU has only up to 188 GB device memory~\cite{h200}.

One native approach to host huge models is to aggregate device memory from many data-center GPUs on high-end clusters like DGX platform~\cite{dgx} to fine-tune a 100B-scale model~\cite{galvatron, megatron, alpa, het, ps, fastflow, sdpipe, flexmoe, guo2021model, flexps, jiang2017heterogeneity, sketchml, miao2021heterogeneity, sandha2019database, zhang2021distributed, memflow, hydra, miao2024demystifying}. 
For example, it takes 32$\times$ \$14177 NVIDIA A100 GPUs with 80 GB of device memory to fine-tune a model with 100B parameters, so accommodating a cluster of high-end GPUs introduces prohibitive costs for most data scientists with tight budgets. 

In this paper, we aim to explore whether it is feasible to efficiently \textit{fine-tune a 100B-scale LLM on a single consumer-grade 4090 GPU (\$1600, up to 24 GB device memory) with limited main memory capacity (256 GB)}. 
Such a solution would be attractive to researchers who seek to minimize LLM fine-tuning costs. To do so, the existing low-cost works~\cite{angel-ptm, capuchin, mpress, zero-offload, vdnn, superneurons} offload 
the tensors during the fine-tuning process 
from GPU memory to NVMe memory to maximize the trainable model size. 
However, we identify that these SSD-equipped systems suffer from two severe issues: low throughput and small maximum trainable model size. 

\begin{figure*}[t]
    \centering
    \vspace{-1ex}
    \hspace*{-6pt}
    \subfloat[ZeRO-Infinity: 1)~executing the CPU optimizer in a separate stage where GPU and GPU-main memory link is idle, 2)~only offloading inter-block activations and recomputing the others, which leads to 5.7s GPU recomputation in the backward stage thus causing only 24\% PCIe utilization, 3)~activations only offloaded to main memory thus limiting the trainable model size.]{
        \label{fig_system_overview_zero}
        \includegraphics[width=0.476\linewidth]{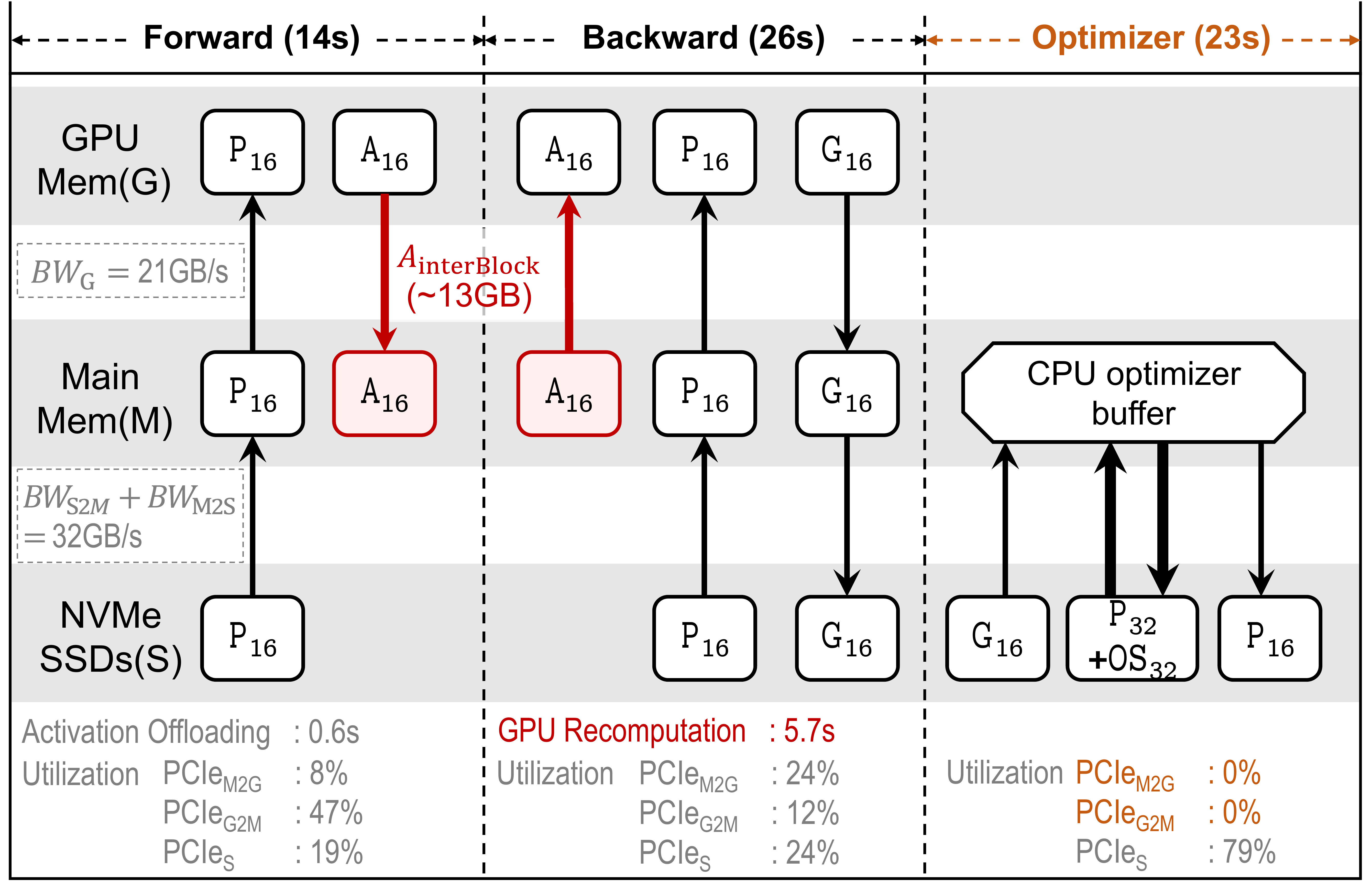}
    }
    \hspace*{1pt}
    \subfloat[G10: 1)~transfering gradients, parameters and model states (182 GB per direction) between GPU and SSDs during the optimizer execution making GPU almost idle, and 2)~offloading almost all activations to unified main/NVMe memory without recomputation thus causing \textasciitilde 213 GB of activation transfer when fine-tuning a 13B model.]{
        \label{fig_system_overview_g10}
        \includegraphics[width=0.447\linewidth]{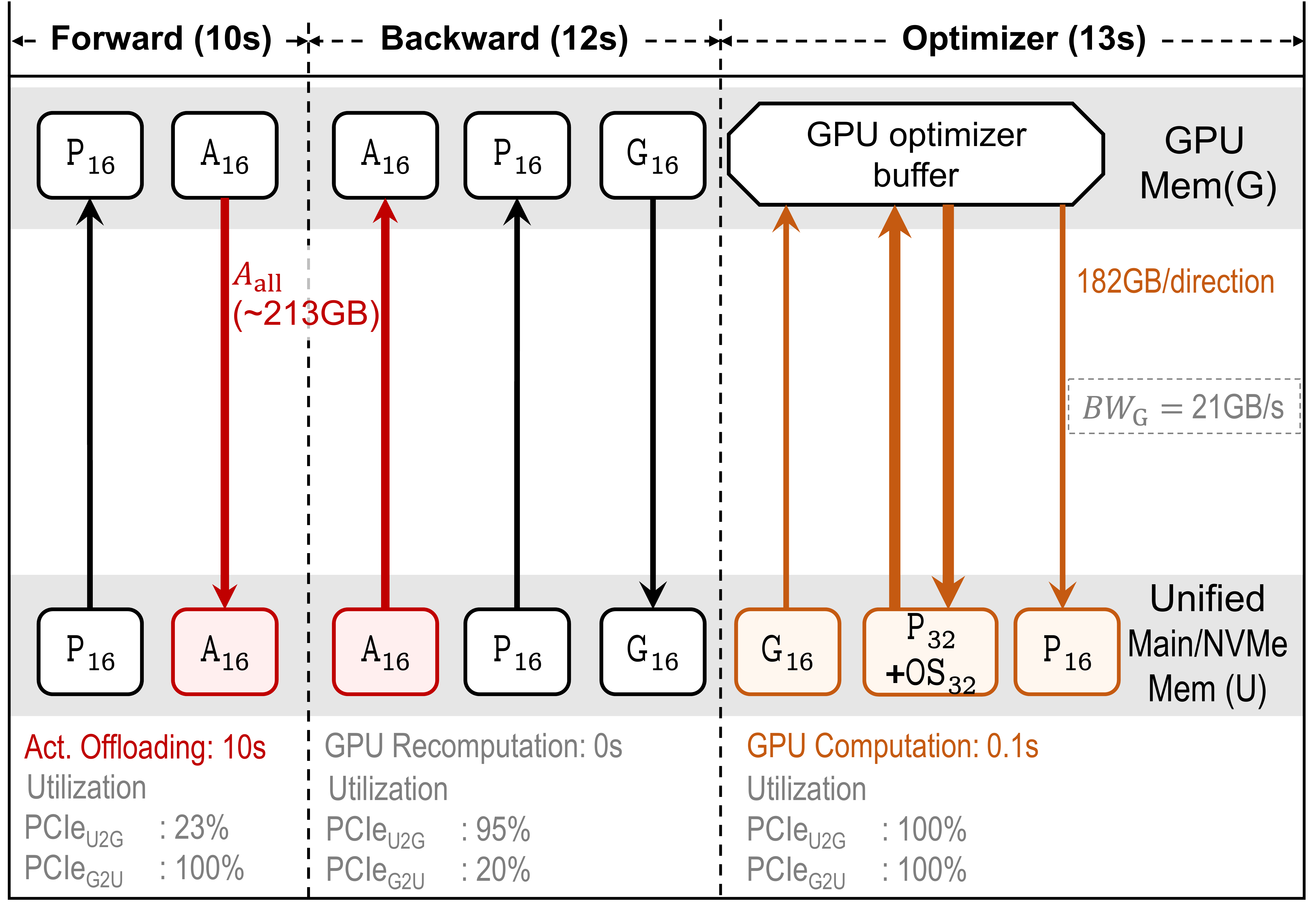}
    }
    \newline
    \vspace{-1ex}
    \subfloat[\SystemName{}: 1)~directly consume gradients in the backward stage, and 2)~determining an optimal amount of activations offloading to main memory and NVMe SSDs so that the total iteration time is minimum. When fine-tuning a 13B model, \SystemName{} only offloads \textasciitilde 34 GB activation with 32\% additional GPU computation in the backward stage.]{
        \label{fig_system_overview_lohan}
        \includegraphics[width=0.97\linewidth]{figure/system_overview_lohan.pdf}
    }
    \caption{Comparison of offloading-based systems. Bandwidth and overhead numbers are obtained on our evaluation server with 12 SSDs when fine-tuning a 13B model with a batch size of 32.}
    \label{fig_system_overview}
    \vspace{-4ex}
\end{figure*}

\begin{itemize}[topsep=0pt, leftmargin=*]
\vspace{-0.5ex}
\item \textbf{Offloading Activation Tensors to SSDs.}
\
LLM training consists of two types of tensors, namely activations and model states.
Existing systems like FlashNeuron~\cite{flashneuron} only offload activations to SSDs and keep model states in GPU memory. We find that keeping model states in GPU memory severely limits the trainable model size. For example, FlashNeuron can only fine-tune a 1.55B model on RTX 4090, while fine-tuning a 175B model (a typical size of 100B-scale models~\cite{opt, mistral}) requires \textasciitilde 2.45~TB of GPU memory, which far exceeds the memory capacity of GPUs. 

\item \textbf{Offloading Model State Tensors to SSDs.}
\
Existing systems like ZeRO-Infinity~\cite{zero-infinity} and Colossal-AI~\cite{colossal-ai} offload model states to NVMe SSDs~\cite{haas2023modern, lerner2021not, petrov2010building} to enlarge the trainable model size. We identify that these systems have three issues in model fine-tuning, as shown in Figure~\ref{fig_system_overview_zero}. First, these systems suffer from low GPU utilization, mainly because they execute the synchronous out-of-core CPU Adam\footnote{Here we refer to the out-of-core optimizer as optimizer executing on CPU instead of GPU (i.e., ``in-core optimizer''). There are also works such as Angel-PTM~\cite{angel-ptm} that present an asynchronous out-of-core optimizer. However, the asynchronous optimizer updating policy might affect model training convergence. Therefore, they are beyond the scope of this paper. }~\cite{adam} in the optimizer stage where the GPU is idle. This stage takes 30\%\textasciitilde 60\% of a training iteration. Second, these systems only offload the inter-transformer block activations (6\% of total activations) to main memory and recompute the rest of the activations. Such an offloading method leads to 5.7 seconds (22\% of the backward stage) of additional GPU recomputation overhead in the backward stage where PCIe bandwidth is underutilized. 
Third, these systems only offload activations to main memory, thus requiring a large amount of main memory to fine-tune an LLM. We estimate that ZeRO-Infinity requires \textasciitilde 1.1 TB main memory to fine-tune the 175B model, while most commodity servers equip only 128 GB\textasciitilde 1 TB main memory.  

\item \textbf{Na\"ively Offloading Both Model State and Activation Tensors to SSDs.}
\
Existing systems like G10~\cite{g10} offload both model states and activations to SSDs and execute the Adam optimizer on GPU. We identify that these systems have three issues, as shown in Figure~\ref{fig_system_overview_g10}. First, executing the Adam optimizer on GPU requires transferring massive model states between GPU and NVMe SSDs, making the 0.1-second GPU computation wait for 13-second model state transfer. Second, these systems offload all the activations (213 GB when fine-tuning a 13B model with a batch size of 32) to SSDs, making 5.9-second GPU computation wait for 10-second activation transfer during the forward stage. Third, these systems rely on GPUDirect~\cite{gpudirect} technology that is not available for consumer-grade GPUs.
\end{itemize}

In summary, they do not have holistic intra-server tensor management that impedes efficient fine-tuning on a 100B model when offloading activations or model states onto SSDs. 

To this end, we present \SystemName{}, a low-cost high-performance deep learning training framework that enables efficient 100B model fine-tuning on a commodity server with a consumer-grade GPU and limited main memory capacity.
The key idea is to add holistic offloading traffic as an optimization dimension. As such, \SystemName{} enables a commodity GPU without GPUDirect to efficiently fine-tune a huge model, whose size is limited by SSD capacity, rather than main memory/GPU memory size, when both model states and activations are offloaded to NVMe SSDs. 
To do so, \SystemName{} consists of two innovations. First, for model states, \SystemName{} presents the first active gradient offloading technology that enables the out-of-core CPU optimizer execution to directly consume the gradients from GPU to CPU, so as to hide CPU optimizer execution behind GPU computation. Second, for activations, \SystemName{} proposes a holistic traffic-aware activation swapping management to automatically determine the amount of swapping activations such that the epoch time is minimized when training on a single GPU in a commodity server. 
To summarize, this paper makes the following contributions:

\begin{itemize}[topsep=0pt, leftmargin=*]
    \item We study the existing offloading strategies and identify the issues of low throughput and small maximum trainable model size due to the lack of holistic intra-server tensor management. 
    \item To optimize the holistic offloading traffic, we design an active gradient offloading technology and a holistic traffic-aware activation swapping management to overlap offloading and computation so as to maximize GPU utilization. 
    \item We implement \SystemName{} on the deep learning framework PyTorch~\cite{pytorch}. Evaluations show that \SystemName{} 1)~is the first to fine-tune a 175B model on an RTX 4090 and 256 GB main memory, 2)~achieves up to 2.32$\times$ throughput than the state-of-the-art baselines when fine-tuning a small 13B model, and 3)~enables a cheap low-end consumer GPU to have higher cost-effectiveness than a DGX-A100 machine.
\end{itemize}
\begin{figure*}[t]
    \subfloat[Largest trainable model size of existing works under \\ different main memory capacities.]{
        \label{fig_motivation_max_model}
        \includegraphics[width=0.360\linewidth]{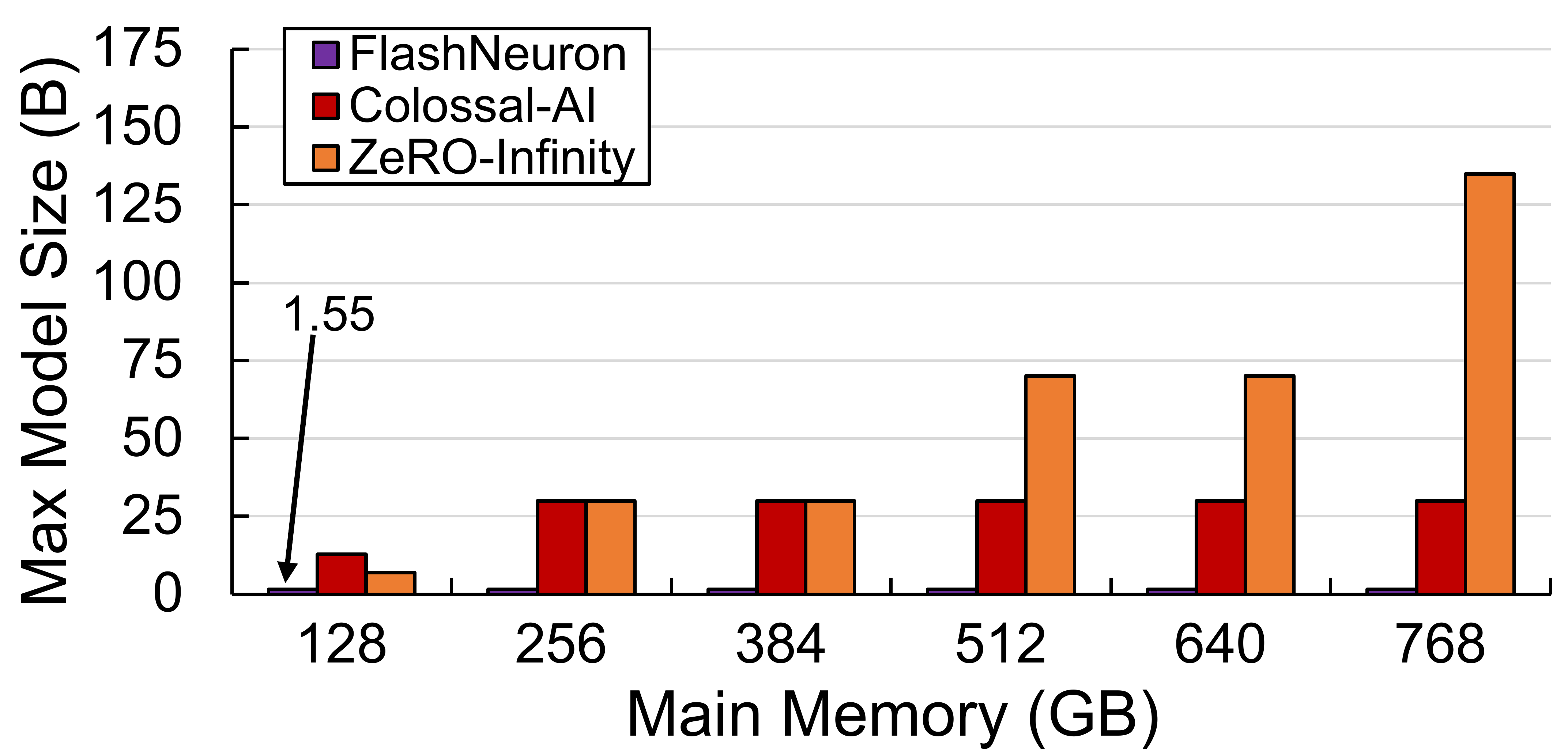}
    }
    \hfill    
    \subfloat[GPU utilization under different batch sizes \\ in ZeRO-Infinity.]{
        \label{fig_motivation_gpu_util}
        \includegraphics[width=0.289\linewidth]{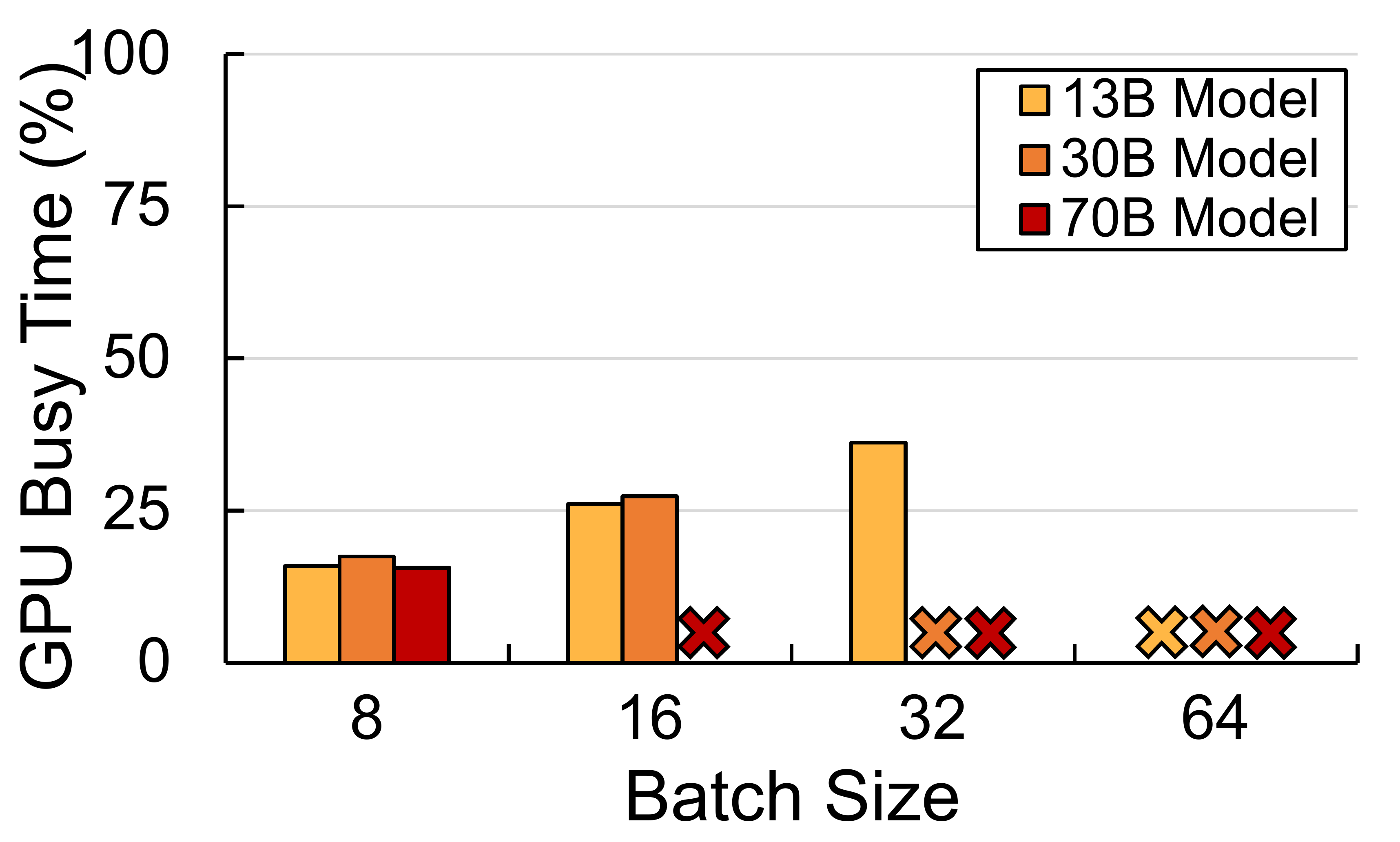}
    }
    \hfill    
    \subfloat[Proportions of optimizer stage in a training step in ZeRO-Infinity.]{
        \label{fig_motivation_optimizer}
        \includegraphics[width=0.289\linewidth]{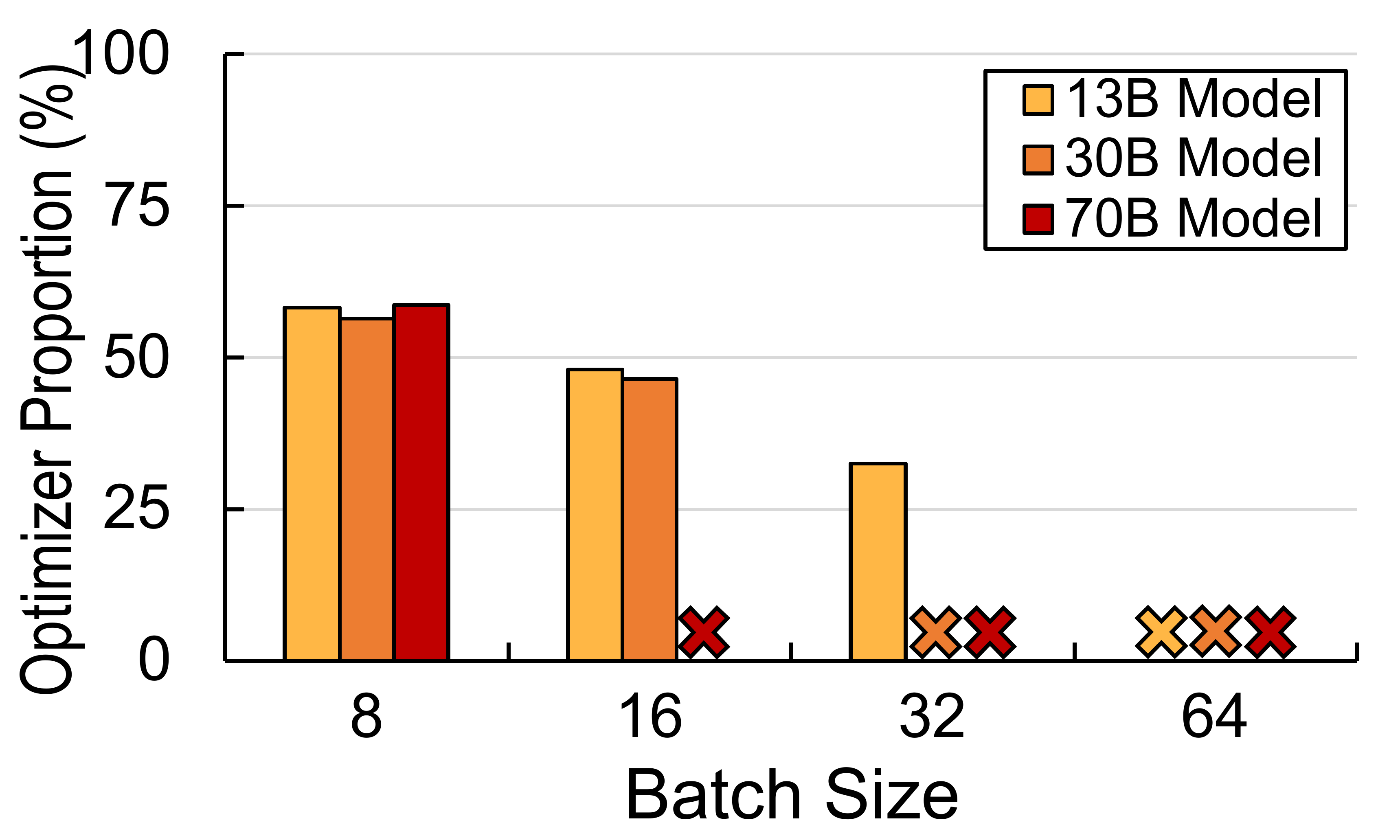}
    }
\vspace{-1ex}
\caption{Issues of SSD-offloading methods that motivate the design of \SystemName{}. We perform the experiments on RTX 4090. } 
    \label{fig_motivation} 
\vspace{-4ex}
\end{figure*} 

\section{Background}
\label{sec:background}

\noindent\textbf{LLM Training Stages.}
\
A model consists of $L$ layers of mathematical functions $f_i\left({\bm x}, {\bm P}_i\right), 1\!\leq\! i\!\leq\!L$, where $i$ denotes layer ID, $\bm x$ denotes input and $\bm P$ denotes its trainable parameters. The training procedure takes multiple training iterations to get the model converged. 
Each iteration consists of three stages: 

\begin{itemize}[topsep=0pt, leftmargin=*]
\item 1)~Forward propagation, where the model takes training data ${\bm a}_0$ as input, computes the activation ${\bm a}_i\!=\! f_i({\bm a}_{i-1}, {\bm P}_i)$ as indimediate values for each layer $i$ successively and gets the loss value ${\bm l}_L={\bm y}-{\bm a}_L$, where ${\bm y}$ denotes expected output.

\item 2)~Backward propagation, where the model performs two computations for each layer $i$ in reverse to get gradients used for model updates: first, each layer takes the loss value ${\bm l}_i$ and computes the loss value delivered to the previous layer ${\bm l}_{i-1}\!=\!\nabla_{{\bm a}_{i-1}}f_i({\bm a}_{i-1}, {\bm P}_i)^{\rm T}{\bm l}_i$; second, it computes the gradients of the layer ${\bm G}_i\!=\!\nabla_{{\bm P}_i}f_i({\bm a}_{i-1}, {\bm P}_i)^{\rm T}{\bm l}_i$.

\item 3)~Optimizer execution, where the parameters are updated according to gradients, i.e., ${\bm P}^{\rm updated}\!\!=\!\!o({\bm G}, {\bm P})$, where $o$ is the optimizer function. When training LLMs, Adam optimizer~\cite{adam} is generally adopted to increase the model convergency, which introduces auxiliary optimizer states to smooth the parameter update process.
\end{itemize}

\noindent
\textbf{Notations.}
\
Table~\ref{tab:notations} lists notations used in the rest of the paper.

\noindent
\textbf{Memory Footprint.}
\
According to the training procedure introduced above, an LLM training iteration requires storing the following tensors: 1)~Model states, including parameters $\tt P_{32}$, optimizer states $\tt {OS}_{32}$, gradients $\tt G_{16}$, and a low-precision parameter copy $\tt P_{16}$ for GPU computation; and 2)~activations $\tt A_{16}$. Table~\ref{tab:memory_footprint} concludes the sizes and life cycles of the tensors within an iteration. The loss values are directly consumed after being produced, and thus are not discussed here.

\noindent
\textbf{Tensor Offloading.}
\
When no memory-saving techniques are applied, all the tensors are produced, stored, and consumed in GPU memory. Tensor offloading is a technique that moves part or all of the tensors from GPU memory to main memory or SSDs after the tensor is produced by GPU, and moves the tensor back to GPU memory before the tensor is consumed by GPU, so as to reduce the GPU memory footprint. The procedure of temporarily offloading a tensor is called swapping.

\noindent
\textbf{CPU Optimizer.}
\
Offloading optimizer execution to CPU~\cite{zero-offload} is a technology to reduce the PCIe traffic of GPU. The CPU optimizer eliminates the heavy parameter and optimizer state transfer between the GPU and main memory because when offloading the model states to NVMe SSDs, $\tt P_{32}$ and $\tt {OS}_{32}$ produced by the CPU optimizer in main memory are directly moved to SSDs. In contrast, $\tt P_{32}$ and $\tt {OS}_{32}$ produced by a GPU optimizer in GPU memory need to first move to main memory, then to SSDs. 

\noindent\textbf{Activation Recomputation.}
\
Activation recomputation~\cite{chen2016training} is a memory-saving technique where only a subset of activations is kept in memory during forward propagation while others are discarded. During the backward propagation, when performing the backward propagation of a layer whose input activations are discarded, extra forward propagation from the last saved activation is performed to get the discarded activation. The extra forward propagation procedure is named recomputation.

\begin{table} [t]
	\centering\footnotesize
    \caption{List of notations.}
    \vspace{-1ex}
	\label{tab:notations}	
    \resizebox{1.03\columnwidth}{!}{
	\begin{tabular}{|>{\centering}p{1cm}||c|}
		\hline
		\textbf{Symbol} & \textbf{Definitions} \\
		\hline
		\hline
            $T_{\rm iter}$ & Elapsed time of a training iteration. \\
		\hline
            $T_{\rm f}, T_{\rm b}$ & Elapsed time of the forward stage and the backward stage. \\ 
        \hline
            \narrow{\makecell{\hspace{-0.2cm}$T_{\rm f}^{\rm G},T_{\rm f}^{\rm G2M},$\\\hspace{-0.2cm}$T_{\rm f}^{\rm M2G},T_{\rm f}^{\rm S}$}} & \vnarrow{\makecell{Elapsed time of GPU computation, GPU to main memory PCIe transfer, \\ main memory to GPU PCIe transfer and SSD I/O during the forward stage.}} \\
        \hline
            \narrow{\makecell{\hspace{-0.2cm}$T_{\rm b}^{\rm G},T_{\rm b}^{\rm G2M},$\\\hspace{-0.2cm}$T_{\rm b}^{\rm M2G},T_{\rm b}^{\rm S}$}} & \vnarrow{\makecell{Elapsed time of GPU computation, GPU to main memory PCIe transfer, \\ main memory to GPU PCIe transfer and SSD I/O during the backward stage.}} \\
        \hline
            $F\!L\!O\!P_{\rm f}$ & \makecell{\narrow{Number of GPU floating point operations during the forward stage.}\\The FLOP during the backward stage is thus $2F\!L\!O\!P_{\rm f}$.} \\
        \hline
            $P$ & \makecell{Number of parameters of a model. } \\
        \hline
            $A_{\rm all}$ & Size of activations in bytes of a model. \\
        \hline
            \narrow{\hspace{-0.24cm} $A_{\rm interBlock}$} & Size of inter-transformer block activations in bytes of a model. \\
        \hline
            $T\!H\!P_{\rm G}$ & Peak GPU throughput in FLOPS measured. \\
        \hline
            $BW_{\rm G}$ & \vnarrow{Maximum unidirectional PCIe bandwidth between GPU and main memory.} \\
        \hline
            $BW_{\rm S2M}$ & Maximum SSD to main memory PCIe bandwidth measured. \\
        \hline
            $BW_{\rm M2S}$ & Maximum main memory to SSD PCIe bandwidth measured. \\
        \hline
            $A_{\rm G2M}$ & Activations size in bytes swapped from GPU. \\
		\hline
            \narrow{\hspace{-0.2cm} $M\!E\!M_{\rm M}^{\rm avail}$} & \narrow{Minimum unallocated main memory in bytes during profiling stage.} \\
        \hline
            $\alpha$ & Proportion of activations swapped to SSDs relative to $A_{\rm G2M}$.\\
        \hline
            $F\!L\!O\!P_{\rm r}$ & \narrow{Number of GPU floating point operations during recomputation.}\\
        \hline
            $O\!B$ & \narrow{Activation offloading benefit of a layer, defined in Subsection~\ref{subsec_act_swap}}.\\
        \hline
	\end{tabular}
    }
    \vspace{-2ex}
\end{table}

\begin{table}
    \centering\footnotesize
    \caption{Tensors in LLM fine-tuning.}
    \vspace{-1ex}
    \label{tab:memory_footprint}
    \begin{tabular}{|c||c|c|c|}
    \hline
        \textbf{Tensor} & \textbf{Produced During} & \textbf{Consumed During} & \textbf{Size} \\
    \hline\hline
        $\tt P_{32}$ & \makecell{optimizer \\ (previous iteration)} & \makecell{optimizer \\ (current iteration)} & $4P$\\
    \hline
        $\tt {OS}_{32}$ & \makecell{optimizer \\ (previous iteration)} & \makecell{optimizer \\ (current iteration)} & $8P$\\
    \hline
        $\tt G_{16}$ & backward & optimizer & $2P$\\
    \hline
        $\tt P_{16}$ & \makecell{optimizer \\ (previous iteration)} & \makecell{forward and backward \\ (current iteration)} & $2P$\\
    \hline
        $\tt A_{16}$ & forward & backward & $A_{\rm all}$\\
    \hline
    \end{tabular}
    \vspace{-4ex}
\end{table}
\vspace{-1ex}
\section{Motivation}
\label{sec:motivation}

In this section, we identify the issues of existing offloading-based works, which motivate the design of \SystemName{}. 

\subsection{Issues of Approaches: Offloading Activations to SSDs}

The existing works such as FlashNeuron~\cite{flashneuron} offload activations to SSDs to train a larger model. However, these systems keep the model states on GPU memory, thus severely limiting the maximum trainable model size.

To illustrate this issue, we implement a prototype of FlashNeuron and fine-tune LLMs on our evaluated server (Detailed implementation and server configurations shown in Section~\ref{sec:exp_setup}). The experimental results in Figure~\ref{fig_motivation_max_model} show that FlashNeuron even fails to fine-tune a 6B model, which is a common scale of today's pre-trained LLMs~\cite{llama2, llama3, mistral}. 

\vspace{-1ex}
\subsection{Issues of Approaches: Offloading Model States to SSDs}
\label{sec:motivation_cpu_optimizer}

Systems like ZeRO-Infinity~\cite{zero-infinity} and Colossal-AI~\cite{colossal-ai} offload model states to SSDs to train a larger model. Meanwhile, they propose a CPU Adam~\cite{adam} that efficiently executes the optimizer in the CPU. These systems perform badly in such a commodity server.\footnote{They are originally designed for high-end DGX servers rather than for a commodity server with a single consumer-grade GPU. } In the following, we identify three concrete issues of these systems. 

\noindent \textbf{1, Heavy Optimizer Execution Overhead.}
\
These systems execute CPU optimizer after the backward propagation of the entire model finishes, as Figure~\ref{fig_system_overview_zero} shows. Thus, the CPU optimizer execution does not overlap with GPU computation, and the GPU is completely idle during the optimizer stage. When training in a high-end DGX cluster with many high-end CPUs, the CPU optimizer contributes a trivial time proportion. However, when fine-tuning on a commodity server with few CPUs, the optimizer execution takes a significant portion of the training time, leading to low GPU utilization during training. 

To illustrate this, we quantitatively analyze the ratio of GPU busy time over the total elapsed time when fine-tuning different models using ZeRO-Infinity. Figure~\ref{fig_motivation_gpu_util} shows that the GPU is busy during only 36\% of an iteration, even when the model is relatively small (such as 13B) and the batch size is large enough to saturate GPU computing resources (such as 32). Colossal-AI achieves even lower GPU utilization (GPU is busy only for 12\% of an iteration, not shown in the figure). 

To show the overhead of the optimizer execution, we measure the time proportion of the optimizer stage in ZeRO-Infinity when fine-tuning different models in Figure~\ref{fig_motivation_optimizer}. The optimizer execution takes 30\%\textasciitilde 60\% of the training step. 

\noindent \textbf{2, Excessive Activation Recomputation Overhead.}
\
These systems adopt a static activation management strategy. Specifically, they offload the inter-transformer block activations (12.5~GB for a 13B model with a batch size of 32) to main memory and recompute all the intra-transformer block activations (200 GB for a 13B model with a batch size of 32). Recomputing activations introduces additional GPU computation during backward propagation because GPU computation generally takes longer than PCIe tensor transfer when fine-tuning with a large batch size. For example, when fine-tuning a 13B model with a batch size of 32, the GPU backward propagation without recomputation takes 1.91$\times$ longer than SSD I/O time and 1.88$\times$ longer than the GPU-CPU activation and gradient transfer, thus GPU computation is the bottleneck. Therefore, this recomputation strategy incurs GPU computation overhead during the backward propagation. 

To illustrate this, we use ZeRO-Infinity to fine-tune a 13B model and break down its training time, as Figure~\ref{fig_system_overview_zero} shows. 
The GPU-main memory tensor transfer $\tt PCIe_{\tt G2M}$ and main memory-SSD tensor transfer $\tt PCIe_{\tt SSD}$ only takes 3.18 and 6.25 seconds respectively during the backward stage, while GPU computation takes 17.6 seconds, indicating that excessive activation recomputation of ZeRO-Infinity introduces GPU computation overhead. 

\noindent \textbf{3, Limited Trainable Model Size under Limited GPU/Main Memory Capacity.}
\
ZeRO-Infinity offloads activation only to main memory, as Figure~\ref{fig_system_overview_zero} shows, and Colossal-AI does not offload activations to either main memory or SSDs. These systems do not offload activations to SSDs because it incurs additional pressure to SSD I/O and introduces additional design complexity. However, hosting activations in GPU and main memory limits the maximum trainable model size when fine-tuning in a server with limited GPU and main memory. 

To illustrate this, we fine-tune LLMs of various sizes with the two systems in our evaluated server (detailed configurations see Subsection~\ref{sec:exp_setup}). We set the batch size to 1 to minimize the effect of activations. Figure~\ref{fig_motivation_max_model} shows the maximum trainable model size with the two systems. They fail to fine-tune a 175B model in our evaluated server with 768GB main memory. 

\vspace{-1ex}
\subsection{Issues of Approaches: Na\"ively Offloading Both Model States and Activations to SSDs}
\label{sec:motivation_gpu_optimizer}

A recent work namely G10~\cite{g10} supports offloading both model states and activations to unified main/NVMe memory, which theoretically supports the 100B-scale model fine-tuning with scarce GPU and main memory. G10 does not consider activation recomputation in model fine-tuning and offloads almost all activations to SSDs. Besides, it executes the Adam optimizer on GPU, as most in-GPU model training systems do. We identify that G10 has three issues in LLM fine-tuning. 

\noindent \textbf{1, Heavy Model States Transfer Overhead.}
\
G10 executes the optimizer on GPU which introduces massive model state transfer (182 GB per direction for a 13B model) on the PCIe link during the optimizer execution, causing heavy PCIe transfer overhead. 

To illustrate this, we simulate the performance of G10 when fine-tuning a 13B model on RTX 4090 with a batch size of 32 assuming the GPUDirect is available and the GPU computation and PCIe transfer are fully pipelined. Figure~\ref{fig_system_overview_g10} shows the result. We observe that the GPU optimizer only takes 0.1 seconds during this stage, while the PCIe transfer takes 13 seconds (37\% of the iteration time). 

\noindent \textbf{2, High Activation Transfer Overhead.} 
\
G10 offloads all activations to main memory and then to SSDs without recomputation, which incurs massive activation transfer (213~GB for a 13B model with a batch size of 32) on GPU's PCIe port, which leads to low GPU utilization during forward propagation. To illustrate this, we break down the training time of the forward stage, where offloading activations takes 10~seconds, far beyond the GPU computation time (5.96s).

This performance overhead would be more severe if we intend to overlap optimizer execution and the backward propagation whose time is bounded by GPU computation,
like some existing works~\cite{stronghold}. Our corresponding simulation shows that PCIe transfer accounts for almost 100\% of both forward and the overlapped backward-optimizer stage time, while
GPU computation only accounts for 59\% of the forward stage time and 69\% of the overlapped backward-optimizer stage time, indicating that the PCIe transfer for model states and activations becomes the bottleneck throughout the whole training process!

\noindent \textbf{3, GPUDirect Requirement.}
\
G10 deeply relies on GPUDirect for tensor offloading. However, consumer-grade GPUs do not support GPUDirect, thus G10 cannot run on consumer-grade GPUs.

\section{Design of \SystemName}

\subsection{Design Overview}

To address the issues of existing works, we present \SystemName, a holistic tensor management system that enables efficient low-cost 100B model fine-tuning on a commodity server with a consumer-grade GPU and limited main memory capacity. 
The key idea is to add holistic tensor offloading management as an optimization dimension. 
As such, \SystemName{} achieves high GPU utilization when fine-tuning a 100B model, even when offloading both model states and activations to NVMe SSDs. 

To do so, \SystemName{} consists of three main components: 1)~hardware-aware profiling that collects essential data for model states and activation management (Subsection~\ref{subsec_profiling_stage}), 2)~active gradient offloading that enables the 
out-of-core CPU optimizer execution to directly consume the gradients from GPU to CPU, so as to hide CPU optimizer execution behind GPU computation (Subsection~\ref{subsec_backward_optimizer}), and 3)~holistic traffic-aware activation swapping management that automatically determines the amount of swapping activations to further minimize epoch time (Subsection~\ref{subsec_act_swap}). 
Figure~\ref{fig_system_overview_lohan} illustrates an overview of \SystemName. 

\vspace{-1ex}
\subsection{Hardware-Aware Profiling}
\label{subsec_profiling_stage}

In the hardware-aware profiling stage, \SystemName{} automatically gathers essential data from both model and hardware settings, which are required by the subsequent components. 

\noindent\textbf{Profiling Goals.}
The profiling stage aims to
provide the minimum unallocated main memory $M\!E\!M^{\rm avail}_{\rm M}$, the total elapsed time of the forward stage $T_{\rm f}$ and backward stage $T_{\rm b}$, number of model parameters $P$, size of model activations $A_{\rm all}$, peak GPU throughput $T\!H\!P_{\rm G}$, maximum PCIe bandwidth between GPU and main memory $BW_{\rm G}$, maximum SSD to main memory PCIe bandwidth $BW_{\rm S2M}$, maximum main memory to SSD PCIe bandwidth $BW_{\rm M2S}$, and the number of GPU floating point operations of each layer, which are required by the holistic traffic-aware activation swapping management.  

\noindent\textbf{Profiling Details. }
\SystemName{} parses the PyTorch model definition during initialization to obtain $P$, $A_{\rm all}$, and the number of GPU floating point operations of each model layer. In the profiling stage, \SystemName{} only offloads inter-layer activations and recomputes the rest of activations, just like ZeRO-Infinity, so as to minimize the activation offloading overhead while ensuring the recomputation process does not exceed the GPU memory limit. Besides, \SystemName{} offloads all activations and model states to NVMe SSDs at this stage without any further optimizations to accurately break down computation and communication costs. we record the computation time of each layer in the model during forward propagation so as to compute $T\!H\!P_{\rm G}$. 
To get $BW_{\rm G}$, $BW_{\rm S2M}$, and $BW_{\rm M2S}$, \SystemName{} gets the system topology from hardware settings during initialization, and monitors the PCIe traffic during the profiling stage, so as to estimate the PCIe bandwidth of each link. 

\noindent\textbf{Profiling Overhead. }
We perform the hardware-aware profiling stage only in the first iteration, which takes about 2\textasciitilde 3$\times$ times longer than that of a subsequent iteration. Fine-tuning an LLM requires thousands of iterations to converge, so the profiling overhead is negligible compared to the whole fine-tuning process. 

\subsection{Active Gradient Offloading} 
\label{subsec_backward_optimizer}

Inspired by \emph{Active Messages}~\cite{active_message_isca92} that allows a sender to specify a user-level handler along with each message and requires a receiver to immediately call the handler on the message arrival with the message body as an argument, we present the active gradient offloading technology that allows the CPU to perform the out-of-core CPU optimizer execution (i.e., user-level handler) upon main memory receiving the offloaded gradients (message body) from GPU, rather than to further offload to SSDs. As such, \SystemName{} has the opportunity to overlap the CPU optimizer execution with GPU backward propagation. In the following, we present the concrete challenge, followed by na\"ive active gradient offloading and optimized active gradient offloading. 

\noindent\textbf{Challenge. } Model states of the 100B model are stored in low-bandwidth and high-latency SSDs, thus exploiting such an opportunity comes with a main challenge. In particular, how to enable the out-of-core CPU optimizer to efficiently consume the offloaded gradients from GPU during backward propagation, because they compete for limited PCIe bandwidth. 

\begin{figure}[t]  
    \centering
    \subfloat[Na\"ive active gradient offloading.]{
        \includegraphics[width=0.94\linewidth]{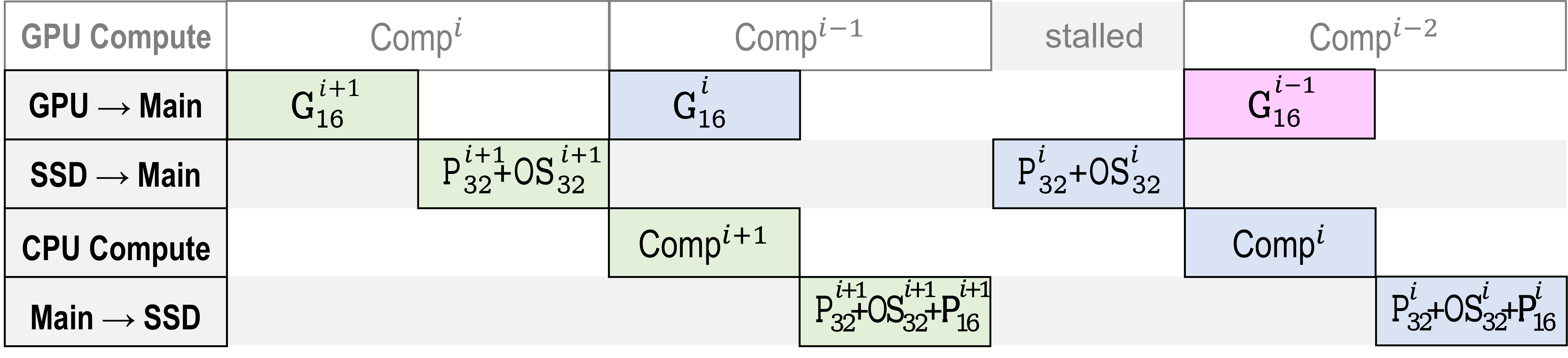}
        \label{fig_design_grad_off_stronghold}
    }
    \vspace{-2ex}
    \\
    \subfloat[Optimized active gradient offloading.]{
        \includegraphics[width=0.94\linewidth]{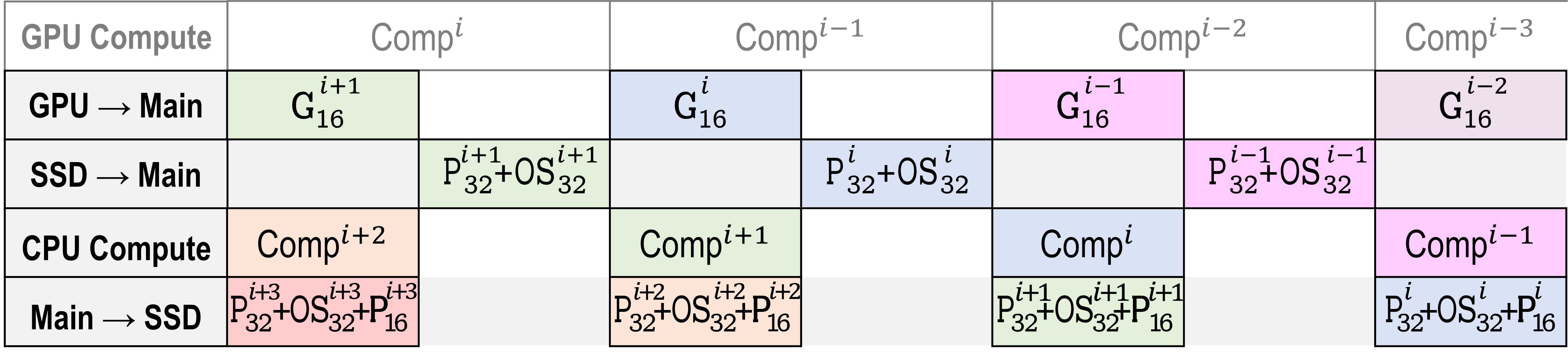}
        \label{fig_design_grad_off_lohan}
    }
    \vspace{-1ex}
    \caption{Comparison of active gradient offloading designs. } 
    \vspace{-4ex}
    \label{fig_design_grad_off} 
\end{figure}

\noindent\textbf{Na\"ive Active Gradient Offloading.}
\
Due to the stacked nature of transformer blocks, we assume the gradient tensors arrive at the CPU optimizer sequentially with a decreasing index during the backward stage. 
When the gradient tensor $i$ arrives at the main memory, its optimizer execution (user-defined handler) consists of three steps, as shown in Figure~\ref{fig_design_grad_off_stronghold}. First, during the \texttt{SSD$\rightarrow$Main} step, the SSD writes the corresponding model states of the model tensor $i$ to main memory. Second, during the \texttt{CPU Compute} step, the CPU updates the model states of the tensor $i$ and produces a 16-bit parameter copy. Third, during the \texttt{Main$\rightarrow$SSD} step, the SSD reads the updated model states and 16-bit parameters back. 

We can easily observe that the na\"ive active gradient offloading mechanism serializes three steps, such that the gradients are slowly consumed because expensive SSD accesses are involved in two steps.

\noindent\textbf{Optimized Active Gradient Offloading.}
\
Our key observation is that GPU's backward propagation, in-core CPU optimizer (\texttt{CPU Compute}), and SSD I/O (\texttt{SSD$\rightarrow$Main} and \texttt{Main$\rightarrow$SSD}) utilize almost different computation and communication resources in a server, as shown in Figure~\ref{fig_system_overview_lohan}. Therefore, these three steps have the potential to overlap to maximize the utilization of GPU, PCIe, and CPU. 

To this end, we are the first to present optimized active gradient offloading that overlaps the SSD I/O, in-core optimizer execution, and GPU's backward propagation to maximize GPU utilization. With optimized scheduling, \texttt{Main$\rightarrow$SSD} of the tensor $i$ is performed after \texttt{SSD$\rightarrow$Main} of the tensor $(i-1)$, as shown in Figure~\ref{fig_design_grad_off_lohan}. By doing so, \texttt{Main$\rightarrow$SSD} of the tensor $i$ can be overlapped with \texttt{CPU Compute} of the tensor $(i-1)$, thus \SystemName{} can overlap the CPU computation and SSD I/O during the out-of-core optimizer execution, increasing the PCIe bandwidth utilization between main memory and SSDs. As such, \SystemName{} keeps synchronous model updating while minimizing the GPU's idle time.\footnote{Active gradient offloading might be confused with the ``one-step delayed update'' optimization of ZeRO-Offload~\cite{zero-offload}. The one-step delayed update postpones the optimizer execution of iteration $i$ until the forward propagation of iteration $(i+1)$, thus introducing the parameter staleness and affecting the model training convergence. In contrast, \SystemName{} does not introduce parameter staleness because forward/backward propagation reads the updated models. } 

\vspace{-1ex}
\subsection{Holistic Traffic-Aware Activation Swapping Management}
\label{subsec_act_swap}

Swapping activations from GPU first to main memory then to SSDs incurs heavy PCIe traffic, while recomputing activations instead of swapping relieves PCIe traffic pressure at the cost of unnecessary GPU computation time. 
Existing activation swapping and recomputation works such as Capuchin~\cite{capuchin} only consider GPU computation time of backward propagation and GPU-main memory PCIe traffic to determine the amount of swapping activations, assuming gradients, parameters, and model states are kept on the GPU memory.

\noindent\textbf{Challenges. }\SystemName{} offloads both model states and activations to the SSDs and adopts the active gradient offloading, thus posing two unique challenges to the activation swapping design: 
1)~During the backward propagation, the GPU needs to swap back not only the activations but also the on-demand activations, incurring complex PCIe traffic. 2)~\SystemName{} adopts active gradient offloading that enables the CPU optimizer to overlap with backward propagation, so that the maximum execution time of backward propagation and the CPU optimizer serves to determine the swapping amount of the activations. 

To address these challenges, \SystemName{} proposes the holistic traffic-aware activation swapping management that automatically determines the amount of swapping activations and thus minimizes each iteration time. In the following, we describe \SystemName{}'s design in detail.

The goal of the activation swapping strategy is to find an $A_{\rm G2M}$ that minimizes the iteration time $T_{\rm iter}$. The high-level observation is that $T_{\rm iter}$ is a convex function of $A_{\rm G2M}$, so that the optimal $A_{\rm G2M}$ can be found by iterating activations to swap, computing the corresponding $A_{\rm G2M}$ and iteration time, and finding the inflection point where $T_{\rm iter}$ is minimized. We first discuss how we compute $T_{\rm iter}$ when given an $A_{\rm G2M}$, then give proof that $T_{\rm iter}$ is convex, and finally show the concrete procedure of choosing activations to swap. 

\noindent\textbf{Iteration Time Computation.}
\
$T_{\rm iter}$ in \SystemName{} is the sum of the forward stage time $T_{\rm f}$ and the backward stage time $T_{\rm b}$, as shown in Equation~\ref{eq_t_iter}.
\begin{equation}\label{eq_t_iter}
    T_{\rm iter} = T_{\rm f} + T_{\rm b} 
\end{equation}

We first evaluate $T_{\rm f}$. When the GPU computation and PCIe tensor transfers are fully overlapped, the forward stage time is the maximum among $T_{\rm f}^{\rm G}$, $T_{\rm f}^{\rm G2M}$, $T_{\rm f}^{\rm M2G}$, and $T_{\rm f}^{\rm S}$, which can be expressed by Equation~\ref{eq_t_f}. Note that the GPU-CPU PCIe link is duplex, while the SSD is simplex, thus we need to compute GPU-CPU transfer time and CPU-GPU PCIe transfer time separately but consider the SSD I/O time as a whole.
\begin{align}\label{eq_t_f}
    T_{\rm f} &= \max\left(T_{\rm f}^{\rm G}, T_{\rm f}^{\rm G2M}, T_{\rm f}^{\rm M2G}, T_{\rm f}^{\rm S}\right) \\
    &= \max\left(
    \tfrac{F\!L\!O\!P_{\rm f}}{T\!H\!P_{\rm G}}, \tfrac{A_{\rm G2M}}{BW_{\rm G}}, \tfrac{2P}{BW_{\rm G}}, \tfrac{2P}{BW_{\rm S2M}}+\tfrac{\alpha A_{\rm G2M}}{BW_{\rm M2S}}
    \right)\nonumber
\end{align}

The last component of $T_{\rm f}$ includes the amount of activations that are swapped to SSDs $\alpha A_{\rm G2M}$. We next describe how $\alpha A_{\rm G2M}$ is decided. When an activation is swapped in \SystemName{}, it is accommodated by either main memory or SSDs. \SystemName{} decides the amount of activations that are accommodated by main memory, $A_{\rm G2M} - \alpha A_{\rm G2M}$, based on the peak main memory usage gathered in the profiling stage. The main memory is first used for storing parameters that are prefetched from SSDs, and model states that are used for optimizer execution, while the rest of the main memory is used for accommodating activations. Therefore, the amount of activations accommodated by SSDs can be expressed as shown in Equation~\ref{eq_alpha_a_g2c}. 
Therefore, the forward stage time $T_{\rm f}$ can be further expressed by Equation~\ref{eq_t_f_2}.

\begin{equation}\label{eq_alpha_a_g2c}
    \alpha A_{\rm G2M} = A_{\rm G2M} - M\!E\!M_{\rm M}^{\rm avail}
\end{equation}
\begin{equation}\label{eq_t_f_2}
    T_{\rm f} = \max\!\left(
    \tfrac{F\!L\!O\!P_{\rm f}}{T\!H\!P_{\rm G}}, \tfrac{A_{\rm G2M}}{BW_{\rm G}}, \tfrac{2P}{BW_{\rm G}}, \tfrac{2P}{BW_{\rm S2M}}\!\!+\!\!\tfrac{A_{\rm G2M}\!-\!M\!E\!M_{\rm C\!P\!U}^{\rm avail}}{BW_{\rm M2S}}
    \right)
\end{equation}

Then we evaluate $T_{\rm b}$. Similar to the forward stage, the backward stage time can be expressed by Equation~\ref{eq_t_b}. Note that we do not consider the CPU Adam execution time because its time is shorter than reading/writing the optimizer states from/to SSDs. 
\begin{align}\label{eq_t_b}
    T_{\rm b} &= \max\left(T_{\rm b}^{\rm G}, T_{\rm b}^{\rm G2M}, T_{\rm b}^{\rm M2G}, T_{\rm b}^{\rm S}\right) \\
    &= \max\left(\!
    \tfrac{2 F\!L\!O\!P_{\rm f}+F\!L\!O\!P_{\rm r}}{T\!H\!P_{\rm G}}, \tfrac{2\!P}{B\!W_{\rm G}}, \tfrac{2\!P+A_{\rm G\!2\!M}}{B\!W_{\rm G}}, \tfrac{1\!4\!P\!+\!\alpha A_{\rm G\!2\!M}}{BW_{\rm S2M}}\!+\!\tfrac{14P}{B\!W_{\rm M2S}}
    \!\right)\nonumber
\end{align}

Since $FLOP_{\rm r}$ can be computed by accumulating the computation quantity of layers that need to be recomputed, we can compute the iteration time once $A_{\rm G2M}$ is provided.

\noindent\textbf{Proving Convexity of Iteration Time.}
\
We first list the mathematical theorems used in our proofs.
\begin{theorem}
\label{theorem_convex_sum}
The sum of convex functions is convex.
\end{theorem}
\begin{theorem}
\label{theorem_convex_max}
The maximum of convex functions is convex.
\end{theorem}
\begin{theorem}
\label{theorem_convex_linear}
A linear function is a convex function.
\end{theorem}
\begin{theorem}
\label{theorem_convex_increase}
$f(y)=af(x)+b$ is convex to $x$ if $f(x)$ is convex to $x$ and $a>0$.
\end{theorem}

To prove that the iteration time $T_{\rm iter}$ is convex to $A_{\rm G2M}$, we prove that $T_{\rm f}$ and $T_{\rm b}$ is convex separately, thus proving $T_{\rm iter}$ is convex according to Theorem~\ref{theorem_convex_sum}.

We first prove that $T_{\rm f}$ is convex to $A_{\rm G2M}$. According to Equation~\ref{eq_t_f_2}, the first and the third components are independent of $A_{\rm G2M}$, while the second and the last components are a linear function of $A_{\rm G2M}$. Thus, all four components are convex according to Theorem~\ref{theorem_convex_linear}. Therefore we conclude that $T_{\rm f}$ is a convex function of $A_{\rm G2M}$ according to Theorem~\ref{theorem_convex_max}. 

Then we prove that $T_{\rm b}$ is convex to $A_{\rm G2M}$. The second component of Equation~\ref{eq_t_b} is independent of $A_{\rm G2M}$, while the third and the last components are a linear function of $A_{\rm G2M}$, thus the last three components are convex functions of $A_{\rm G2M}$ according to Theorem~\ref{theorem_convex_linear}.

To prove that the first component $T_{\rm b}^{\rm G}$ is also convex, we first assume that activations of a layer can be partially offloaded while its recomputation overhead is proportional to discarded activations.\footnote{This assumption is only for interpolation, and we don't actually offload part of a layer's activation in reality.} 
Next, we introduce \SystemName{}'s activation swapping order that is necessary for the proving process. For each layer's activations, \SystemName{} assigns different swapping priorities. A layer's activation is more suitable for swapping rather than recomputing if it requires 1)~more time to recompute or 2)~less time to swap. Since the recomputing time of a layer is proportional to its operation quantity, and the offloading time is proportional to the activation size, we define the offloading benefit of a layer ($O\!B_{\rm layer}$) as its floating point operations in recomputation ($F\!L\!O\!P_{\rm layer}$) over its activation tensor volume ($A_{\rm layer}$), as shown in Equation~\ref{eq_ob_layer}. 

\begin{equation}\label{eq_ob_layer}
    O\!B_{\rm layer}=\frac{F\!L\!O\!P_{\rm layer}}{A_{\rm layer}}
\end{equation}

A layer's activations that have higher offloading benefits have higher priority in swapping rather than recomputing. 

Let $A_{i}$ denote the activation size of the layer $i$, $F\!L\!O\!P_{i}$ denote its number of operations required in recomputation, and $O\!B_{i}$ denote its offloading benefit. For each $i$ that satisfies $\sum\limits_{k=1}^{i}A_k \le A_{\rm G2M} \le \sum\limits_{k=1}^{i+1}A_k$ (That is, $A_{\rm G2M}$ includes the first $i$ layers and part of layer $(i+1)$), the recomputation overhead of a model can be expressed by Equation~\ref{eq_flop_r}.
\vspace{-2ex}
\begin{align}\label{eq_flop_r}
    F\!L\!O\!P_{\rm r} =& F\!L\!O\!P_{\rm f} - \sum\limits_{m=1}^{i}\!F\!L\!O\!P_{\rm m}\!-\!F\!L\!O\!P_{\rm i+1} \!\times\!\tfrac{A_{\rm G\!2\!M}\!-\!\sum\limits_{n=1}^{i}\!A_n}{A_{i+1}}\\
    =& F\!L\!O\!P_{\rm f}-\sum\limits_{m=1}^{i}\!O\!B_m A_m-O\!B_{i+1}A_{i+1}\times \tfrac{A_{\rm G\!2\!M}\!-\!\sum\limits_{n=1}^{i}\!A_n}{A_{i+1}}\nonumber
\end{align}
\vspace{-2ex}

Thus the derivative of $F\!L\!O\!P_{\rm r}$ is expressed by Equation~\ref{eq_d_flop_r}.
\begin{equation}\label{eq_d_flop_r}
    \tfrac{{\rm d}F\!L\!O\!P_{\rm r}}{{\rm d}A_{\rm G2M}} = -O\!B_{i+1}
\end{equation}

Since $O\!B_{i}$ is an decreasing function of $i$, $\tfrac{{\rm d}F\!L\!O\!P_{\rm r}}{{\rm d}A_{\rm G2M}}$ is an increasing function of $A_{\rm G2M}$, thus $F\!L\!O\!P_{\rm r}$ is convex. Therefore, $T_{\rm b}^{\rm G}$ is also convex according to Theorem~\ref{theorem_convex_increase}. According to Theorem~\ref{theorem_convex_max}, we conclude that the backward stage time $T_{\rm b}$ is a convex function of $A_{\rm G2M}$.

Accordingly, the sum of $T_{\rm f}$ and $T_{\rm b}$, which is the iteration time $T_{\rm iter}$, is convex according to Theorem~\ref{theorem_convex_sum}. 

\noindent\textbf{Concrete Procedure of Activation Swapping Strategy.}
\
Now we describe the concrete procedure of the activation swapping strategy. From the convexity of $T_{\rm iter}$, we deduce three possible cases for the iteration time with regard to the offloaded activation size.

Case 1: The iteration time increases as $A_{\rm G2M}$ increases, indicating that the PCIe transfer is the training bottleneck even without recomputation. In this case, it is better to reduce the offloaded activation size as long as the recomputation does not exceed the GPU memory capacity. In \SystemName{}, we choose $A_{\rm interBlock}$ as the minimum safe swapped activation amount by default. 

Case 2: The iteration time decreases as $A_{\rm G2M}$ increases, indicating that GPU computation is the training bottleneck even when offloading all activations. In this case, all activations should be offloaded (i.e., $A_{\rm G2M}=A_{\rm all}$). 

Case 3: As $A_{\rm G2M}$ increases, the iteration time decreases when $A_{\rm G2M}$ is smaller than a $A_{\rm optimal}$, and increases when $A_{\rm G2M}$ is larger than the $A_{\rm optimal}$, thus the $A_{\rm optimal}$ is the optimal offloaded activation size. 

From the analysis, we can find the $A_{\rm optimal}$ by computing the iteration time when iterating different $A_{\rm G2M}$ and detecting the inflection point of $T_{\rm iter}$ with regard to $A_{\rm G2M}$ (Case 3). If no inflection point is detected, \SystemName{} decides $A_{\rm G2M}$ by matching the pattern of $T_{\rm iter}$ to Cases 1 and 2. Algorithm~\ref{alg_scheduling} describes the details of this procedure. 

\begin{algorithm}\small
\caption{\label{alg_scheduling}\narrow{Finding Optimal Activation Swapping Strategy.}}
\KwData{layer\_list: List of all layers in the LLM.}
\KwData{swap\_list: List of swapped activations.}

swap\_list $\leftarrow$ []\;
$T_{\rm min} \leftarrow \infty$\; 
$A_{\rm G2M} \leftarrow 0$\; 
$F\!L\!O\!P_{\rm r} \leftarrow F\!L\!O\!P_{\rm f}$ \tcp*{\narrow{Full recomputation}}
$i \leftarrow 0$\; 
layer\_list.sortByOffloadingBenefit()\;
\For{layer {\rm \textbf{in}} layer\_list}
{
    $A_{\rm G2M} \leftarrow A_{\rm G2M} + $layer.actSize\; 
    $F\!L\!O\!P_{\rm r} \leftarrow F\!L\!O\!P_{\rm r} - $layer.flop\; 
    $T_{\rm iter} \leftarrow $ computeIterTime($A_{\rm G2M}$, $F\!L\!O\!P_{\rm r}$)\;
    \eIf {$T_{\rm iter} \geq T_{\rm min}$} {
        \If {$A_{\rm G2M} \geq A_{\rm interBlock}$} {
            \textbf{break} \tcp*{\narrow{Ensure $A_{\rm G2M} \geq A_{\rm interBlock}$ to avoid OOM}}
        }
    }{
        $T_{\rm min} \leftarrow T_{\rm iter}$\;
    }
    swap\_list.append(layer.activation)
}
\end{algorithm}

\begin{figure}[t]  
    \vspace{-5ex}
    \subfloat[PyTorch.]{
        \label{fig_design_interface_torch}
        \includegraphics[width=0.4114\linewidth]{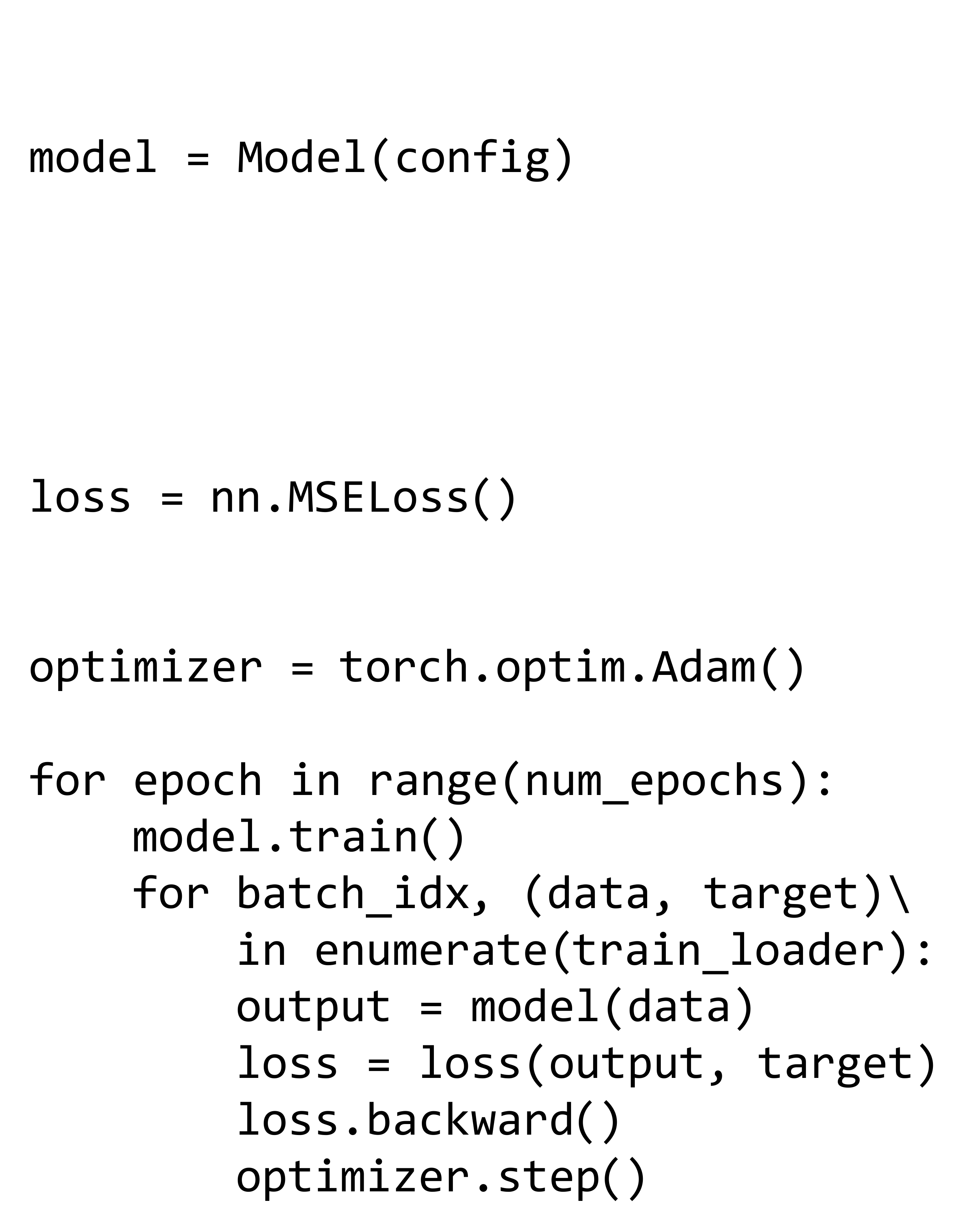}
    }
    \hfill  
    \subfloat[\SystemName{}. Highlighted functions are APIs provided by \SystemName{}.]{
        \label{fig_design_interface_lohan}
        \includegraphics[width=0.5485\linewidth]{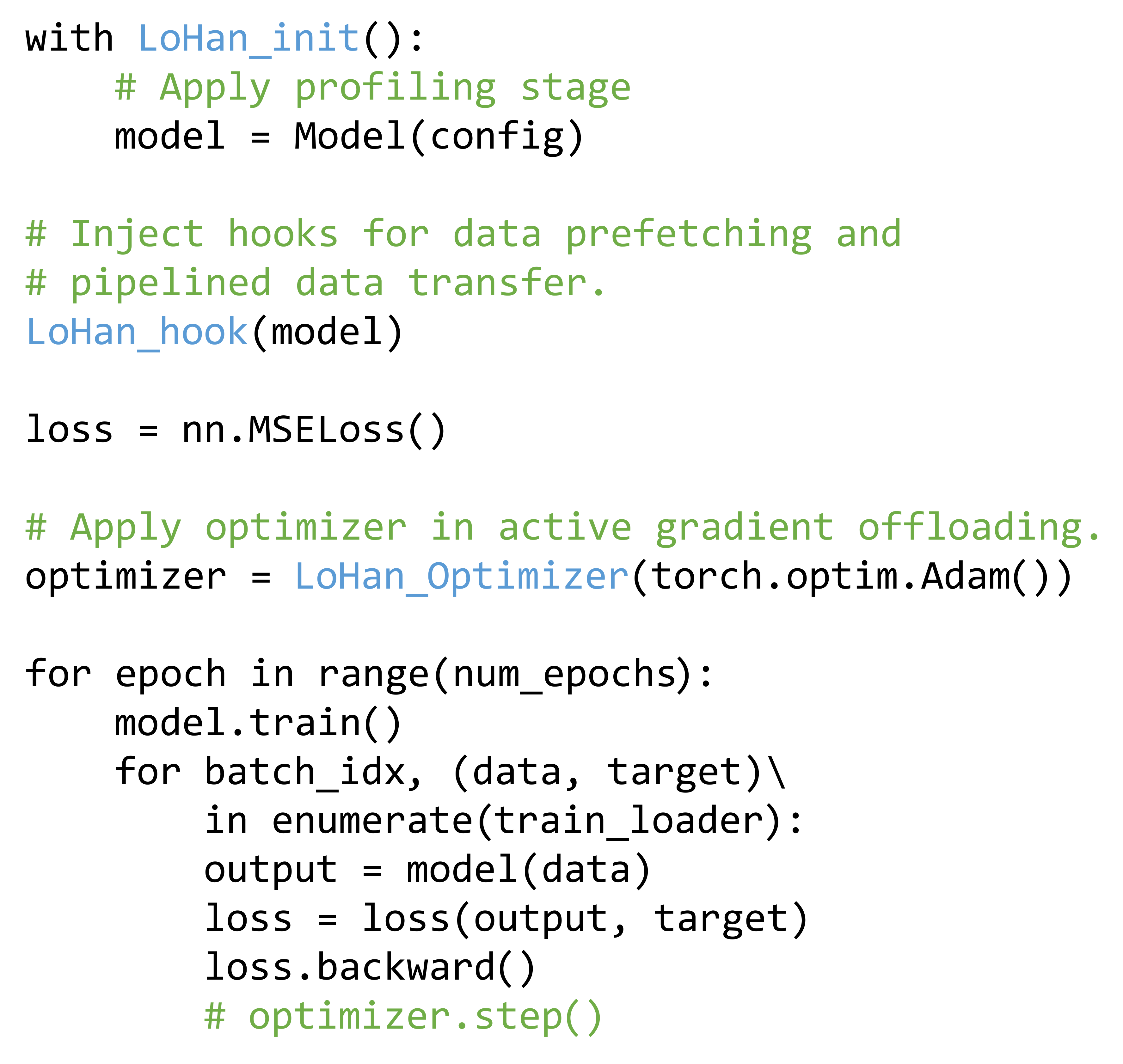}
    }
    \vspace{-1ex}
    \caption{User interface comparison of PyTorch and \SystemName{}.} 
    \vspace{-3ex}
    \label{fig_design_interface} 
\end{figure} 

\vspace{-1ex}
\subsection{Framework Integration}
We implement \SystemName{} on the top of the popular deep-learning framework PyTorch~\cite{pytorch}. \SystemName{} provides a set of wrappers to hide the implementation details, so that users can enable efficient model fine-tuning via \SystemName{} with only a few lines of code changes. Figure~\ref{fig_design_interface} shows the user interface comparison between PyTorch and \SystemName{}. \SystemName{} performs the profiling stage via the \narrow{\texttt{\SystemName{}\_init}} wrapper. Compared to PyTorch, \SystemName{} removes the optimizer execution from the serialized execution flow. Besides, \SystemName{}'s hooking the operators in the model enables the automatic activation management without explicit code change by users.

\begin{table}[t]
    \centering\footnotesize
    \vspace{-1ex}  
    \caption{Configurations of the evaluation server.}  
    \vspace{-1ex}  
    \label{table:Setup} 
    \begin{tabular}{|c|c|}   
        \hline
        \textbf{CPU}  & \makecell{Dual Intel Xeon Gold 5320 CPU @ 2.20GHz}  \\ \hline
        \textbf{Main Memory} & 768 GB 3200MHz DDR4 \\ \hline
        \textbf{PCIe} &  PCIe Gen 4  \\ \hline
        \textbf{GPU} & \makecell{NVIDIA GeForce RTX 3090/4080/4090} \\ \hline
        \textbf{SSD}  & 12$\times$ 3.84TB Intel P5510 SSDs \\ \hline 
        \textbf{CUDA Toolkit}  & 11.8 \\ \hline 
        \textbf{PyTorch}  & 2.0.0+cu118 \\ \hline 
    \end{tabular}  
    \vspace{-1ex} 
\end{table}

\begin{table} [t]
	\centering\footnotesize
    \vspace{-1ex}
    \caption{LLM for evaluation.}
    \vspace{-1ex}  
	\label{tab:modelsize}	
	\begin{tabular}{|c||c|c|c|}
		\hline
		\textbf{Model Size} & \textbf{\#Layers} & \textbf{\#Heads} & \textbf{Hidden Dimension}\\
		\hline
		\hline
            6B & 28 & 32 & 4096 \\
		\hline
            13B & 40 & 40 & 5120 \\
		\hline
		  30B & 48 & 56 & 7168 \\
		\hline
		  70B & 80 & 64 & 8192 \\
		\hline
		  135B & 88 & 88 & 11264 \\
		\hline
		  175B & 96 & 96 & 12288 \\
		\hline
            276B & 112 & 112 & 14336 \\
        \hline
            412B & 128 & 128 & 16384 \\
        \hline
	\end{tabular}
    \vspace{-4ex}
\end{table}

\begin{figure*}[t]
    \vspace{-1ex}
    \subfloat[Fine-tuning 13B model on RTX 4090]{
        \label{fig_exp_thpt_13b_4090}
        \includegraphics[width=0.33\linewidth]{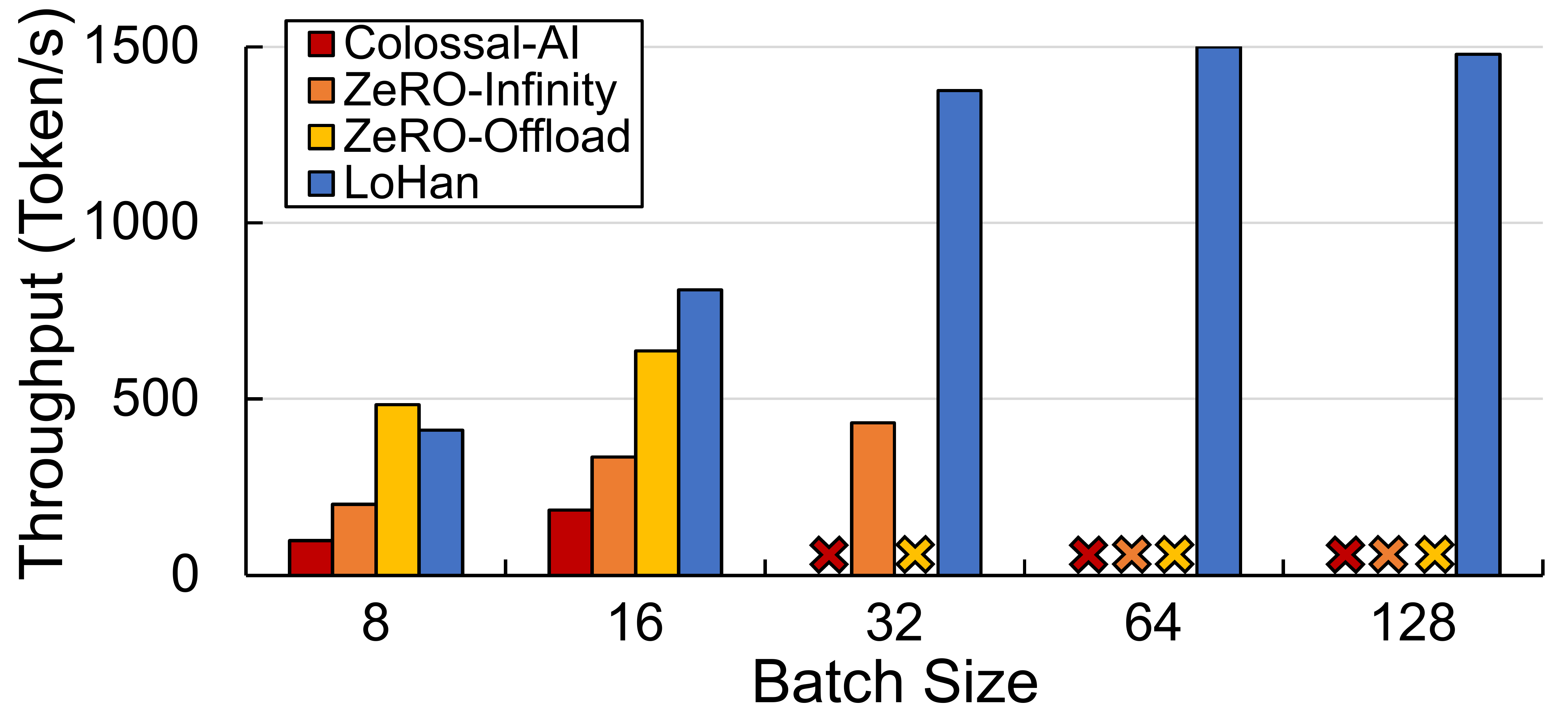}
    }
    \hfill
    \subfloat[Fine-tuning 13B model on RTX 3090]{
        \label{fig_exp_thpt_13b_3090}
        \includegraphics[width=0.28\linewidth]{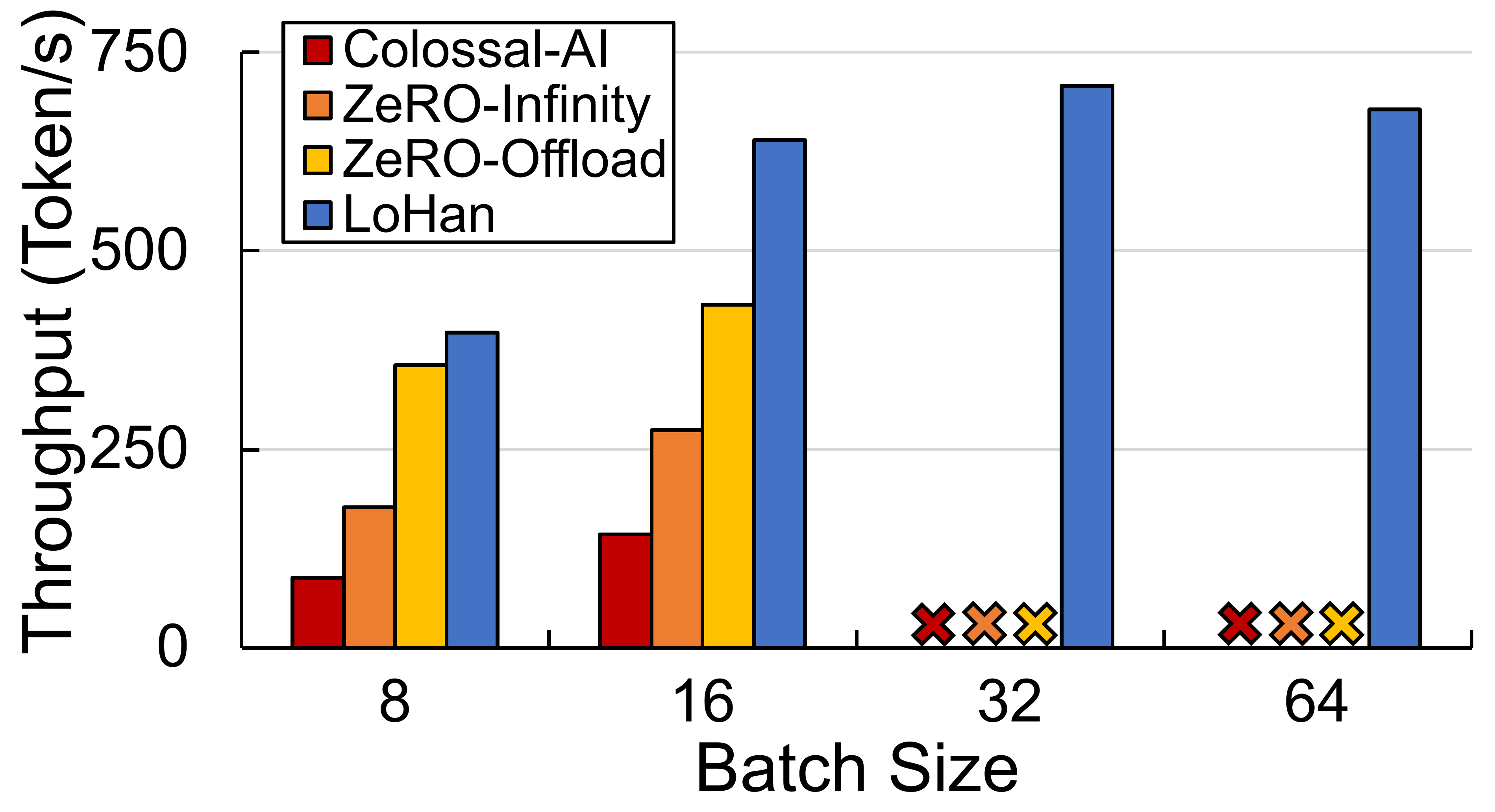}
    }
    \hfill  
    \subfloat[Throughput vs model size on RTX 4090]{
        \label{fig_exp_thpt_wrt_model_size}
        \includegraphics[width=0.28\linewidth]{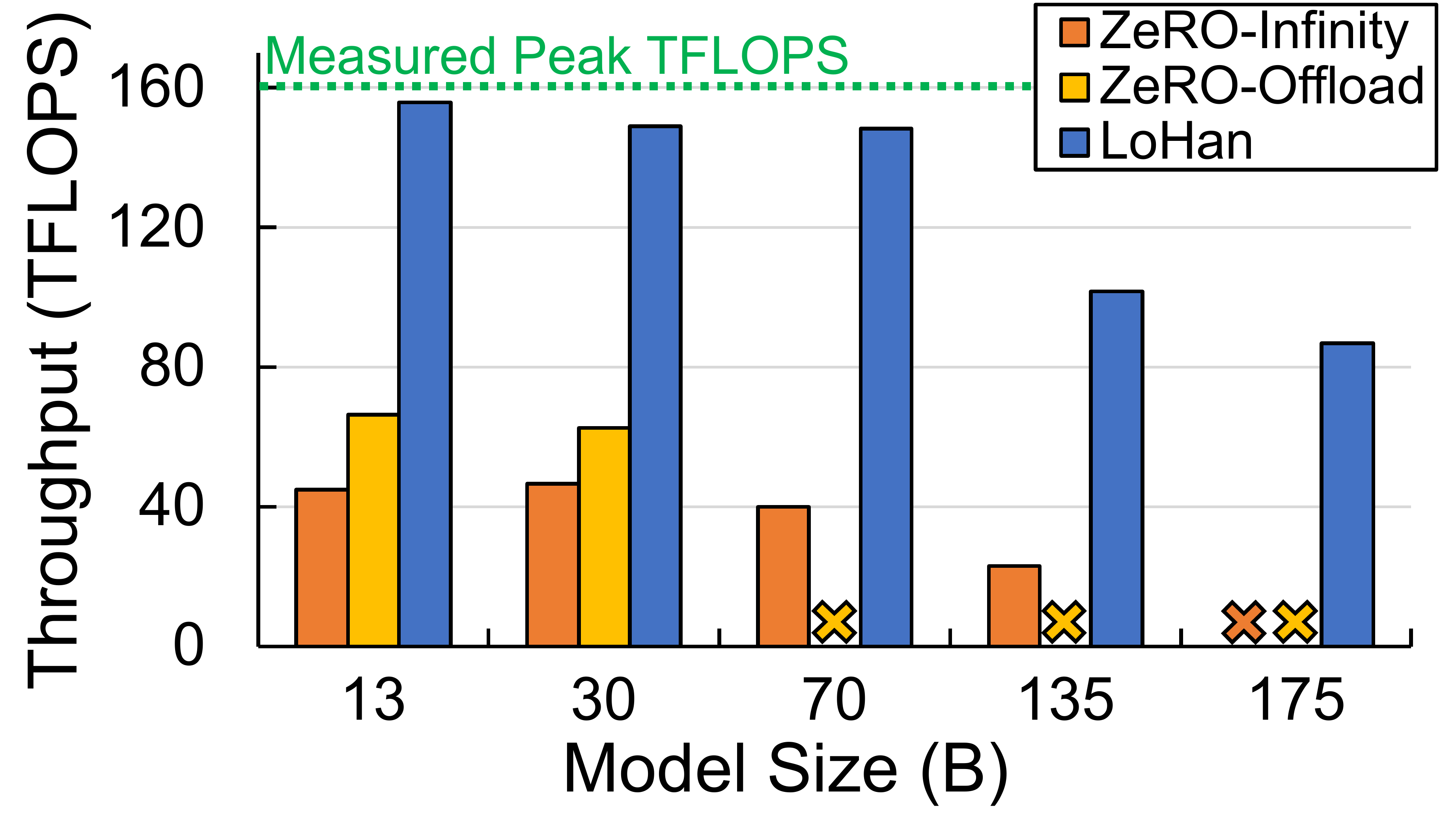}
    }
    \vspace{-1ex}
    \caption{End-to-end GPU throughput comparison between \SystemName{} and baselines with different batch sizes.} 
    \label{fig:all_compare} 
    \vspace{-2ex}
\end{figure*} 

\begin{figure}[t]
    \vspace{-2ex}
    \begin{minipage}{0.86\linewidth}
        \includegraphics[width=\linewidth]{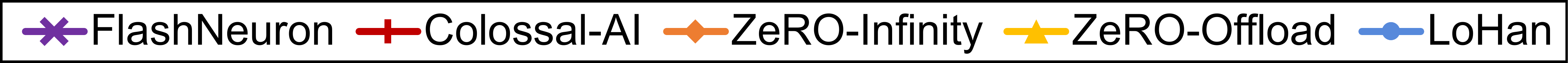}
        \vspace{-6ex}
    \end{minipage}
	\centering
     \subfloat[on RTX 4090 and 3090]{
        \label{fig_exp_max_model_4090}
        \includegraphics[width=0.465\linewidth]{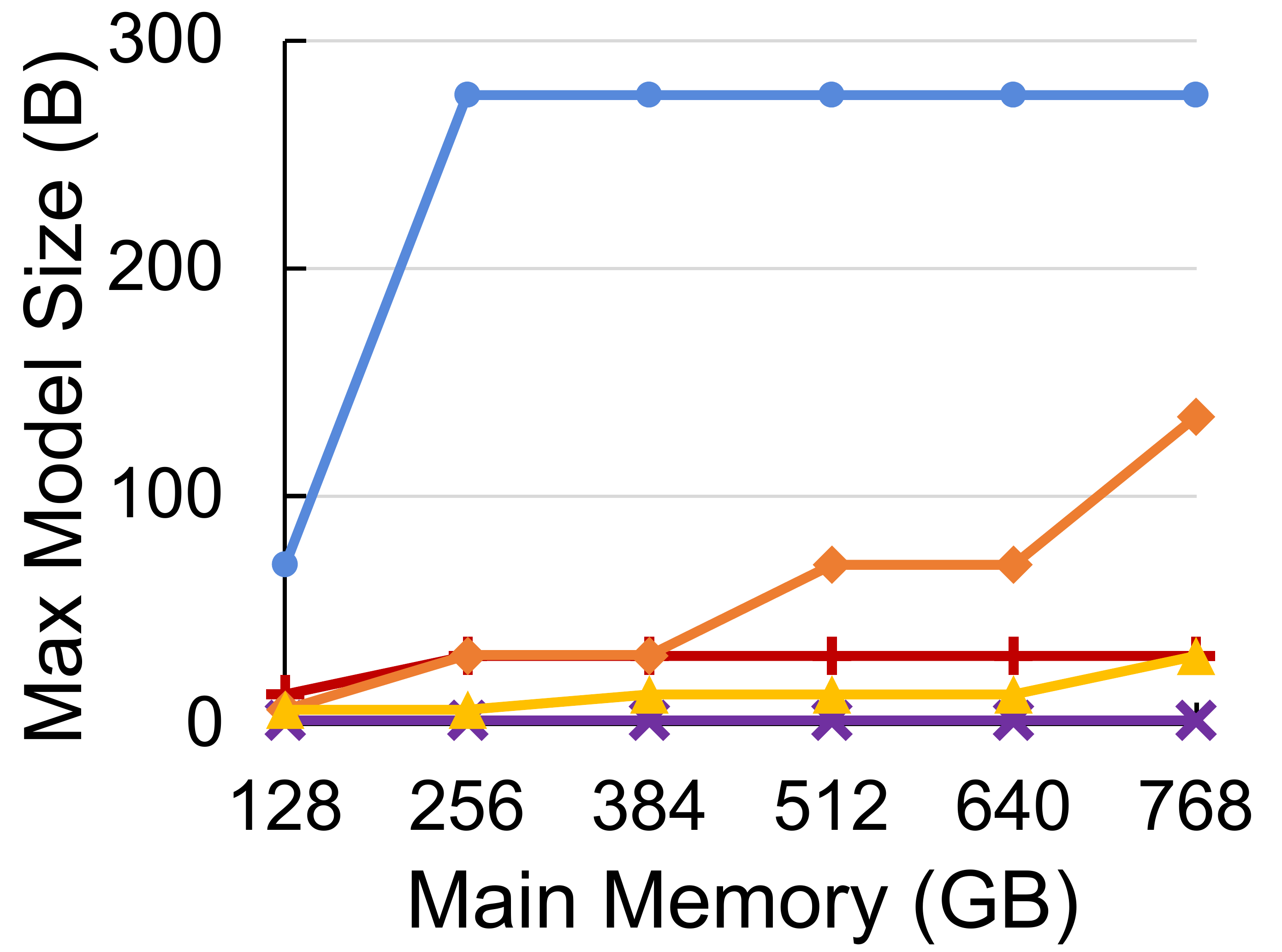}
    }
    \hfill    
    \subfloat[on RTX 4080]{
        \label{fig_exp_max_model_4080}
        \includegraphics[width=0.465\linewidth]{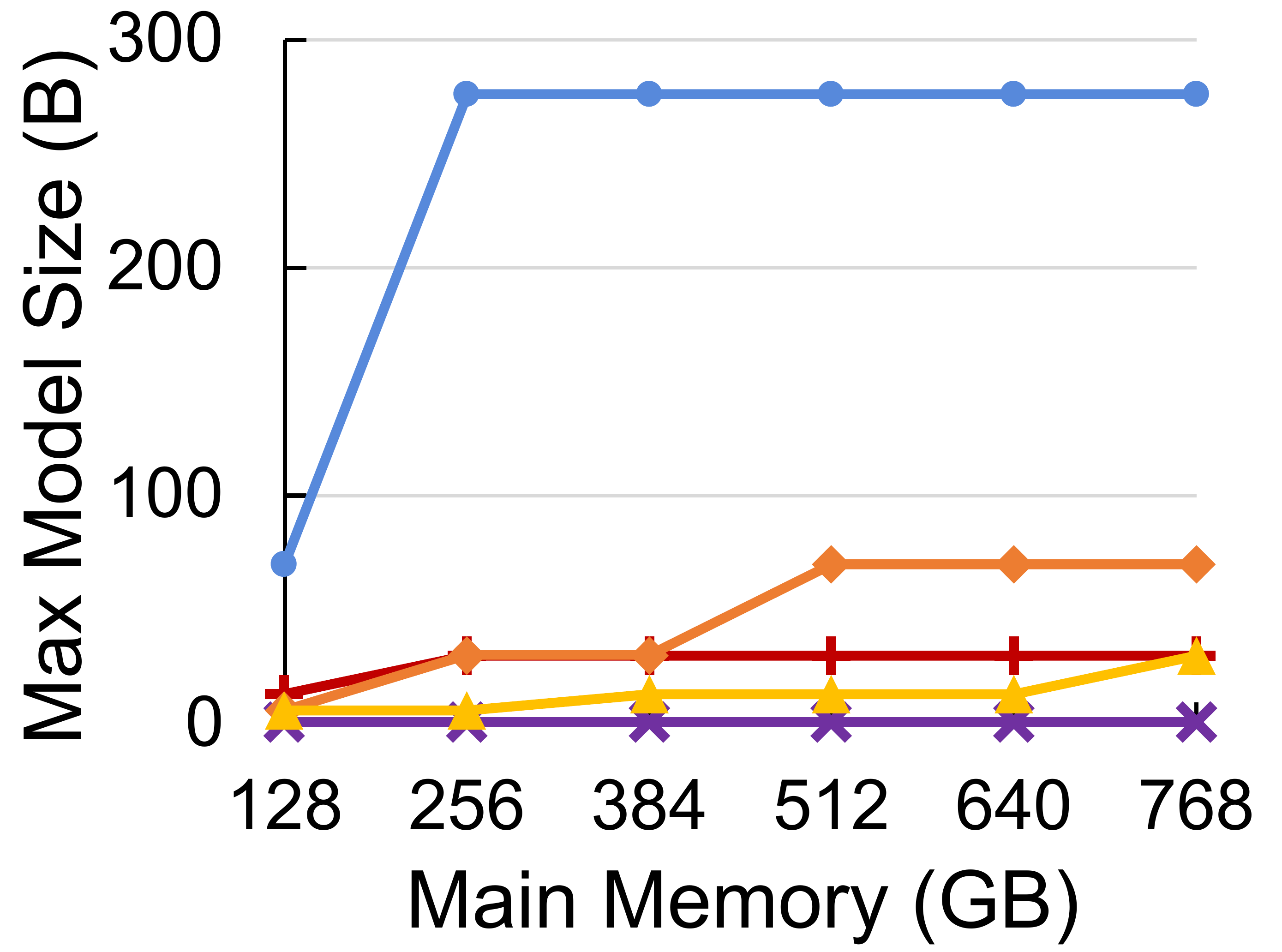}
    }
    \vspace{-1ex}
	\caption{Maximum trainable model size of \SystemName{} and baselines under different main memory capacities.}
    \vspace{-4ex}
	\label{fig_max_model_size}
\end{figure}

\section{Evaluation}

\subsection{Experimental Setup}
\label{sec:exp_setup}

\noindent\textbf{Evaluated Machine. } We perform all the experiments on a server whose configurations are summarized in Table~\ref{table:Setup}.

\noindent\textbf{Workloads.} We choose a series of decoder-only models for our experiments. The hyperparameter choice of the models follows GPT-3~\cite{gpt3} and open-source pre-trained models like OPT~\cite{opt} and are listed in Table~\ref{tab:modelsize}.  
We simply randomly initialize model parameters and datasets for evaluations that do not require model convergence. We train the models in mixed precision that is widely adopted in LLM fine-tuning. In our experiments, the sequence length is set to 1024 and the vocabulary size is 50257. 

\noindent\textbf{Baseline Configurations. }
We choose the following open-source baselines for evaluation. 

The first baseline is ZeRO-Infinity~\cite{zero-infinity} and ZeRO-Offload~\cite{zero-offload} from DeepSpeed. The former offloads model states to SSDs while the latter offloads model states to main memory. Both systems swap the inter-transformer block activations to main memory and recompute the intra-block activations. We evaluate with DeepSpeed version 0.9.3 and disable the one-step delayed optimizer of ZeRO-Offload since it introduces parameter staleness. 

The second baseline is Colossal-AI~\cite{colossal-ai}, a popular billion-scale model training solution. Colossal-AI keeps the inter-block activations in GPU memory and recomputes the intra-block activations. We evaluate with Colossal-AI version 0.3.5 and enable the Gemini memory manager~\cite{colossal-gemini, patrickstar}.

The third baseline is FlashNeuron~\cite{flashneuron}, which only offloads activations to SSDs. We implement a prototype of FlashNeuron using POSIX file API instead of GPUDirect to offload activations to main memory, and then to SSDs, so that FlashNeuron can run on our consumer-grade GPUs. 

\vspace{-1ex}
\subsection{Maximum Trainable Model Size}
\label{sec:exp_model_size}
We first compare the maximum trainable model size of \SystemName{} and the baselines by fine-tuning the models on three consumer-grade GPUs, namely RTX 4090, 3090 (24GB device memory), and 4080 (16 GB device memory), with different main memory capacities. We set the batch size to 1 to minimize effect of the batch size. To limit main memory capacity, we pin a certain amount of memory so that the evaluated systems cannot utilize the pinned memory. We further disable Linux swap partition. 

Figure~\ref{fig_max_model_size} illustrates the comparison results.  
\SystemName{} is able to fine-tune significantly larger models than the baselines under any GPU and main memory capacities, because \SystemName{} fully leverages the memory capacities of main memory and GPU to its best by holistically offloading model states and activations. \SystemName{} enables the fine-tuning of a 276B model under 768 GB main memory on RTX 4090, which is 2.04$\times$ larger than that of ZeRO-Infinity. \SystemName{} succeeds in training a 175B model even with only 256~GB main memory and RTX 4080, which is reachable by most researchers. 

\vspace{-1ex}
\subsection{End-to-End Throughput Comparison}
\label{sec:exp_endtoend}
\noindent\textbf{Throughput w.r.t. Batch Size. }
To demonstrate the efficiency of \SystemName{}, we first compare the end-to-end training throughput of \SystemName{} and the three baselines. We employ \SystemName{} and the baselines to fine-tune the 13B model on both RTX 4090 and 3090 with different batch sizes. 

Figure~\ref{fig_exp_thpt_13b_4090} shows the throughput when fine-tuning the 13B model on RTX 4090. We observe that \SystemName{} achieves 2.32$\times$, 3.46$\times$, and 8.02$\times$ higher throughput over ZeRO-Offload, ZeRO-Infinity, and Colossal-AI, respectively. The figure does not include FlashNeuron which fails to fine-tune the model on RTX 4090, because it only offloads activation checkpoints to SSDs while keeping massive model states in GPU memory, thus requiring much larger GPU memory space than the 24GB memory capacity of RTX 4090. 

Figure~\ref{fig_exp_thpt_13b_3090} shows the throughput when fine-tuning the 13B model on RTX 3090. \SystemName{} achieves 1.57$\times$, 2.48$\times$, and 4.72$\times$ improvements over ZeRO-Offload, ZeRO-Infinity, and Colossal-AI, respectively, showing a similar trend as on 4090.  

\noindent\textbf{Throughput w.r.t. Model Size. }
We compare the maximum TFLOPS of \SystemName{}, ZeRO-Infinity, and ZeRO-Offload fine-tuning different models on 4090, as shown in Figure~\ref{fig_exp_thpt_wrt_model_size}, where the green line indicates the peak FLOPS measured by benchmarking a transformer block inside the GPU without any PCIe traffic. 

We observe that \SystemName{} achieves 90\%\textasciitilde 95\% of peak FLOPS when the model size is smaller than 70B, while the baselines achieve only 40\% at most. \SystemName{} maintains a relatively small 53\% of peak FLOPS when fine-tuning a 175B model, because a single layer of a large model has a large size of parameters and activations, thus the allowable batch size is small to fit in limited GPU memory. However, this is still significantly higher than ZeRO-Infinity at its maximum FLOPS. 

\noindent\textbf{Conclusion. }
\SystemName{} can fine-tune the 175B model on RTX 4090 while the baselines cannot. Besides, \SystemName{} achieves significantly higher throughput than the baselines, indicating that \SystemName{} enables efficient fine-tuning on large-scale models. 

\begin{figure}[t]
    \vspace{-5ex}
    \subfloat[Fine-tuning 13B Model]{
        \label{fig_exp_grad_off_13b}
        \includegraphics[width=0.545\linewidth]{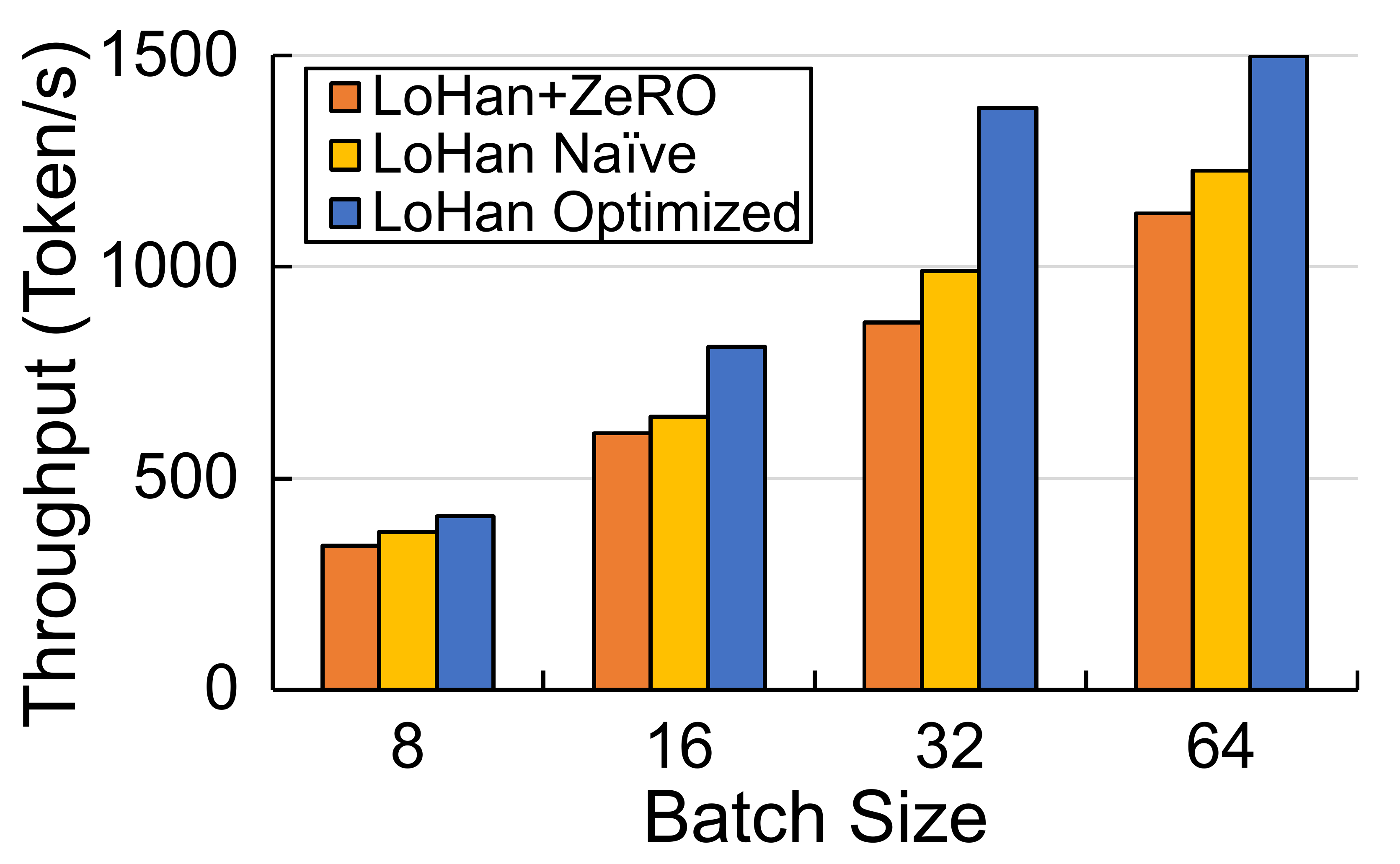}
    }
    \hfill    
    \subfloat[Fine-tuning 175B Model]{
        \label{fig_exp_grad_off_175b}
        \includegraphics[width=0.385\linewidth]{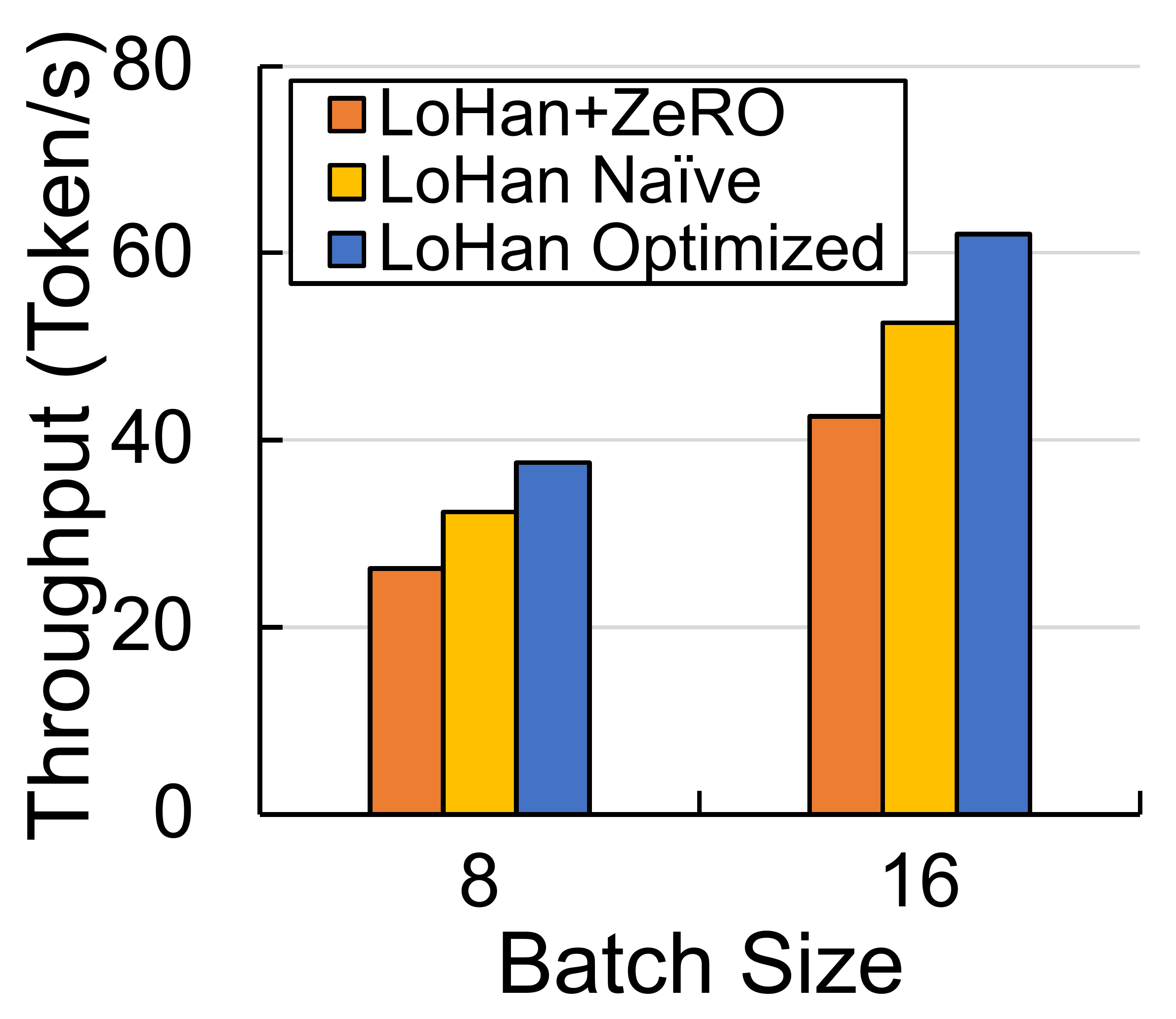}
    }
    \vspace{-1ex}
    \caption{Effect of active gradient offloading.} 
    \vspace{-4ex}
    \label{fig_exp_grad_off} 
\end{figure} 

    \vspace{-1ex}
\subsection{Effect of Active Gradient Offloading}
\label{sec:exp_parallization}
To demonstrate the benefits of active gradient offloading (Subsection~\ref{subsec_backward_optimizer}), we test \SystemName{} with three implementations: 1)~\SystemName{} Optimized, the implementation with the optimized active gradient offloading, 2)~\SystemName{} Na\"ive that implements a na\"ive active gradient offloading, and 3)~\SystemName{}+ZeRO that does not overlap backward and optimizer execution as ZeRO-Infinity does. All implementations follow the same training procedure except the gradient offloading strategy. 

We test the implementations by fine-tuning the 13B and 175B models on RTX 4090 GPU, as shown in Figure~\ref{fig_exp_grad_off}. We make two observations. 

First, the optimized active gradient offloading generally achieves higher performance gain than the na\"ive active gradient offloading. For example, \SystemName{} Optimized achieves 1.22$\times$ throughput than \SystemName{} Na\"ive and 1.33$\times$ throughput than \SystemName{}+ZeRO when fine-tuning 13B model with a batch size of 64. This is because the optimized active gradient offloading completely overlaps CPU computation and SSD I/O, thus minimizing the GPU's idle time.

Second, the throughput gain of the active gradient offloading over serializing backward stage and the CPU optimizer drops when the batch size is too small (e.g., 8), because GPU backward propagation costs significantly less time than CPU optimizer, thus resulting in fewer overlapping opportunities.

\begin{figure}[t] 
    \subfloat[With 128 GB main memory.]{
        \label{fig_exp_act_ssd_128gb}
        \includegraphics[width=0.465\linewidth]{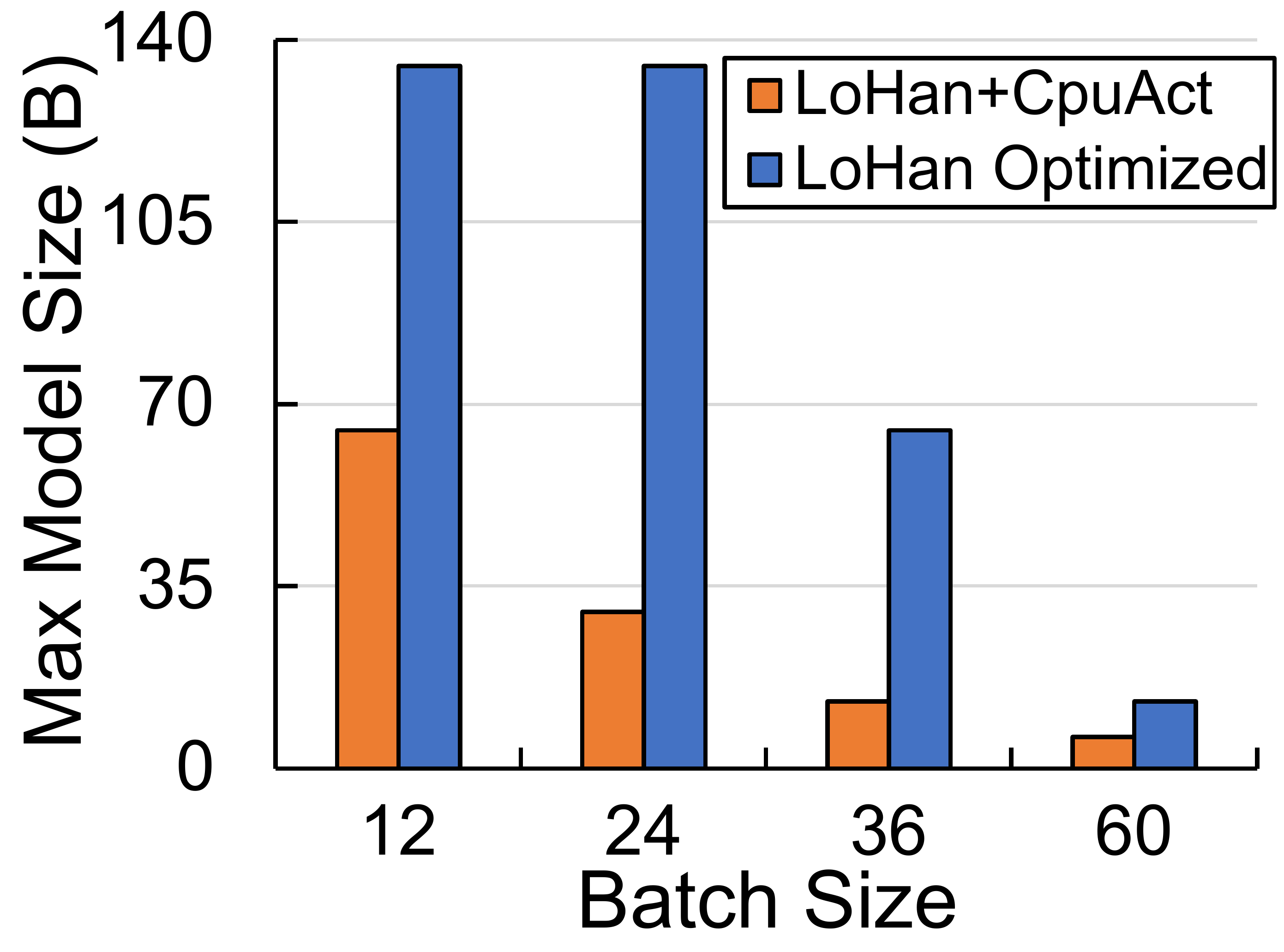}
    }
    \subfloat[With 256 GB main memory.]{
        \label{fig_exp_act_ssd_256gb}
        \includegraphics[width=0.465\linewidth]{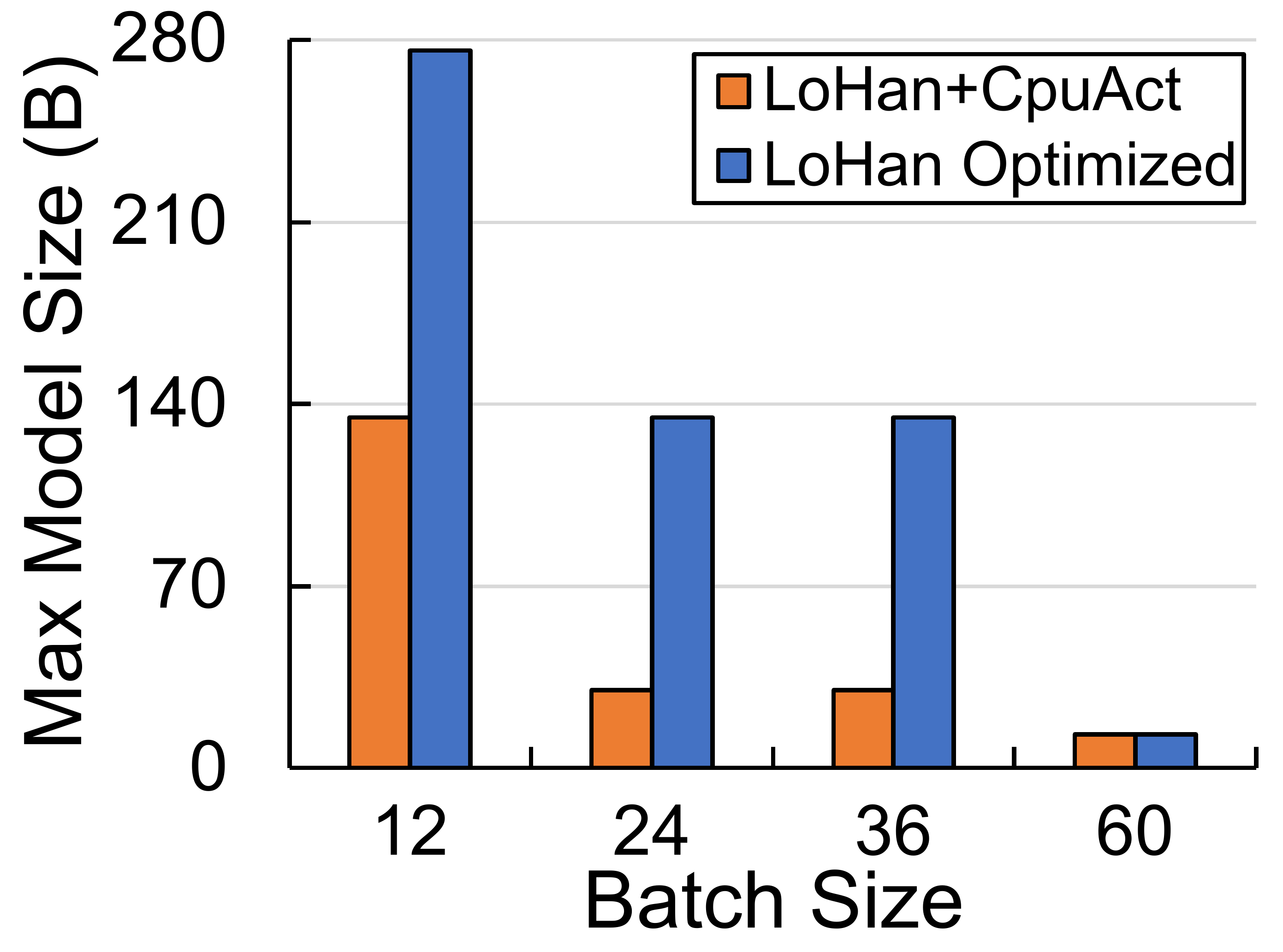}
    }
    \vspace{-1ex}
    \caption{Effect of swapping activations to SSDs.}
    \label{fig_exp_act_ssd}
    \vspace{-3ex}
\end{figure}

  \vspace{-1ex}
\subsection{Effect of Holistic Traffic-Aware Activation Management}
\label{sec:exp_swapping}

To demonstrate the benefit of the holistic traffic-aware activation swapping management (Subsection~\ref{subsec_act_swap}), we first show the benefit of swapping activations to SSDs rather than to main memory in maximum trainable model size, then show the throughput gain of the activation management strategy.

\noindent\textbf{Benefit of Swapping Activations to SSDs.}
\
To show the benefit of swapping activations to both main memory and SSDs rather than only to main memory, we evaluate \SystemName{} (\SystemName{} Optimized) and an implementation \SystemName{}+CpuAct, which follows the same training procedure as \SystemName{} except that \SystemName{}+CpuAct swaps activations only to main memory rather than to SSDs. We measure the maximum trainable model sizes of two implementations fine-tuning on RTX 4090 with different main memory and batch sizes. 

Figure~\ref{fig_exp_act_ssd} illustrates the comparison results. We observe that 1)~swapping activations to SSDs significantly enlarges the trainable model size in the single-GPU commodity server with scarce main/GPU memory. For example, \SystemName{} Optimized can fine-tune 2$\times$\textasciitilde 5$\times$ larger model than \SystemName{}+CpuAct with 128~GB main memory. 2)~The difference in trainable model size is not significant when the batch size is too large, e.g., the maximum model size of two implementations is the same with 256 GB main memory and batch size of 60, because when the batch size is too large, the maximum trainable model size is bounded by accommodating activations of a single layer in limited GPU memory capacity, rather than main memory. 

\noindent\textbf{Benefit of the Activation Management Strategy.}
\
To validate the effectiveness of the activation management strategy, we evaluate \SystemName{} on the 70B model with five implementations: 1)~\SystemName{}+Optimized that uses holistic traffic-aware activation management to swap activations, 2)~\SystemName{}+ZeRO that statically swaps the inter-layer activations of each transformer block to main memory and recompute the rest, 3)~\SystemName{}+Cap that smartly keeps, recomputes or swaps the activations to CPU by profiling the overhead of activation swapping and recomputation as proposed by Capuchin~\cite{capuchin}, 4)~\SystemName{}+G10 that smartly keeps or swaps activations to SSDs based on inactive time measurement as proposed by G10~\cite{g10}, and 5)~\SystemName{}+CM that smartly recomputes or offloads activations to main memory with a cost-model and MILP solver proposed by Checkmate~\cite{checkmate}.
All the implementations offload model states to SSDs and execute optimizer in CPU, which is necessary for 70B model fine-tuning. 

\begin{table}[t]
	\begin{center}
        \footnotesize
        \caption{Batch size adopted by different activation management strategies fine-tuning the 70B model.}
        \vspace{-1ex}
        \label{tab:act_swap_batch_size}
        \centering
        \begin{tabular}{|c||c|c|c|} \hline 
             \textbf{Main Memory Size}&  \textbf{128 GB}&  \textbf{256 GB}& \textbf{512 GB}\\ \hline \hline
             \textbf{\SystemName{}+DS}&  16&  24& 32\\ \hline 
             \textbf{\SystemName{}+Cap}&  16&  24& 32\\ \hline 
             \textbf{\SystemName{}+G10}&  32&  32& 32\\ \hline
             \textbf{\SystemName{}+CM}&  Failed&  24& 32\\ \hline
             \textbf{\SystemName{}+Optimized}&  32&  32& 32\\ \hline
        \end{tabular}
    \end{center}
    \vspace{-6ex}
\end{table}

\begin{figure}[t]  
    \vspace{-1ex}
    \subfloat[Throughput of \SystemName{} integrating different activation management strategies.]{
        \includegraphics[width=0.465\linewidth]{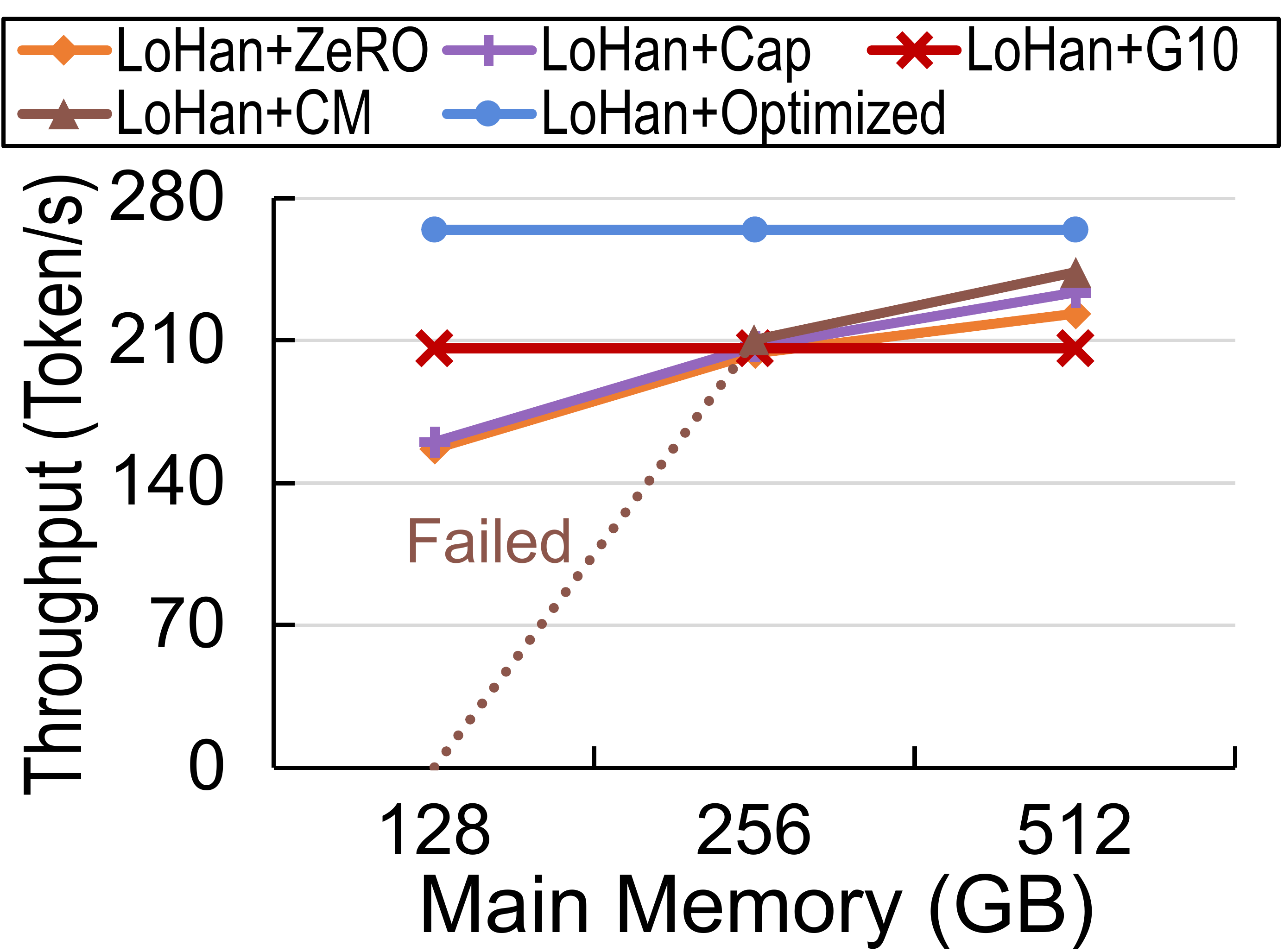}
        \label{fig_exp_swap_comparison}
    }
    \hfill  
    \subfloat[Iteration time of \SystemName{} with different amounts of swapped activations. Stars are predicted optimal size.]{
        \includegraphics[width=0.465\linewidth]{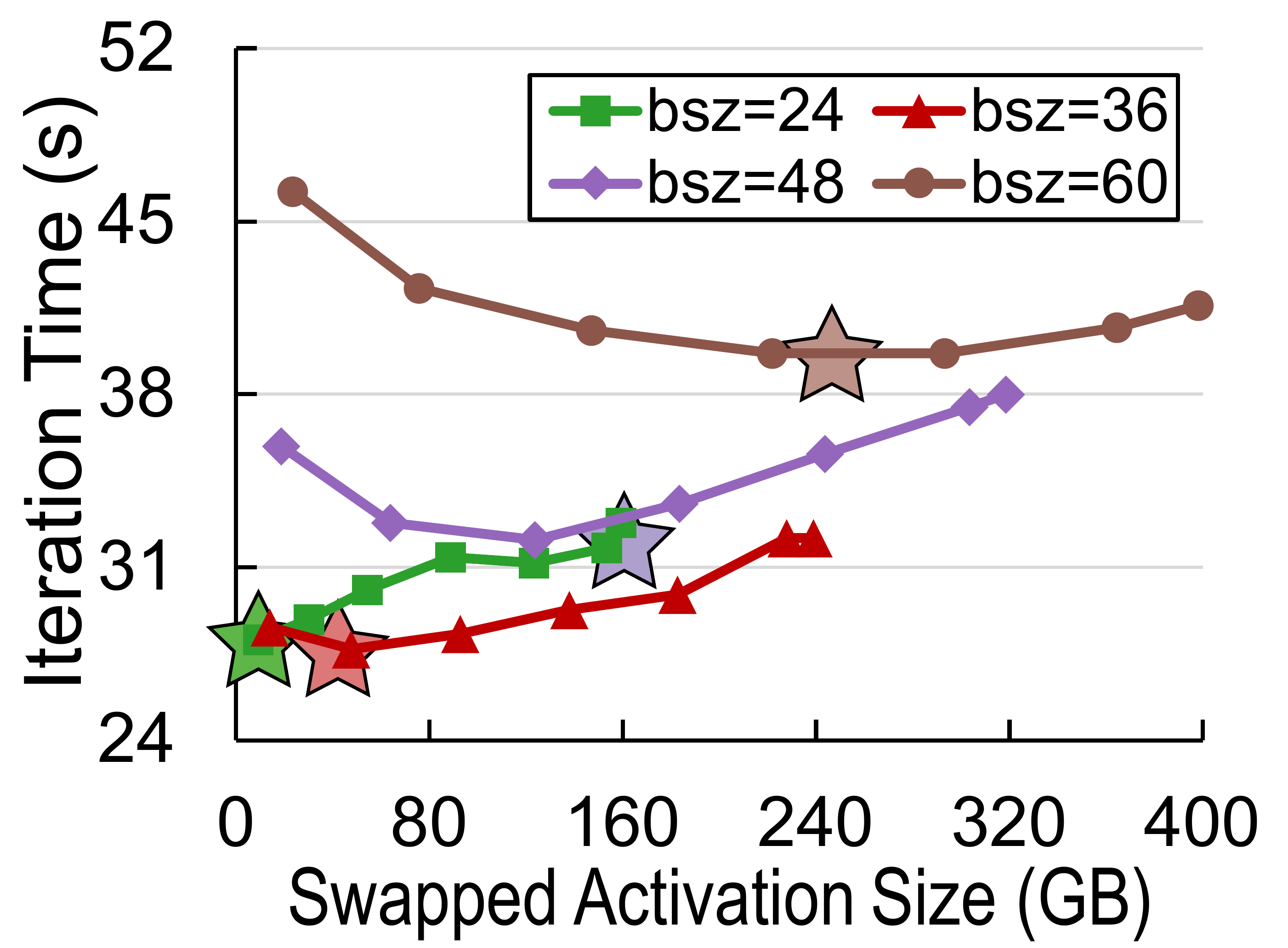}
        \label{fig_exp_swap_strategy}
    }
    \vspace{-1ex}
    \caption{Effect of activation management strategy.} 
    \vspace{-4ex}
    \label{fig:swap} 
\end{figure} 

Figure~\ref{fig_exp_swap_comparison} shows the throughput comparison and Table~\ref{tab:act_swap_batch_size} shows the respective batch sizes. We observe that 1)~the performance of all baselines except G10 drops when with less main memory capacity because these systems swap activations only to main memory, thus limiting the achievable batch size with scarce memory capacity. In contrast, \SystemName{} achieves steady throughput by offloading activations to SSDs; and 2)~With the same batch size (e.g., the batch size is 32 with 512 GB main memory), \SystemName{} achieves higher throughput than all the baselines, because \SystemName{}’s offloading strategy is holistic, considering the traffic from both activations and model states.

Further, we show that \SystemName{} predicts the optimal amount of swapped activations. To illustrate this, we test the iteration time of \SystemName{} when fine-tuning the 13B model with different amounts of swapped activation. Figure~\ref{fig_exp_swap_strategy} illustrates the results with batch sizes of 24, 36, 48, and 60, where stars indicate the predicted optimal amount of swapped activations. We observe that 1)~for all batch sizes, \SystemName{}'s iteration time model produces nearly optimal predictions according to the experimental results. 2)~The iteration time increases as the swapped activation amount increases at a batch size of 24, which fits case 1 deduced in Subsection~\ref{subsec_act_swap}. Meanwhile, the trend of iteration time with regard to swapped activation amount fits well with deduced case 3 at batch sizes of 36, 48, and 60, showing the correctness of \SystemName{}'s iteration time model and the preciseness of the profiling stage.

\begin{figure}[t]  
    \vspace{-1ex}
    \subfloat[Maximum throughput of \SystemName{} and ZeRO-Infinity when fine-tuning 135B model.]{
        \includegraphics[width=0.465\linewidth]{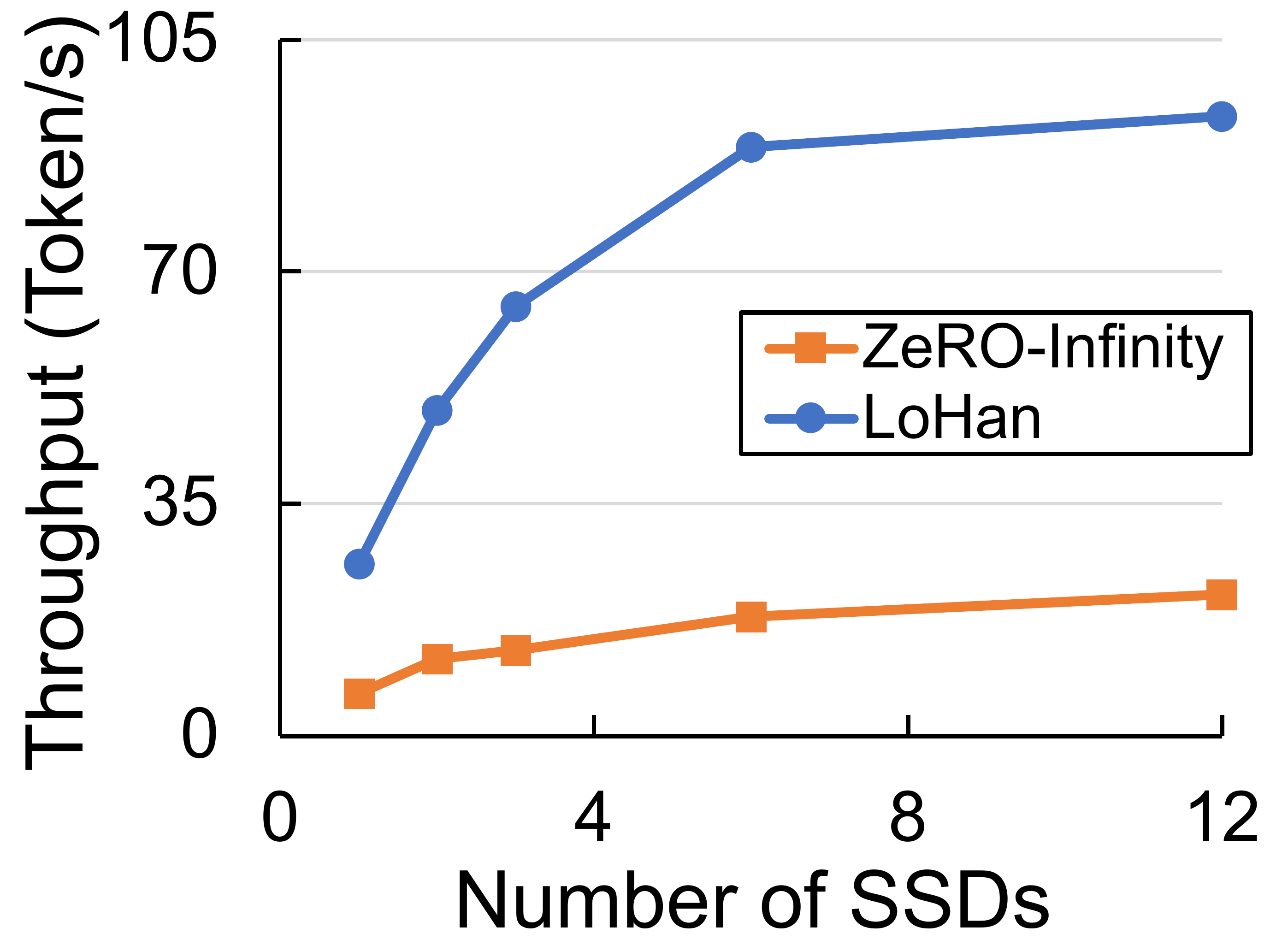}
        \label{fig_exp_scalability_135b}
    }
    \hfill  
    \subfloat[Throughput of \SystemName{} when fine-tuning 13B model with different batch sizes.]{
        \includegraphics[width=0.465\linewidth]{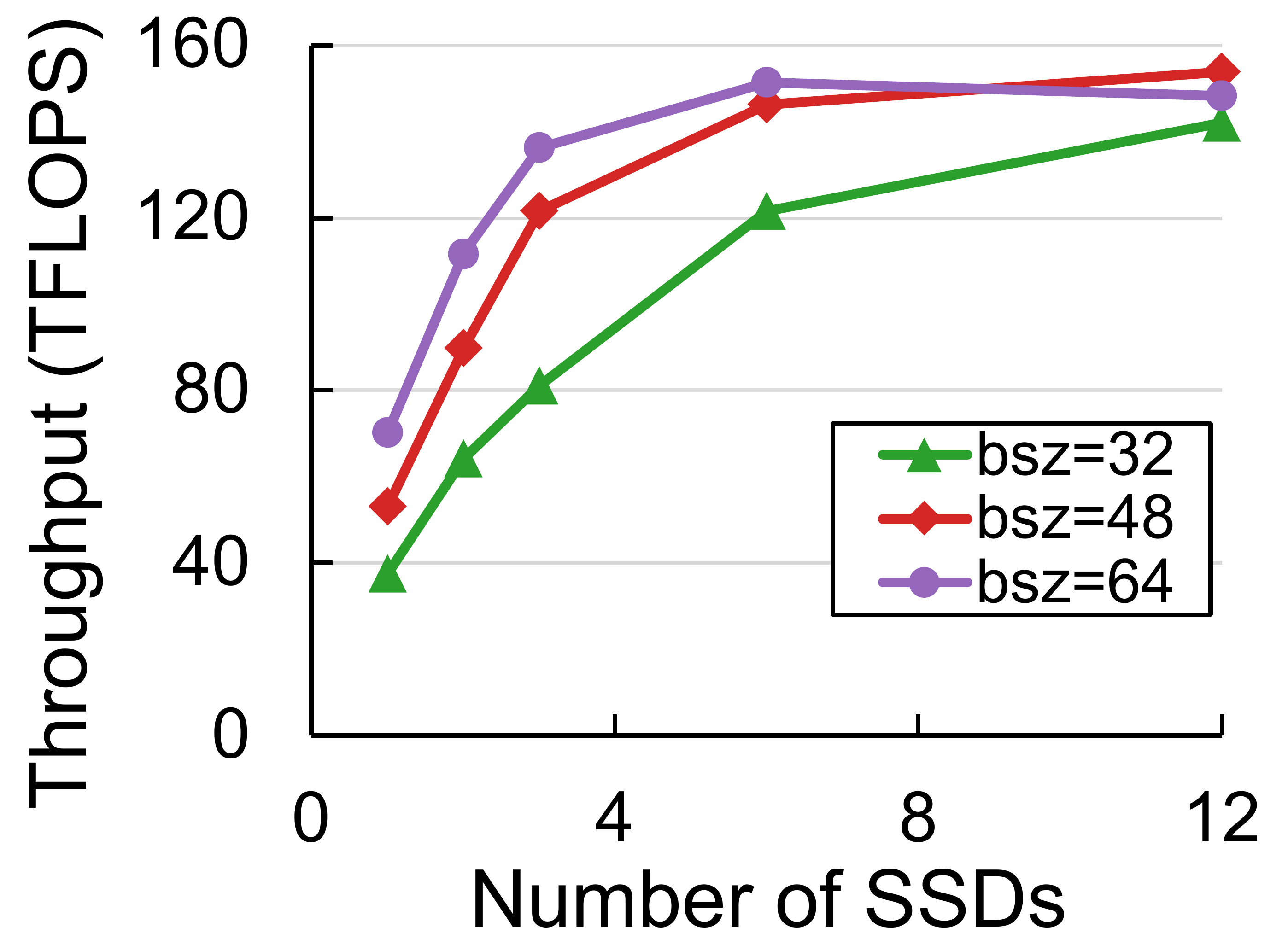}
        \label{fig_exp_scalability_13b}
    }
      \vspace{-1ex}
    \caption{Effect of the Number of SSDs.} 
    \vspace{-4ex}
    \label{fig_exp_scalability} 
\end{figure} 

\vspace{-1ex}
\subsection{Effect of the Number of SSDs}
\label{sec:exp_num_ssd}
We study \SystemName{}'s throughput w.r.t. number of SSDs. We evaluate the maximum training throughput of \SystemName{} and ZeRO-Infinity when fine-tuning the 135B model (the largest model ZeRO-Infinity can fine-tune) on RTX 4090 with different numbers of SSDs. We adopt the largest batch size the two systems can fine-tune. Figure~\ref{fig_exp_scalability_135b} illustrates the effect of the number of SSDs on the achievable throughput. 

We make three observations. First, \SystemName{} achieves near linear scalability when the SSD number increases from 1 to 3, indicating that SSD I/O is the training bottleneck in this case and \SystemName{} aggregates the bandwidth of multiple SSDs well. Second, \SystemName{}'s throughput gain is small as the SSD number increases from 6 to 12. This is because the system bottleneck with adequate I/O bandwidth has shifted to GPU computation and GPU-main memory PCIe transfer. Third, ZeRO-Infinity's throughput grows slowly as the number of SSDs increases. This is because ZeRO-Infinity almost serializes GPU computation, CPU computation, and SSD access, thus the I/O bandwidth of SSDs is not well utilized.

To further study the throughput characteristics w.r.t. SSD number, we measure the training throughput of \SystemName{} when fine-tuning the 13B model on RTX 4090 with different SSD numbers and batch sizes.
Figure~\ref{fig_exp_scalability_13b} illustrates the result. 

We make two observations. First, \SystemName{}'s throughput is maximized under sufficient cheap SSDs. With more than 6 SSDs for batch sizes of 32 and 48 or 12 SSDs for batch size of 64, \SystemName{} achieves the near maximum throughput. Second, \SystemName{} needs fewer SSDs to reach maximum throughput when using a larger batch size. For example, It requires 12 SSDs to achieve 135 TFLOPS with a batch size of 32, while requiring 6 and 3 SSDs for batch sizes 48 and 64 respectively.

\begin{figure}[t]  
    \vspace{-1ex}
    \subfloat[13B model with 2 GPUs.]{
        \includegraphics[width=0.535\linewidth]{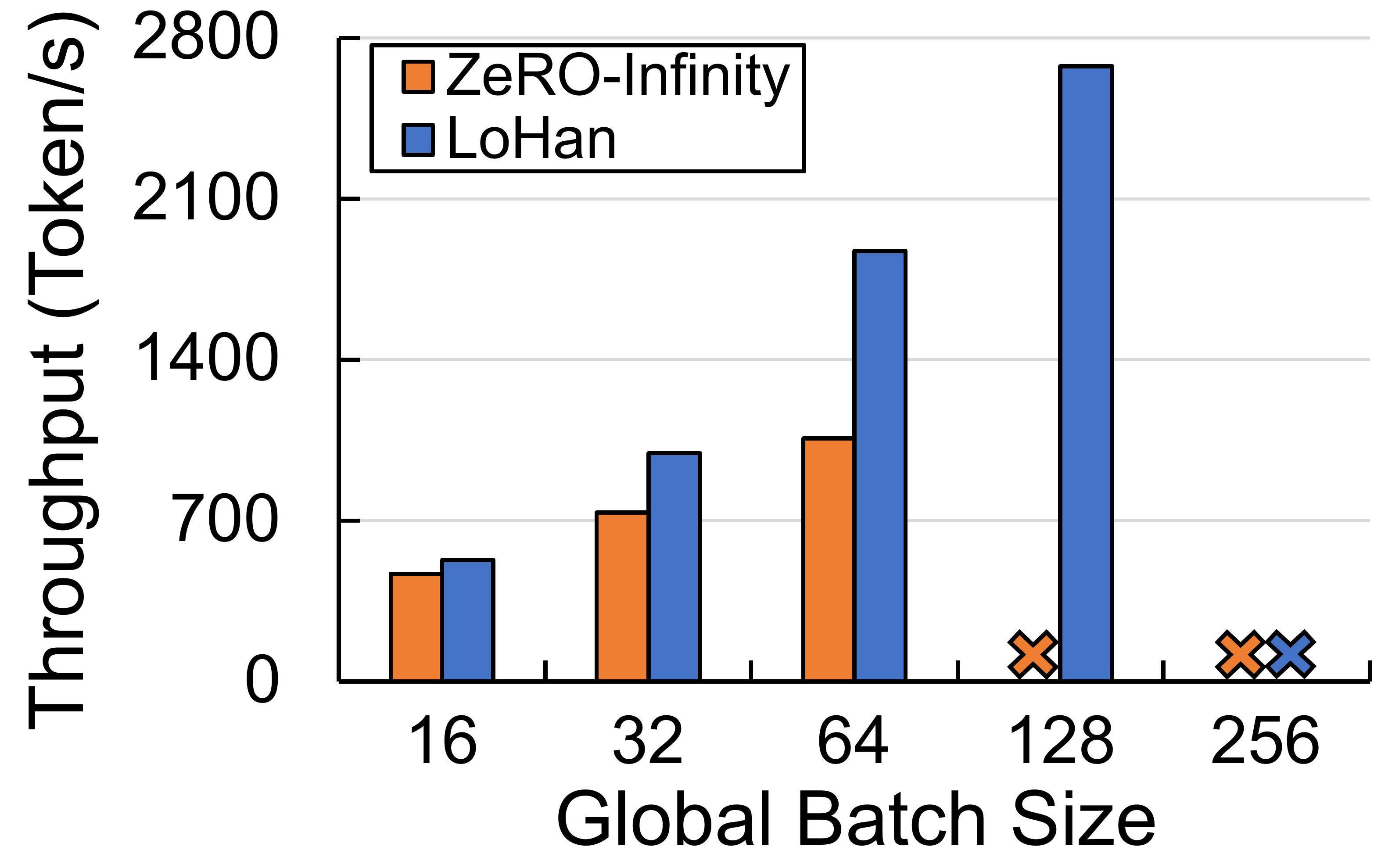}
    }
    \hfill
    \subfloat[70B model with 2 GPUs.]{
        \includegraphics[width=0.395\linewidth]{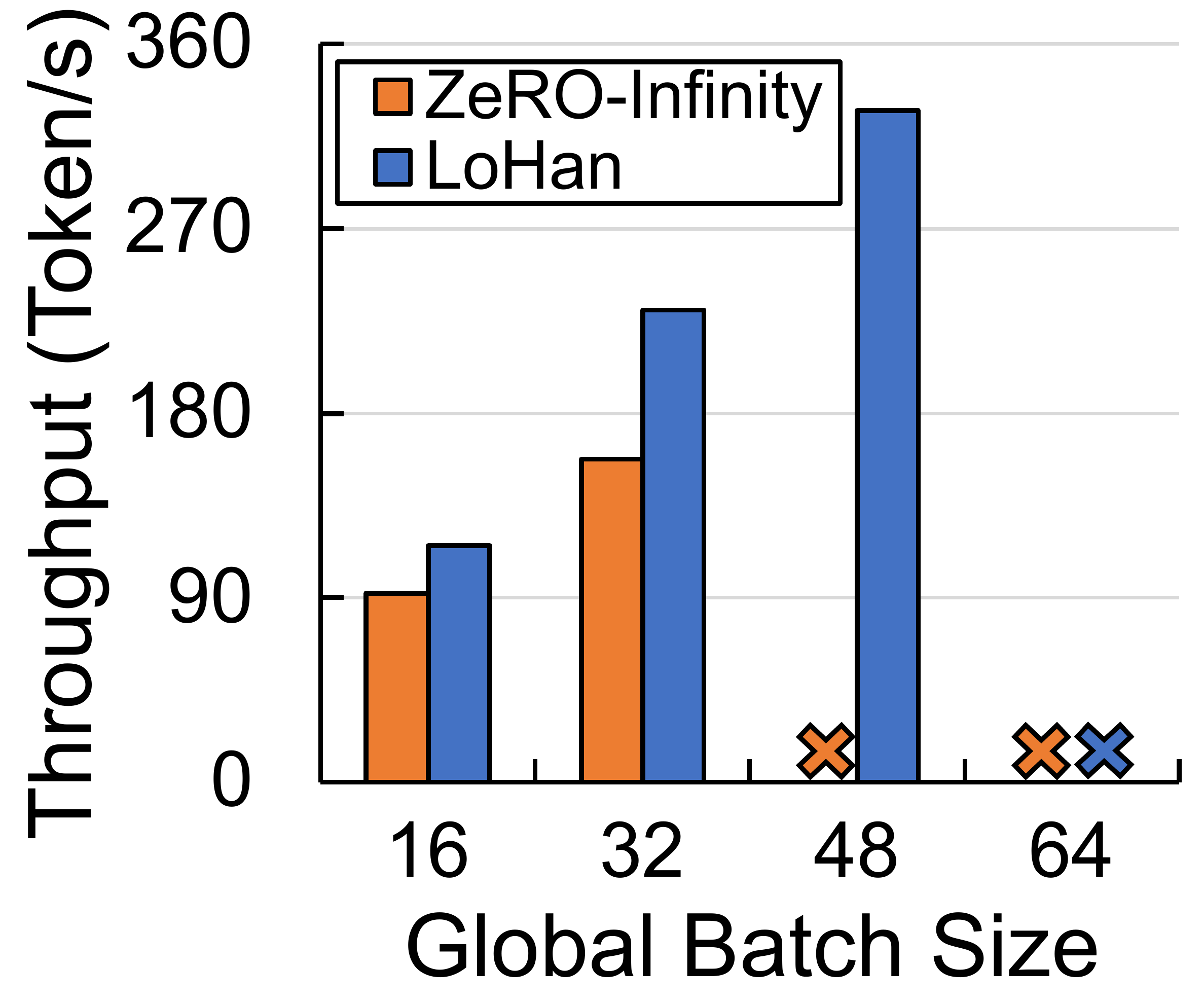}
    }
    \vspace{-2ex}\\
    \subfloat[13B model with 4 GPUs.]{
        \includegraphics[width=0.535\linewidth]{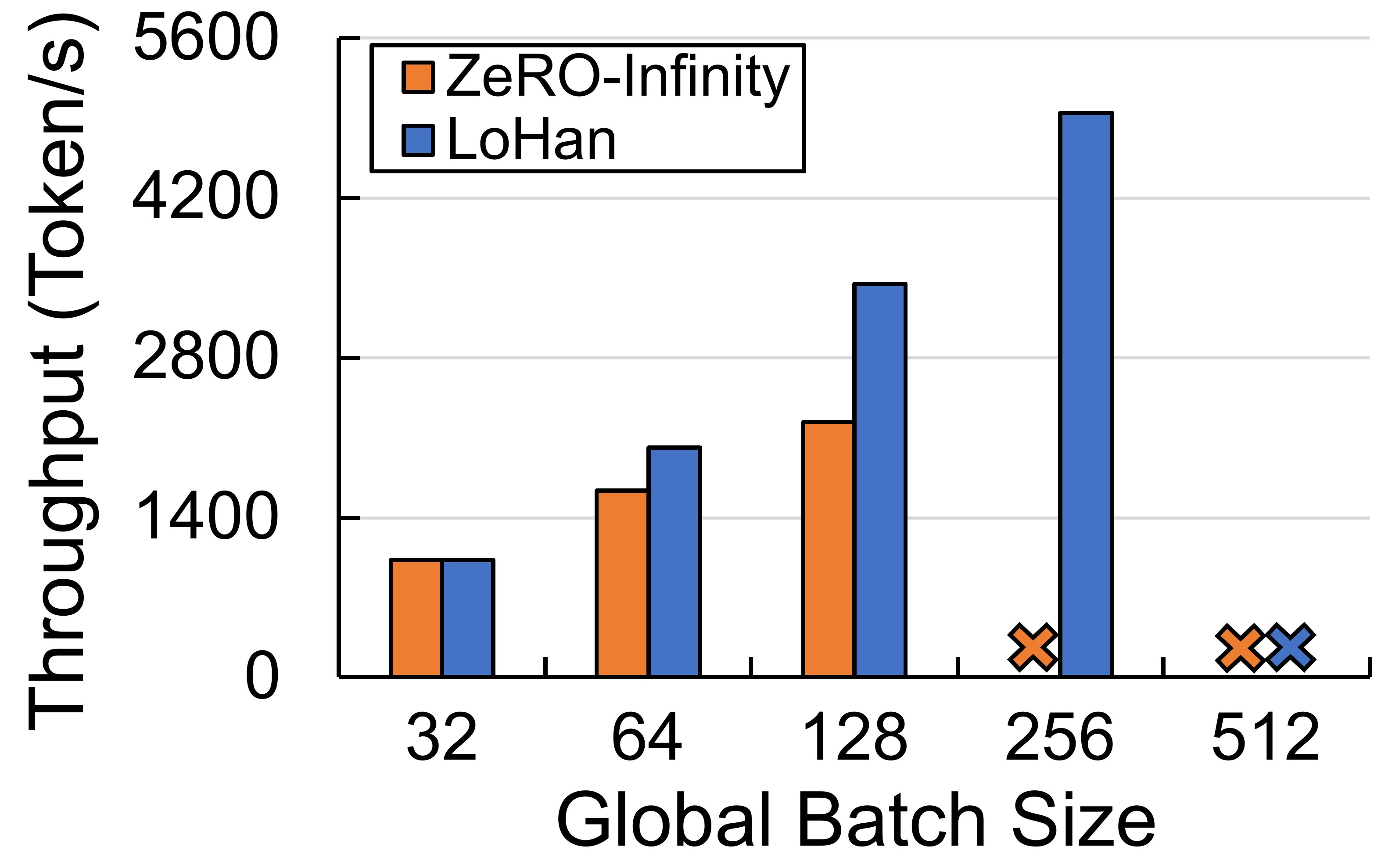}
    }
    \hfill
    \subfloat[70B model with 4 GPUs.]{
        \includegraphics[width=0.395\linewidth]{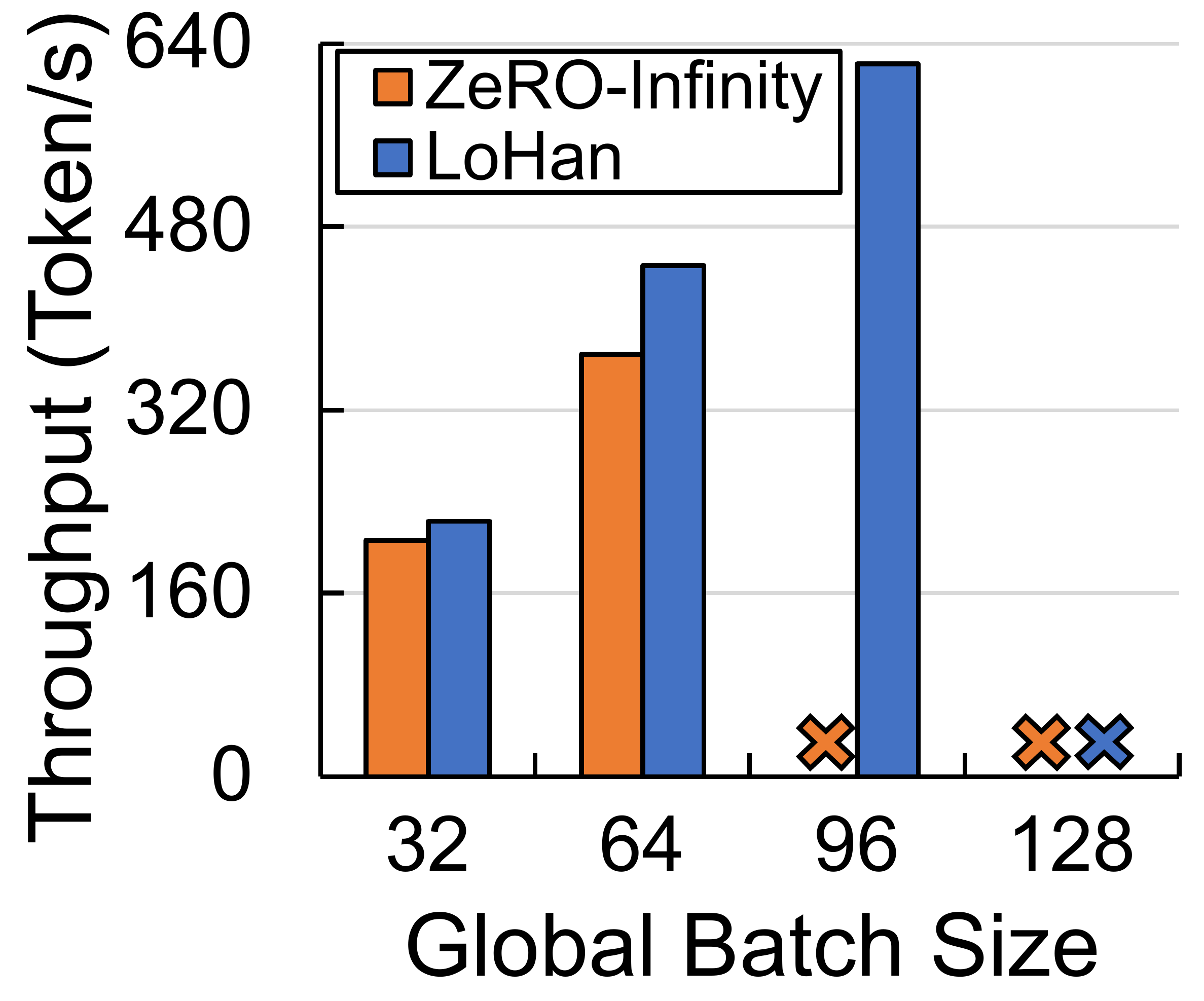}
    }
    \vspace{-1ex}
    \caption{Throughput comparison between \SystemName{} and ZeRO-Infinity on 4-GPU machine.} 
    \vspace{-4ex}
    \label{fig_exp_multigpu} 
\end{figure} 

\subsection{Performance on Multi-GPU Server}
\label{sec:exp_multi_gpu}

Many data scientists might own a commodity server with multiple consumer-grade GPUs. We show that \SystemName{}'s optimizations also work for the multi-GPU servers. We evaluate the training throughput of \SystemName{} and ZeRO-Infinity on a multi-GPU server, whose configurations are the same as the single 4090 server, except that the multi-GPU server equips 4 NVIDIA RTX 4090 GPUs (The maximum supported number within the server's power supply). We use the two systems to fine-tune the 13B model and the 70B model (The largest model ZeRO-Infinity can fine-tune\footnote{Even though ZeRO-Infinity can fine-tune the 135B model with a single RTX 4090, it can only fine-tune the 70B model on the multi-GPU server because of the additional GPU and main memory overhead introduced by multi-GPU synchronization and multiprocessing.}) at different global batch sizes. Figure~\ref{fig_exp_multigpu} shows the global throughput of the two systems fine-tuning on 2 and 4 RTX 4090 GPUs. 

The experimental result shows that \SystemName{} achieves higher throughput than ZeRO-Infinity. \SystemName{} achieves 2.21$\times$ and 1.69$\times$ throughput than ZeRO-Infinity when fine-tuning the 13B and 70B models on 4 GPUs respectively. The underlying reason is two-fold. First, \SystemName{} offloads the activations to SSDs, thus allowing the fine-tuning with a larger batch size. Second, \SystemName{} considers holistic offloading traffic as an optimization dimension, and thus achieves higher throughput than ZeRO-Infinity even with the same batch size. In conclusion, \SystemName{} still benefit the fine-tuning on a multi-GPU server. 

\begin{table} [t]
	\centering\footnotesize
    \caption{Diffusion models for evaluation.}
    \vspace{-2ex}  
	\label{tab:modelsize_dit}	
	\begin{tabular}{|c||c|c|c|}
		\hline
		\textbf{Model Size} & \textbf{\#Layers} & \textbf{\#Heads} & \textbf{Hidden Dimension}\\
		\hline
		\hline
            0.67B & 28 & 16 & 1152 \\
		\hline
            0.90B & 30 & 16 & 1280 \\
		\hline
		  1.4B & 32 & 16 & 1536 \\
		\hline
		  10B & 28 & 32 & 4096 \\
		\hline
		  20B & 40 & 40 & 5120 \\
		\hline
		  40B & 48 & 56 & 7168 \\
        \hline
	\end{tabular}
    \vspace{-1ex}
\end{table}

\begin{figure}[t]
    \centering
    \vspace{-2ex}
    \begin{minipage}{0.465\linewidth}
        \begin{center}
            \includegraphics[width=\linewidth]{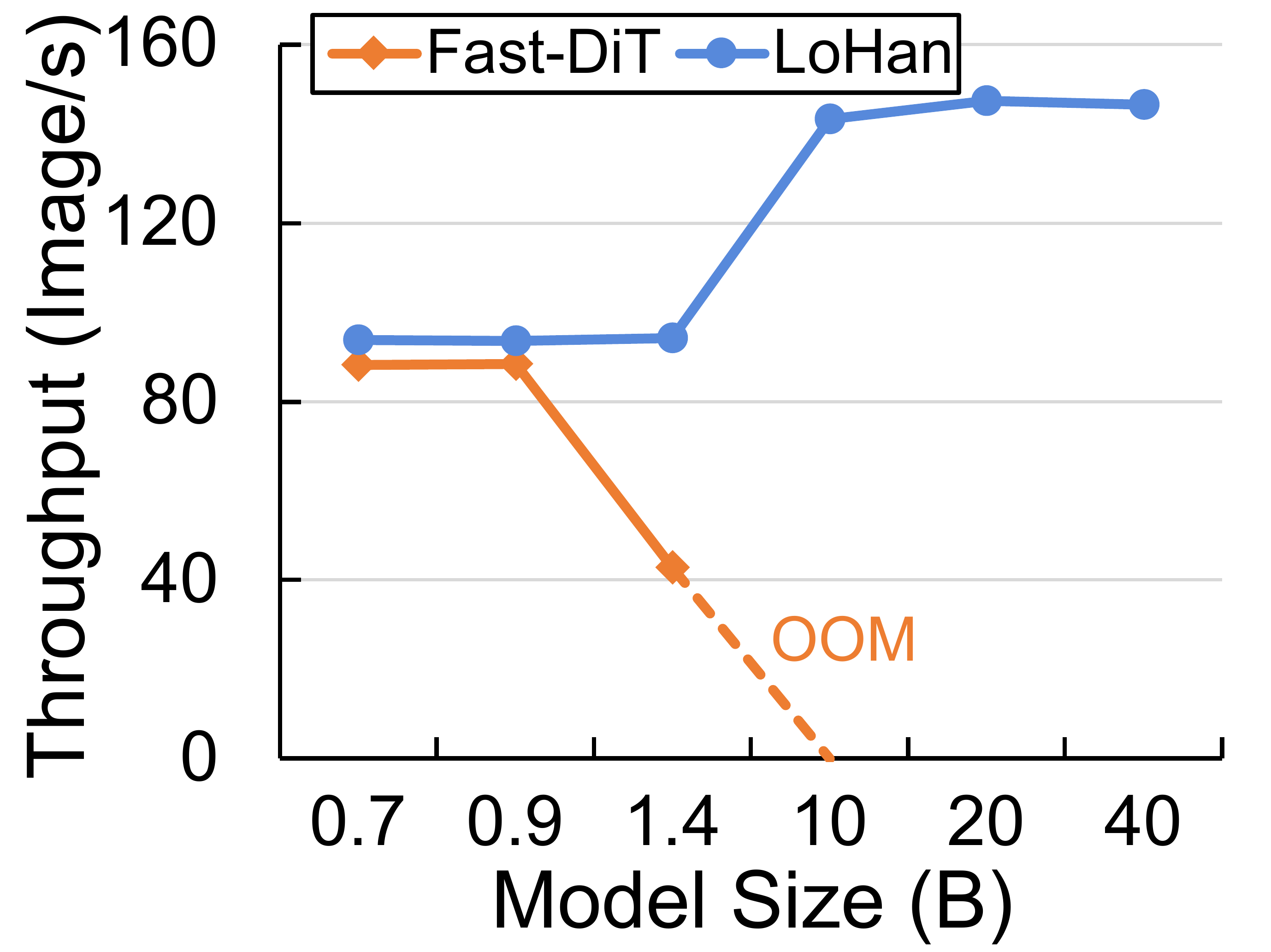}
        \end{center}
        \vspace{-2ex}
        \caption{\label{fig_exp_general_model_dit} Throughput on diffusion models: \SystemName{} vs. Fast-DiT.}
        \vspace{-4ex}
    \end{minipage} 
    \hfill
    \begin{minipage}{0.465\linewidth}
        \begin{center}
            \includegraphics[width=\linewidth]{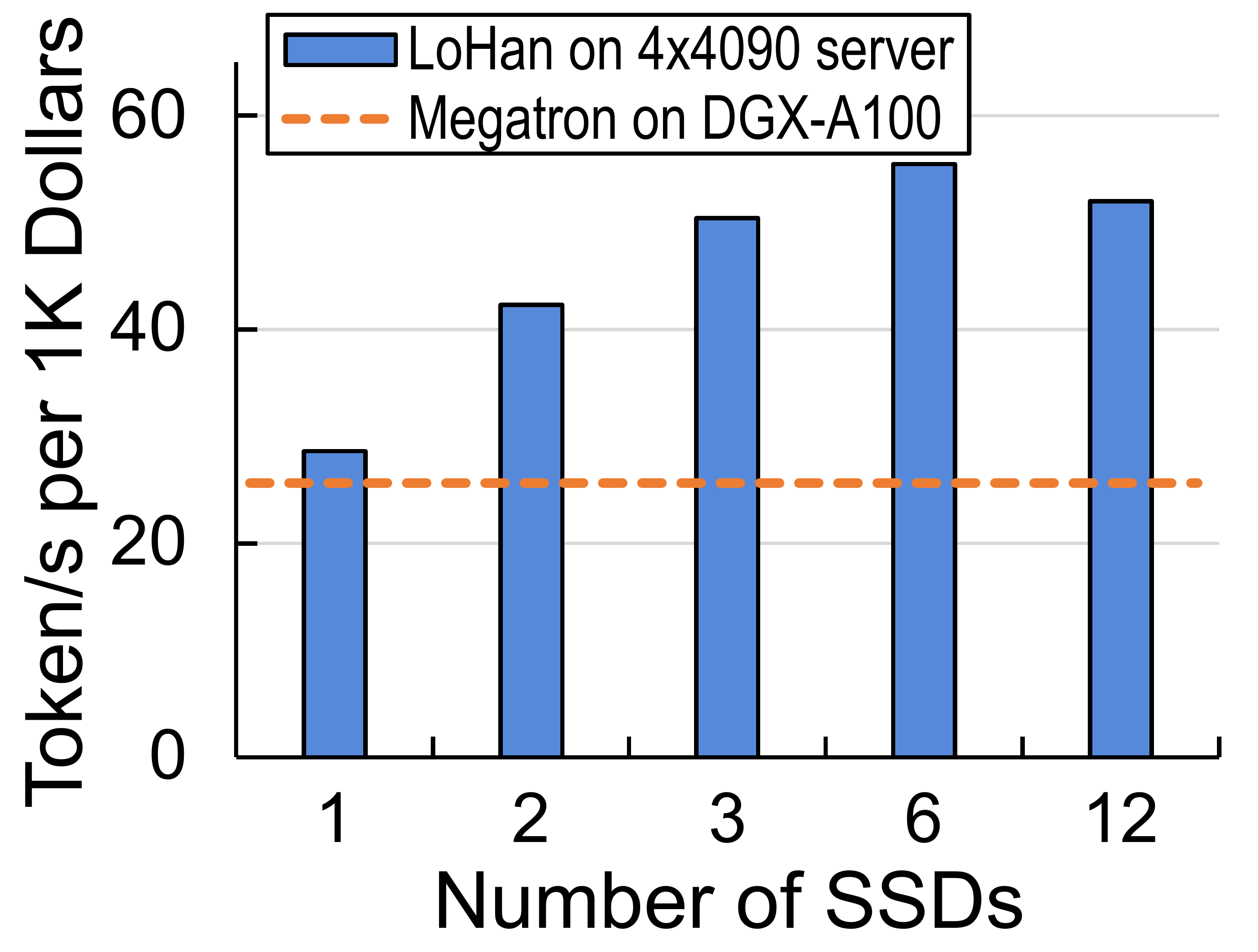}
        \end{center}
        \vspace{-2ex}
        \caption{\label{fig_exp_cost_efficiency} Comparison of Cost-effectiveness: \SystemName{} vs. Megatron on DGX-A100.}
        \vspace{-4ex}
    \end{minipage} 
\end{figure}

\vspace{-1ex}
\subsection{Performance on Other Large-Scale Models}
\label{sec:exp_general_models}

Large-scale models not only exist in language models (LMs) but also in models for tasks such as vision and image generation~\cite{dit, vit, swin-transformer, dalle}. 
We take fine-tuning diffusion models as an example to show that \SystemName{}'s optimizations are beneficial for more general large-scale deep-learning models. 

To evaluate performance on large diffusion models, we adopt the model architecture of the DiT-XL/2~\cite{dit} model and scale the layer number, attention head number, and hidden dimension of the backbone, as listed in Table~\ref{tab:modelsize_dit}. The input image size is 512$\times$512. We compare the throughput of \SystemName{} and Fast-DiT~\cite{fast-dit}, the state-of-the-art open-source training framework for DiT models. Figure~\ref{fig_exp_general_model_dit} illustrates the result. We make two observations. 

First, \SystemName{} enables fine-tuning much larger models than Fast-DiT because \SystemName{} offloads activations and model states to main memory and SSDs. In contrast, Fast-DiT keeps the tensors in GPU memory.

Second, \SystemName{} achieves higher throughput than Fast-DiT when fine-tuning the same model. The reason is two-fold: 1)~Fast-DiT suffers from low throughput due to small trainable batch size as the model size grows (e.g., 1.4B), while \SystemName{} allows the fine-tuning with high batch size; and 2)~\SystemName{}'s activation management strategy reduces the fine-tuning time compared to Fast-DiT's static activation swapping strategy, thus enabling \SystemName{} to achieve higher throughput even when both two systems fine-tune at the same large batch size.

\subsection{Cost-Effectiveness Comparison}
\label{sec:throughput_per_dollar}
\begin{table}[t]
	\centering\footnotesize
    \caption{Estimated price of components.}
    \vspace{-1ex}
    \label{tab:server-price}	
    \begin{tabular}{|c|c|}
        \hline
        \textbf{Machines and Components} & \textbf{Price (\$)}\\
        \hline
        \hline
            \narrow{\textbf{DGX-A100 server with 8 A100-80G NVLink GPUs}} & 200,000~\cite{mobius} \\
        \hline
            \narrow{\textbf{Commodity 4U server, without GPUs and SSDs}} & 14,098~\cite{supermicro-price} \\
        \hline
            \textbf{NVIDIA RTX 4090} & 1,600~\cite{4090} \\
        \hline
          \textbf{Intel P5510 SSD} & 308~\cite{supermicro-price} \\
        \hline
    \end{tabular}
    \vspace{-5ex}
\end{table}

To show the cost-effectiveness of utilizing holistic offloading in improving training throughput, we compare the throughput over the server price of \SystemName{} on the 4$\times$ RTX 4090 GPU server and a baseline: Megatron-LM~\cite{megatronlm} on an NVLink-enhanced DGX-A100~\cite{dgx} server using tensor parallelism. Megatron-LM does not rely on data offloading. We fine-tune the 30B model (the largest model Megatron-LM can fine-tune on the DGX machine) on the two systems. The prices of server components are estimated as shown in Table~\ref{tab:server-price}. We evaluate \SystemName{} on RTX 4090 GPUs with different SSD numbers when fine-tuning the 30B model.

Figure~\ref{fig_exp_cost_efficiency} illustrates the comparison results. We observe that \SystemName{} on RTX 4090 achieves at most 2.17$\times$ cost-effectiveness over Megatron-LM on a DGX-A100 machine. This shows that for large-scale training, \SystemName{} enables a commodity GPU to achieve higher cost-effectiveness than high-end data-center clusters that do not rely on offloading to train a huge model. 
Here cost-effectiveness decreases when \SystemName{}'s number of SSDs is increased from 6 to 12 because adding SSDs beyond the optimal number of SSDs has only a small performance gain but raises costs.
\section{Related Works}
\label{sec:related_works}

\noindent\textbf{Offloading Optimization in Data Management Tasks.}
\
Prior works~\cite{haas2023modern, kroviakov2024heterogeneous, lerner2024data, maschi2023difficult, alonso2023future, von2022you, wei2022much} study the characteristics of modern storage devices such as SSDs and provide guidelines for optimizing the data offloading in data management tasks.

Following these works, existing practices have optimized the storage I/O of several data management fields such as buffer management~\cite{lv2011operation, do2013fast, lru-c, papon2023aceing, hippogriffdb}, indexing~\cite{thonangi2017log, bushstore}, query scheduling~\cite{hippogriffdb, latte}, and transaction logging~\cite{latte, haubenschild2020rethinking} of rational databases, or other fields such as distributed databases~\cite{ziegler2022scalestore, fan2024tengine}, object databases~\cite{shedge2022extended}, key-value stores~\cite{chai2019ldc, dotori, do2021better, leaderkv, wang2024boosting}, vector data processing~\cite{huang2024neos}, graph data processing~\cite{papon2024enhancing, ginex, gts, gids, helios_ppopp25, helios_icde25, torchgt_sc24, legion_atc23}, and information retrieval~\cite{wang2021evaluating}.

Compared to these systems, \SystemName{} targets the LLM fine-tuning tasks that have different application characteristics.

\noindent\textbf{Tensor Management Methods for LLM Fine-tuning.}
\
Many existing works~\cite{chen2016training, checkmate, reforwarding, gruslys2016memory, herrmann2019optimal, beaumont2020optimal, kusumoto2019graph, dtr} consider the optimal recomputation strategies while keeping the rest of activations and the entire model states on GPU, thus these works fail to fine-tune even a 1B model. In contrast, \SystemName{} is the first to schedule activation offloading and activation recomputation under holistic activations and model states offloading. 

Many existing works~\cite{vdnn, superneurons, beaumont2021efficient, str, tflms, layrub, zhang2019efficient, vdnn++, beaumont2020optimal2, sentinel, defines, deepum} offloads the activations to main memory to train models that cannot fit in GPU memory. FlashNeuron~\cite{flashneuron} further offloads activations to NVMe SSDs. However, these systems do not offload model states, thus they even fail to fine-tune a 6B model, and only considering the optimal activation swapping and recomputation is not optimal when offloading both activations and model states. 
Further, some works~\cite{swapadvisor, capuchin, tsplit, poet} offloads both activations and model states to main memory, and G10~\cite{g10} further offloads these tensors to SSDs. However, these systems execute optimizer in GPU, thus incurring heavy model state transfer overhead on PCIe interconnect as discussed in Subsection~\ref{sec:motivation_gpu_optimizer}. 
In contrast, \SystemName{} considers both activation management and model states offloading to SSDs as optimization dimensions, thus enabling 100B-scale LLM fine-tuning on a single GPU while keeping efficiency.

Some prior works~\cite{zero-offload, stronghold, patrickstar, l2l, elixir} introduce CPU Adam and offload model states to main memory so as to enlarge the trainable model sizes of LLMs and ZeRO-Infinity~\cite{zero-infinity} offloads the model states to SSDs. However, these systems do not holistically manage activations swapping, activation swapping, and model state offloading. In particular, they either do not adopt activation recomputation thus incurring heavy activation transfer overhead on PCIe interconnect, or only adopts a naive activation recomputation strategy. Angel-PTM~\cite{angel-ptm} adopts asynchronous weight updates which leads to parameter staleness. In contrast, \SystemName{} presents the first active gradient offloading that overlaps CPU's SSD accesses, optimizer execution, and GPU's backward propagation to maximize GPU utilization. In contrast, \SystemName{} is the first to propose holistic traffic-aware activation swapping management that achieves optimal activation management under large model sizes, as shown in Subsection~\ref{sec:exp_swapping}.

\vspace{-1ex}
\section{Conclusion}
In this paper, we propose \SystemName{}, a low-cost high-performance training framework that enables efficient 100B huge model fine-tuning on a low-end server with a low-end GPU and limited main memory capacity. The key idea is to add holistic offloading traffic as an optimization dimension. 
The experimental results show that 1)~\SystemName{} is the first to fine-tune a 175B model on an RTX 4090 and 256 GB main memory, 
and 2)~\SystemName{} enables a cheap low-end consumer GPU to have higher cost-effectiveness than a DGX-A100 cluster when fine-tuning a 175B model. \SystemName{}'s artifact is available at \url{https://github.com/RC4ML/LoHan}.

\noindent\textbf{Acknowledgement.}
\
The work is supported by the following grants: 
the National Key R\&D Program of China (Grant No. 2022ZD0119301), 
the National Natural Science Foundation of China under the grant number (62472384, 62441605), 
Starry Night Science Fund of Zhejiang University Shanghai Institute for Advanced Study (SN-ZJU-SIAS-0010). Zeke Wang is the corresponding author. 








\bibliographystyle{IEEEtran}

\bibliography{references}

\begin{thebibliography}{100}
\providecommand{\url}[1]{#1}
\csname url@samestyle\endcsname
\providecommand{\newblock}{\relax}
\providecommand{\bibinfo}[2]{#2}
\providecommand{\BIBentrySTDinterwordspacing}{\spaceskip=0pt\relax}
\providecommand{\BIBentryALTinterwordstretchfactor}{4}
\providecommand{\BIBentryALTinterwordspacing}{\spaceskip=\fontdimen2\font plus
\BIBentryALTinterwordstretchfactor\fontdimen3\font minus \fontdimen4\font\relax}
\providecommand{\BIBforeignlanguage}[2]{{%
\expandafter\ifx\csname l@#1\endcsname\relax
\typeout{** WARNING: IEEEtran.bst: No hyphenation pattern has been}%
\typeout{** loaded for the language `#1'. Using the pattern for}%
\typeout{** the default language instead.}%
\else
\language=\csname l@#1\endcsname
\fi
#2}}
\providecommand{\BIBdecl}{\relax}
\BIBdecl

\bibitem{bert}
J.~Devlin, M.-W. Chang, K.~Lee, and K.~Toutanova, ``Bert: Pre-training of deep bidirectional transformers for language understanding,'' in \emph{NAACL-HLT}, 2019.

\bibitem{opt}
S.~Zhang, S.~Roller, N.~Goyal, M.~Artetxe, M.~Chen, S.~Chen, C.~Dewan, M.~Diab, X.~Li, X.~V. Lin, T.~Mihaylov, M.~Ott, S.~Shleifer, K.~Shuster, D.~Simig, P.~S. Koura, A.~Sridhar, T.~Wang, and L.~Zettlemoyer, ``Opt: Open pre-trained transformer language models,'' \emph{arXiv preprint}, 2022.

\bibitem{gpt2}
A.~Radford, J.~Wu, R.~Child, D.~Luan, D.~Amodei, and I.~Sutskever, ``Language models are unsupervised multitask learners,'' \emph{OpenAI blog}, 2019.

\bibitem{gpt3}
T.~Brown, B.~Mann, N.~Ryder, M.~Subbiah, J.~D. Kaplan, P.~Dhariwal, A.~Neelakantan, P.~Shyam, G.~Sastry, A.~Askell, S.~Agarwal, A.~Herbert-Voss, G.~Krueger, T.~Henighan, R.~Child, A.~Ramesh, D.~Ziegler, J.~Wu, C.~Winter, C.~Hesse, M.~Chen, E.~Sigler, M.~Litwin, S.~Gray, B.~Chess, J.~Clark, C.~Berner, S.~McCandlish, A.~Radford, I.~Sutskever, and D.~Amodei, ``Language models are few-shot learners,'' in \emph{NeurIPS}, 2020.

\bibitem{bloom}
T.~L. Scao, A.~Fan, C.~Akiki, E.~Pavlick, S.~Ili{\'c}, D.~Hesslow, R.~Castagn{\'e}, A.~S. Luccioni, F.~Yvon, M.~Gall{\'e}, J.~Tow, A.~M. Rush, S.~Biderman, A.~Webson, P.~S. Ammanamanchi, T.~Wang, B.~Sagot, N.~Muennighoff, A.~V. del Moral, O.~Ruwase, R.~Bawden, S.~Bekman, A.~McMillan-Major, T.~Wolf, I.~Beltagy, H.~Nguyen, L.~Saulnier, S.~Tan, P.~O. Suarez, V.~Sanh, H.~Lauren{\c{c}}on, Y.~Jernite, J.~Launay, M.~Mitchell, and C.~Raffel, ``Bloom: A 176b-parameter open-access multilingual language model,'' \emph{arXiv preprint}, 2022.

\bibitem{trummer2021case}
I.~Trummer, ``The case for nlp-enhanced database tuning: towards tuning tools that" read the manual",'' in \emph{VLDB}, 2021.

\bibitem{fernandez2023large}
R.~C. Fernandez, A.~J. Elmore, M.~J. Franklin, S.~Krishnan, and C.~Tan, ``How large language models will disrupt data management,'' in \emph{VLDB}, 2023.

\bibitem{sft}
L.~Ouyang, J.~Wu, X.~Jiang, D.~Almeida, C.~Wainwright, P.~Mishkin, C.~Zhang, S.~Agarwal, K.~Slama, A.~Ray, J.~Schulman, J.~Hilton, F.~Kelton, L.~Miller, M.~Simens, A.~Askell, P.~Welinder, P.~F. Christiano, J.~Leike, and R.~Lowe, ``Training language models to follow instructions with human feedback,'' in \emph{NeurIPS}, 2022.

\bibitem{llama2}
H.~Touvron, L.~Martin, K.~Stone, P.~Albert, A.~Almahairi, Y.~Babaei, N.~Bashlykov, S.~Batra, P.~Bhargava, S.~Bhosale, D.~Bikel, L.~Blecher, C.~C. Ferrer, M.~Chen, G.~Cucurull, D.~Esiobu, J.~Fernandes, J.~Fu, W.~Fu, B.~Fuller, C.~Gao, V.~Goswami, N.~Goyal, A.~Hartshorn, S.~Hosseini, R.~Hou, H.~Inan, M.~Kardas, V.~Kerkez, M.~Khabsa, I.~Kloumann, A.~Korenev, P.~S. Koura, M.-A. Lachaux, T.~Lavril, J.~Lee, D.~Liskovich, Y.~Lu, Y.~Mao, X.~Martinet, T.~Mihaylov, P.~Mishra, I.~Molybog, Y.~Nie, A.~Poulton, J.~Reizenstein, R.~Rungta, K.~Saladi, A.~Schelten, R.~Silva, E.~M. Smith, R.~Subramanian, X.~E. Tan, B.~Tang, R.~Taylor, A.~Williams, J.~X. Kuan, P.~Xu, Z.~Yan, I.~Zarov, Y.~Zhang, A.~Fan, M.~Kambadur, S.~Narang, A.~Rodriguez, R.~Stojnic, S.~Edunov, and T.~Scialom, ``Llama 2: Open foundation and fine-tuned chat models,'' \emph{arXiv preprint}, 2023.

\bibitem{llama3}
LlamaTeam, ``Thellama3herdofmodels,'' \url{https://ai.meta.com/research/publications/the-llama-3-herd-of-models/}, 2024.

\bibitem{h200}
NVIDIA, ``Nvidia h200 tensor core gpu,'' \url{https://www.nvidia.com/en-us/data-center/h200/}, 2023.

\bibitem{dgx}
------, ``Nvidia dgx platform,'' \url{https://www.nvidia.com/en-us/data-center/dgx-platform/}, 2023.

\bibitem{galvatron}
X.~Miao, Y.~Wang, Y.~Jiang, C.~Shi, X.~Nie, H.~Zhang, and B.~Cui, ``Galvatron: Efficient transformer training over multiple gpus using automatic parallelism,'' in \emph{VLDB}, 2022.

\bibitem{megatron}
M.~Shoeybi, M.~Patwary, R.~Puri, P.~LeGresley, J.~Casper, and B.~Catanzaro, ``Megatron-lm: Training multi-billion parameter language models using model parallelism,'' \emph{arXiv preprint}, 2020.

\bibitem{alpa}
L.~Zheng, Z.~Li, H.~Zhang, Y.~Zhuang, Z.~Chen, Y.~Huang, Y.~Wang, Y.~Xu, D.~Zhuo, E.~P. Xing, J.~E. Gonzalez, and I.~Stoica, ``Alpa: Automating inter- and {Intra-Operator} parallelism for distributed deep learning,'' in \emph{OSDI}, 2022.

\bibitem{het}
X.~Miao, H.~Zhang, Y.~Shi, X.~Nie, Z.~Yang, Y.~Tao, and B.~Cui, ``Het: Scaling out huge embedding model training via cache-enabled distributed framework,'' in \emph{VLDB}, 2022.

\bibitem{ps}
M.~Li, D.~G. Andersen, J.~W. Park, A.~J. Smola, A.~Ahmed, V.~Josifovski, J.~Long, E.~J. Shekita, and B.-Y. Su, ``Scaling distributed machine learning with the parameter server,'' in \emph{OSDI}, 2014.

\bibitem{fastflow}
T.~Um, B.~Oh, B.~Seo, M.~Kweun, G.~Kim, and W.-Y. Lee, ``Fastflow: Accelerating deep learning model training with smart offloading of input data pipeline,'' in \emph{VLDB}, 2023.

\bibitem{sdpipe}
X.~Miao, Y.~Shi, Z.~Yang, B.~Cui, and Z.~Jia, ``Sdpipe: A semi-decentralized framework for heterogeneity-aware pipeline-parallel training,'' in \emph{VLDB}, 2023.

\bibitem{flexmoe}
X.~Nie, X.~Miao, Z.~Wang, Z.~Yang, J.~Xue, L.~Ma, G.~Cao, and B.~Cui, ``Flexmoe: Scaling large-scale sparse pre-trained model training via dynamic device placement,'' \emph{PACMMOD}, 2023.

\bibitem{guo2021model}
Y.~Guo, Z.~Zhang, J.~Jiang, W.~Wu, C.~Zhang, B.~Cui, and J.~Li, ``Model averaging in distributed machine learning: a case study with apache spark,'' \emph{VLDBJ}, 2021.

\bibitem{flexps}
Y.~Huang, T.~Jin, Y.~Wu, Z.~Cai, X.~Yan, F.~Yang, J.~Li, Y.~Guo, and J.~Cheng, ``Flexps: Flexible parallelism control in parameter server architecture,'' in \emph{VLDB}, 2018.

\bibitem{jiang2017heterogeneity}
J.~Jiang, B.~Cui, C.~Zhang, and L.~Yu, ``Heterogeneity-aware distributed parameter servers,'' in \emph{SIGMOD}, 2017.

\bibitem{sketchml}
J.~Jiang, F.~Fu, T.~Yang, and B.~Cui, ``Sketchml: Accelerating distributed machine learning with data sketches,'' in \emph{SIGMOD}, 2018.

\bibitem{miao2021heterogeneity}
X.~Miao, X.~Nie, Y.~Shao, Z.~Yang, J.~Jiang, L.~Ma, and B.~Cui, ``Heterogeneity-aware distributed machine learning training via partial reduce,'' in \emph{SIGMOD}, 2021.

\bibitem{sandha2019database}
S.~S. Sandha, W.~Cabrera, M.~Al-Kateb, S.~Nair, and M.~Srivastava, ``In-database distributed machine learning: demonstration using teradata sql engine,'' in \emph{VLDB}, 2019.

\bibitem{zhang2021distributed}
Y.~Zhang, F.~Mcquillan, N.~Jayaram, N.~Kak, E.~Khanna, O.~Kislal, D.~Valdano, and A.~Kumar, ``Distributed deep learning on data systems: a comparative analysis of approaches,'' in \emph{VLDB}, 2021.

\bibitem{memflow}
N.~Band, ``Memflow: Memory-aware distributed deep learning,'' in \emph{SIGMOD}, 2020.

\bibitem{hydra}
K.~Nagrecha, ``Model-parallel model selection for deep learning systems,'' in \emph{SIGMOD}, 2021.

\bibitem{miao2024demystifying}
X.~Miao, Z.~Jia, and B.~Cui, ``Demystifying data management for large language models,'' in \emph{SIGMOD}, 2024.

\bibitem{angel-ptm}
X.~Nie, Y.~Liu, F.~Fu, J.~Xue, D.~Jiao, X.~Miao, Y.~Tao, and B.~Cui, ``Angel-ptm: A scalable and economical large-scale pre-training system in tencent,'' in \emph{VLDB}, 2023.

\bibitem{capuchin}
X.~Peng, X.~Shi, H.~Dai, H.~Jin, W.~Ma, Q.~Xiong, F.~Yang, and X.~Qian, ``Capuchin: Tensor-based gpu memory management for deep learning,'' in \emph{ASPLOS}, 2020.

\bibitem{mpress}
Q.~Zhou, H.~Wang, X.~Yu, C.~Li, Y.~Bai, F.~Yan, and Y.~Xu, ``Mpress: Democratizing billion-scale model training on multi-gpu servers via memory-saving inter-operator parallelism,'' in \emph{HPCA}, 2023.

\bibitem{zero-offload}
J.~Ren, S.~Rajbhandari, R.~Y. Aminabadi, O.~Ruwase, S.~Yang, M.~Zhang, D.~Li, and Y.~He, ``Zero-offload: Democratizing billion-scale model training,'' in \emph{ATC}, 2021.

\bibitem{vdnn}
M.~Rhu, N.~Gimelshein, J.~Clemons, A.~Zulfiqar, and S.~W. Keckler, ``vdnn: Virtualized deep neural networks for scalable, memory-efficient neural network design,'' in \emph{MICRO}, 2016.

\bibitem{superneurons}
L.~Wang, J.~Ye, Y.~Zhao, W.~Wu, A.~Li, S.~L. Song, Z.~Xu, and T.~Kraska, ``Superneurons: Dynamic gpu memory management for training deep neural networks,'' in \emph{PPoPP}, 2018.

\bibitem{flashneuron}
J.~Bae, J.~Lee, Y.~Jin, S.~Son, S.~Kim, H.~Jang, T.~J. Ham, and J.~W. Lee, ``Flashneuron: Ssd-enabled large-batch training of very deep neural networks,'' in \emph{FAST}, 2021.

\bibitem{mistral}
A.~Q. Jiang, A.~Sablayrolles, A.~Mensch, C.~Bamford, D.~S. Chaplot, D.~de~las Casas, F.~Bressand, G.~Lengyel, G.~Lample, L.~Saulnier, L.~R. Lavaud, M.-A. Lachaux, P.~Stock, T.~L. Scao, T.~Lavril, T.~Wang, T.~Lacroix, and W.~E. Sayed, ``Mistral 7b,'' \emph{arXiv preprint}, 2023.

\bibitem{zero-infinity}
S.~Rajbhandari, O.~Ruwase, J.~Rasley, S.~Smith, and Y.~He, ``Zero-infinity: Breaking the gpu memory wall for extreme scale deep learning,'' in \emph{SC}, 2021.

\bibitem{colossal-ai}
Z.~Bian, H.~Liu, B.~Wang, H.~Huang, Y.~Li, C.~Wang, F.~Cui, and Y.~You, ``Colossal-ai: A unified deep learning system for large-scale parallel training,'' in \emph{ICPP}, 2023.

\bibitem{haas2023modern}
G.~Haas and V.~Leis, ``What modern nvme storage can do, and how to exploit it: High-performance i/o for high-performance storage engines,'' in \emph{VLDB}, 2023.

\bibitem{lerner2021not}
A.~Lerner and P.~Bonnet, ``Not your grandpa's ssd: The era of co-designed storage devices,'' in \emph{SIGMOD}, 2021.

\bibitem{petrov2010building}
I.~Petrov, G.~G. Almeida, A.~P. Buchmann, and U.~Gr{\"a}f, ``Building large storage based on flash disks,'' in \emph{ADMS@VLDB}, 2010.

\bibitem{adam}
D.~P. Kingma and J.~Ba, ``Adam: A method for stochastic optimization,'' \emph{arXiv preprint}, 2014.

\bibitem{g10}
H.~Zhang, Y.~Zhou, Y.~Xue, Y.~Liu, and J.~Huang, ``G10: Enabling an efficient unified gpu memory and storage architecture with smart tensor migrations,'' in \emph{MICRO}, 2023.

\bibitem{gpudirect}
NVIDIA, ``Nvidia gpudirect: Enhancing data movement and access for gpus,'' \url{https://developer.nvidia.com/gpudirect}, 2011.

\bibitem{pytorch}
A.~Paszke, S.~Gross, S.~Chintala, G.~Chanan, E.~Yang, Z.~DeVito, Z.~Lin, A.~Desmaison, L.~Antiga, and A.~Lerer, ``Automatic differentiation in pytorch,'' in \emph{NIPS Autodiff Workshop}, 2017.

\bibitem{chen2016training}
T.~Chen, B.~Xu, C.~Zhang, and C.~Guestrin, ``Training deep nets with sublinear memory cost,'' \emph{arXiv preprint}, 2016.

\bibitem{stronghold}
X.~Sun, W.~Wang, S.~Qiu, R.~Yang, S.~Huang, J.~Xu, and Z.~Wang, ``Stronghold: fast and affordable billion-scale deep learning model training,'' in \emph{SC}, 2022.

\bibitem{active_message_isca92}
T.~Eicken, D.~Culler, S.~Goldstein, and K.~Schauser, ``Active messages: A mechanism for integrated communication and computation,'' in \emph{ISCA}, 1992.

\bibitem{colossal-gemini}
J.~Fang and Y.~You, ``Meet gemini: The heterogeneous memory manager of colossal-ai,'' \url{https://colossalai.org/docs/advanced_tutorials/meet_gemini/}, 2022.

\bibitem{patrickstar}
J.~Fang, Z.~Zhu, S.~Li, H.~Su, Y.~Yu, J.~Zhou, and Y.~You, ``Parallel training of pre-trained models via chunk-based dynamic memory management,'' \emph{TPDS}, 2022.

\bibitem{checkmate}
P.~Jain, A.~Jain, A.~Nrusimha, A.~Gholami, P.~Abbeel, J.~Gonzalez, K.~Keutzer, and I.~Stoica, ``Checkmate: Breaking the memory wall with optimal tensor rematerialization,'' in \emph{MLSys}, 2020.

\bibitem{dit}
W.~Peebles and S.~Xie, ``Scalable diffusion models with transformers,'' in \emph{ICCV}, 2023.

\bibitem{vit}
A.~Dosovitskiy, L.~Beyer, A.~Kolesnikov, D.~Weissenborn, X.~Zhai, T.~Unterthiner, M.~Dehghani, M.~Minderer, G.~Heigold, S.~Gelly, J.~Uszkoreit, and N.~Houlsby, ``An image is worth 16x16 words: Transformers for image recognition at scale,'' in \emph{ICLR}, 2021.

\bibitem{swin-transformer}
Z.~Liu, Y.~Lin, Y.~Cao, H.~Hu, Y.~Wei, Z.~Zhang, S.~Lin, and B.~Guo, ``Swin transformer: Hierarchical vision transformer using shifted windows,'' in \emph{ICCV}, 2021.

\bibitem{dalle}
M.~D.~M. Reddy, M.~S.~M. Basha, M.~M.~C. Hari, and M.~N. Penchalaiah, ``Dall-e: Creating images from text,'' \emph{UGC Care Group I Journal}, 2021.

\bibitem{fast-dit}
C.~Jin and S.~Xie, ``Fast-dit: Fast diffusion models with transformers,'' \url{https://github.com/chuanyangjin/fast-DiT}, 2024.

\bibitem{mobius}
Y.~Feng, M.~Xie, Z.~Tian, S.~Wang, Y.~Lu, and J.~Shu, ``Mobius: Fine tuning large-scale models on commodity gpu servers,'' in \emph{ASPLOS}, 2023.

\bibitem{supermicro-price}
Supermicro, ``Supermicro sys-420gp-tnr dual xeon scalable 4u gpu superserver,'' \url{https://store.supermicro.com/us_en/4u-gpu-superserver-sys-420gp-tnr.html}, 2023.

\bibitem{4090}
NVIDIA, ``Geforce rtx 4090,'' \url{https://www.nvidia.com/en-us/geforce/graphics-cards/40-series/rtx-4090/}, 2022.

\bibitem{megatronlm}
D.~Narayanan, M.~Shoeybi, J.~Casper, P.~LeGresley, M.~Patwary, V.~Korthikanti, D.~Vainbrand, P.~Kashinkunti, J.~Bernauer, B.~Catanzaro, A.~Phanishayee, and M.~Zaharia, ``Efficient large-scale language model training on gpu clusters using megatron-lm,'' in \emph{SC}, 2021.

\bibitem{kroviakov2024heterogeneous}
A.~Kroviakov, P.~Kurapov, C.~Anneser, and J.~Giceva, ``Heterogeneous intra-pipeline device-parallel aggregations,'' in \emph{DaMoN}, 2024.

\bibitem{lerner2024data}
A.~Lerner and G.~Alonso, ``Data flow architectures for data processing on modern hardware,'' in \emph{ICDE}, 2024.

\bibitem{maschi2023difficult}
F.~Maschi and G.~Alonso, ``The difficult balance between modern hardware and conventional cpus,'' in \emph{DaMoN}, 2023.

\bibitem{alonso2023future}
G.~Alonso, N.~Ailamaki, S.~Krishnamurthy, S.~Madden, S.~Sivasubramanian, and R.~Ramakrishnan, ``Future of database system architectures,'' in \emph{SIGMOD}, 2023.

\bibitem{von2022you}
L.~Von~Merzljak, P.~Fent, T.~Neumann, and J.~Giceva, ``What are you waiting for? use coroutines for asynchronous i/o to hide i/o latencies and maximize the read bandwidth!'' in \emph{ADMS@VLDB}, 2022.

\bibitem{wei2022much}
J.~Wei and X.~Zhang, ``How much storage do we need for high performance server,'' in \emph{ICDE}, 2022.

\bibitem{lv2011operation}
Y.~Lv, B.~Cui, B.~He, and X.~Chen, ``Operation-aware buffer management in flash-based systems,'' in \emph{SIGMOD}, 2011.

\bibitem{do2013fast}
J.~Do, D.~Zhang, J.~M. Patel, and D.~J. DeWitt, ``Fast peak-to-peak behavior with ssd buffer pool,'' in \emph{ICDE}, 2013.

\bibitem{lru-c}
B.~Lee, M.~An, and S.-W. Lee, ``Lru-c: Parallelizing database i/os for flash ssds,'' in \emph{VLDB}, 2023.

\bibitem{papon2023aceing}
T.~I. Papon and M.~Athanassoulis, ``Aceing the bufferpool management paradigm for modern storage devices,'' in \emph{ICDE}, 2023.

\bibitem{hippogriffdb}
J.~Li, H.-W. Tseng, C.~Lin, Y.~Papakonstantinou, and S.~Swanson, ``Hippogriffdb: Balancing i/o and gpu bandwidth in big data analytics,'' in \emph{VLDB}, 2016.

\bibitem{thonangi2017log}
R.~Thonangi and J.~Yang, ``On log-structured merge for solid-state drives,'' in \emph{ICDE}, 2017.

\bibitem{bushstore}
Z.~Wang, L.~Shou, K.~Chen, and X.~Zhou, ``Bushstore: Efficient b+ tree group indexing for lsm-tree in non-volatile memory,'' in \emph{ICDE}, 2024.

\bibitem{latte}
J.~Chu, Y.~Tu, Y.~Zhang, and C.~Weng, ``Latte: A native table engine on nvme storage,'' in \emph{ICDE}, 2020.

\bibitem{haubenschild2020rethinking}
M.~Haubenschild, C.~Sauer, T.~Neumann, and V.~Leis, ``Rethinking logging, checkpoints, and recovery for high-performance storage engines,'' in \emph{SIGMOD}, 2020.

\bibitem{ziegler2022scalestore}
T.~Ziegler, C.~Binnig, and V.~Leis, ``Scalestore: A fast and cost-efficient storage engine using dram, nvme, and rdma,'' in \emph{ICDE}, 2022.

\bibitem{fan2024tengine}
X.~Fan, S.~Yan, Y.~Huang, and C.~Weng, ``Tengine: A native distributed table storage engine,'' in \emph{ICDE}, 2024.

\bibitem{shedge2022extended}
S.~Shedge, N.~Sharma, A.~Agarwal, M.~Abouzour, and G.~Alu{\c{c}}, ``An extended ssd-based cache for efficient object store access in sap iq,'' in \emph{ICDE}, 2022.

\bibitem{chai2019ldc}
Y.~Chai, Y.~Chai, X.~Wang, H.~Wei, N.~Bao, and Y.~Liang, ``Ldc: a lower-level driven compaction method to optimize ssd-oriented key-value stores,'' in \emph{ICDE}, 2019.

\bibitem{dotori}
C.~Duffy, J.~Shim, S.-H. Kim, and J.-S. Kim, ``Dotori: A key-value ssd based kv store,'' in \emph{VLDB}, 2023.

\bibitem{do2021better}
J.~Do, I.~L. Picoli, D.~Lomet, and P.~Bonnet, ``Better database cost/performance via batched i/o on programmable ssd,'' \emph{VLDBJ}, 2021.

\bibitem{leaderkv}
Y.~Wang, J.~Yuan, S.~Wu, H.~Liu, J.~Chen, C.~Ma, and J.~Qin, ``Leaderkv: Improving read performance of kv stores via learned index and decoupled kv table,'' in \emph{ICDE}, 2024.

\bibitem{wang2024boosting}
Y.~Wang, J.~He, K.~Sun, Y.~Dong, J.~Chen, C.~Ma, A.~C. Zhou, and R.~Mao, ``Boosting write performance of kv stores: An nvm-enabled storage collaboration approach,'' in \emph{ICDE}, 2024.

\bibitem{huang2024neos}
Y.~Huang, X.~Fan, S.~Yan, and C.~Weng, ``Neos: A nvme-gpus direct vector service buffer in user space,'' in \emph{ICDE}, 2024.

\bibitem{papon2024enhancing}
T.~I. Papon, ``Enhancing data systems performance by exploiting ssd concurrency \& asymmetry,'' in \emph{ICDE}, 2024.

\bibitem{ginex}
Y.~Park, S.~Min, and J.~W. Lee, ``Ginex: Ssd-enabled billion-scale graph neural network training on a single machine via provably optimal in-memory caching,'' in \emph{VLDB}, 2022.

\bibitem{gts}
M.-S. Kim, K.~An, H.~Park, H.~Seo, and J.~Kim, ``Gts: A fast and scalable graph processing method based on streaming topology to gpus,'' in \emph{SIGMOD}, 2016.

\bibitem{gids}
J.~B. Park, V.~S. Mailthody, Z.~Qureshi, and W.-m. Hwu, ``Accelerating sampling and aggregation operations in gnn frameworks with gpu initiated direct storage accesses,'' in \emph{VLDB}, 2024.

\bibitem{helios_ppopp25}
J.~Sun, Z.~Shi, L.~Su, W.~Shen, Z.~Wang, Y.~Li, W.~Yu, W.~Lin, F.~Wu, J.~Zhou, and B.~He, ``Helios: Efficient distributed dynamic graph sampling for online gnn inference,'' in \emph{PPoPP}, 2025.

\bibitem{helios_icde25}
J.~Sun, M.~Sun, Z.~Zhang, J.~Xie, Z.~Shi, Z.~Yang, J.~Zhang, F.~Wu, and Z.~Wang, ``Hyperion: Optimizing ssd access is all you need to enable cost-efficient out-of-core gnn training,'' in \emph{ICDE}, 2025.

\bibitem{torchgt_sc24}
M.~Zhang, J.~Sun, Q.~Hu, P.~Sun, Z.~Wang, Y.~Wen, and T.~Zhang, ``Torchgt: A holistic system for large-scale graph transformer training,'' in \emph{SC}, 2024.

\bibitem{legion_atc23}
J.~Sun, L.~Su, Z.~Shi, W.~Shen, Z.~Wang, L.~Wang, J.~Zhang, Y.~Li, W.~Yu, J.~Zhou, and F.~Wu, ``Legion: Automatically pushing the envelope of {Multi-GPU} system for {Billion-Scale} {GNN} training,'' in \emph{ATC}, 2023.

\bibitem{wang2021evaluating}
J.~Wang, C.~Lin, Y.~Papakonstantinou, and S.~Swanson, ``Evaluating list intersection on ssds for parallel i/o skipping,'' in \emph{ICDE}, 2021.

\bibitem{reforwarding}
J.~Feng and D.~Huang, ``Optimal gradient checkpoint search for arbitrary computation graphs,'' in \emph{CVPR}, 2021.

\bibitem{gruslys2016memory}
A.~Gruslys, R.~Munos, I.~Danihelka, M.~Lanctot, and A.~Graves, ``Memory-efficient backpropagation through time,'' in \emph{NIPS}, 2016.

\bibitem{herrmann2019optimal}
J.~Herrmann, O.~Beaumont, L.~Eyraud-Dubois, J.~Hermann, A.~Joly, and A.~Shilova, ``Optimal checkpointing for heterogeneous chains: how to train deep neural networks with limited memory,'' \emph{TOMS}, 2024.

\bibitem{beaumont2020optimal}
O.~Beaumont, J.~Herrmann, G.~Pallez, and A.~Shilova, ``Optimal memory-aware backpropagation of deep join networks,'' \emph{Philosophical Transactions of the Royal Society A}, 2020.

\bibitem{kusumoto2019graph}
M.~Kusumoto, T.~Inoue, G.~Watanabe, T.~Akiba, and M.~Koyama, ``A graph theoretic framework of recomputation algorithms for memory-efficient backpropagation,'' in \emph{NeurIPS}, 2019.

\bibitem{dtr}
M.~Kirisame, S.~Lyubomirsky, A.~Haan, J.~Brennan, M.~He, J.~Roesch, T.~Chen, and Z.~Tatlock, ``Dynamic tensor rematerialization,'' \emph{arXiv preprint}, 2020.

\bibitem{beaumont2021efficient}
O.~Beaumont, L.~Eyraud-Dubois, and A.~Shilova, ``Efficient combination of rematerialization and offloading for training dnns,'' in \emph{NeurIPS}, 2021.

\bibitem{str}
Z.~Zong, L.~Lin, L.~Lin, L.~Wen, and Y.~Sun, ``Str: Hybrid tensor re-generation to break memory wall for dnn training,'' \emph{TPDS}, 2023.

\bibitem{tflms}
T.~D. Le, H.~Imai, Y.~Negishi, and K.~Kawachiya, ``Tflms: Large model support in tensorflow by graph rewriting,'' \emph{arXiv preprint}, 2018.

\bibitem{layrub}
H.~Jin, B.~Liu, W.~Jiang, Y.~Ma, X.~Shi, B.~He, and S.~Zhao, ``Layer-centric memory reuse and data migration for extreme-scale deep learning on many-core architectures,'' \emph{TACO}, 2018.

\bibitem{zhang2019efficient}
J.~Zhang, S.~H. Yeung, Y.~Shu, B.~He, and W.~Wang, ``Efficient memory management for gpu-based deep learning systems,'' \emph{arXiv preprint}, 2019.

\bibitem{vdnn++}
S.~Shriram, A.~Garg, and P.~Kulkarni, ``Dynamic memory management for gpu-based training of deep neural networks,'' in \emph{IPDPS}, 2019.

\bibitem{beaumont2020optimal2}
O.~Beaumont, L.~Eyraud-Dubois, and A.~Shilova, ``Optimal gpu-cpu offloading strategies for deep neural network training,'' in \emph{Euro-Par}, 2020.

\bibitem{sentinel}
J.~Ren, J.~Luo, K.~Wu, M.~Zhang, H.~Jeon, and D.~Li, ``Sentinel: Efficient tensor migration and allocation on heterogeneous memory systems for deep learning,'' in \emph{HPCA}, 2021.

\bibitem{defines}
L.~Mei, K.~Goetschalckx, A.~Symons, and M.~Verhelst, ``Defines: Enabling fast exploration of the depth-first scheduling space for dnn accelerators through analytical modeling,'' in \emph{HPCA}, 2023.

\bibitem{deepum}
J.~Jung, J.~Kim, and J.~Lee, ``Deepum: Tensor migration and prefetching in unified memory,'' in \emph{ASPLOS}, 2023.

\bibitem{swapadvisor}
C.-C. Huang, G.~Jin, and J.~Li, ``Swapadvisor: Pushing deep learning beyond the gpu memory limit via smart swapping,'' in \emph{ASPLOS}, 2020.

\bibitem{tsplit}
X.~Nie, X.~Miao, Z.~Yang, and B.~Cui, ``Tsplit: Fine-grained gpu memory management for efficient dnn training via tensor splitting,'' in \emph{ICDE}, 2022.

\bibitem{poet}
S.~G. Patil, P.~Jain, P.~Dutta, I.~Stoica, and J.~Gonzalez, ``Poet: Training neural networks on tiny devices with integrated rematerialization and paging,'' in \emph{ICML}, 2022.

\bibitem{l2l}
B.~Pudipeddi, M.~Mesmakhosroshahi, J.~Xi, and S.~Bharadwaj, ``Training large neural networks with constant memory using a new execution algorithm,'' \emph{arXiv preprint}, 2020.

\bibitem{elixir}
H.~Huang, J.~Fang, H.~Liu, S.~Li, and Y.~You, ``Elixir: Train a large language model on a small gpu cluster,'' \emph{arXiv preprint}, 2022.

\end{thebibliography}

\end{document}